\documentclass[a4paper,11pt]{article}
\pdfoutput=1

\usepackage{jheppub} 

\usepackage[T1]{fontenc} 

\usepackage{amssymb}\usepackage[normalem]{ulem} 
\usepackage{ulem}

\usepackage{graphicx} 
\usepackage{graphics}
\usepackage{epsfig}
\usepackage{color}
\usepackage{amssymb}
\usepackage{amsmath}
\usepackage{doi}
\usepackage{dsfont}
\usepackage{setspace}
\usepackage{float}
\usepackage{caption}
\usepackage{subcaption}
\usepackage{placeins} 
\usepackage{slashed}
\usepackage{booktabs}
\usepackage{multirow}
\usepackage{soul}
\usepackage{colortbl}
\usepackage{slashed}

\usepackage{tikz}

\def\WWbb{e^+\nu_e\,\mu^-\bar{\nu}_\mu\, b\bar{b}}

\title{SMEFT everywhere: a NLO study of $\boldsymbol{pp \to t\bar{t}H}$ with decaying tops}

\author[\, a]{Giuseppe Bevilacqua}
\author[\, b]{Minos Reinartz}
\author[\, b]{and Malgorzata Worek}

\affiliation[a]{Institute of Nuclear and Particle Physics, NCSR Demokritos, 15341 Agia Paraskevi, Greece}
\affiliation[b]{Institute for Theoretical Particle Physics and Cosmology, RWTH Aachen University, 52056 Aachen, Germany}

\emailAdd{bevilacqua@inp.demokritos.gr}
\emailAdd{minos.reinartz@rwth-aachen.de}
\emailAdd{worek@physik.rwth-aachen.de}

\abstract{
We present  the computation of the next-to-leading order QCD corrections to the $pp\to t\bar{t} H+X$ process in the di-lepton channel at the LHC, including relevant dimension-6 operators $({\cal O}_{t\phi}, \, {\cal O}_{\phi G},\, {\cal O}_{tG}, \, {\cal O}_{tW})$ from the Standard Model Effective Field Theory. In our studies, higher-order corrections and effective operators are consistently included in the production part of the process as well as in the top-quark decays. We perform a detailed study of linear, cross, and quadratic contributions and their uncertainties, including renormalisation group effects. Our findings are presented at the integrated and differential cross-section level for the LHC Run III center-of-mass energy of $\sqrt{s}=13.6$ TeV. Finally, we provide predictions for $pp\to t\bar{t} H+X$ with stable top quarks and compare them with the results in which top quarks are reconstructed from their decay products. We show that kinematic cuts, as well as higher-order effects and SMEFT operators in top-quark decays, are important and should be consistently considered together, because they have a significant impact on the shape of the standard observables measured for the $pp\to t\bar{t}H+X$ process at the LHC.}

\dedicated{\rm TTK-26-12, P3H-26-035}


\begin{document} 

\maketitle
\flushbottom

%
\section{Introduction}
\label{sec:introduction}
%

The discovery of the Higgs boson at the LHC completed the particle content of the Standard Model (SM) \cite{ATLAS:2012yve,CMS:2012qbp}. The Standard Model, while being a very successful theory, seems fundamentally incomplete when considered in a larger context. It cannot explain several phenomena in the universe, such as dark matter, dark energy and the matter-antimatter asymmetry. Many proposed extensions of the SM predict new particles at or above the TeV scale, often with strong couplings to the Higgs boson and top quark. A central goal of the LHC program is therefore to search for such \textit{new physics} beyond the SM (BSM). So far, however, no unambiguous new resonances have been observed, which suggests that if new particles exist, they are either too heavy to be directly produced or interact too weakly to show up as clear peaks in measured invariant‑mass spectra. In this situation, the most sensible way forward is to look for signs of new physics indirectly, through precise measurements of the SM processes. In this respect, the processes involving Higgs-boson and/or top-quarks might be the best candidates for revealing the small deviations caused by the heavy and yet undiscovered new physics.

An appropriate theoretical tool to describe such indirect new-physics effects in a model-independent manner is Standard Model Effective Field Theory (SMEFT). SMEFT assumes that at energies below some very high scale denoted by $\Lambda$, the only light degrees of freedom are the SM fields, and that the SM gauge symmetry is still realized. New physics effects, if present, are encoded in higher‑dimensional operators, which are  built from the SM fields respecting the SM symmetry. Assuming lepton and baryon number conservation, the SMEFT Lagrangian can be generally written as
\begin{equation}
{\cal L}_{\rm SMEFT} = {\cal L}_{\rm SM} + \sum_i \frac{C_i^{(6)}}{\Lambda^2}{{\cal O}_i^{(6)}}+\sum_j \frac{C_j^{(8)}}{\Lambda^4} {\cal O}_j^{(8)} + \cdots \,,
\end{equation}
where $\mathcal{O}_i^{(6)}$ and $\mathcal{O}_i^{(8)}$ are operators of dimension-6 and dimension-8, respectively, and $C_i^{(d)}$ with $d>4$ are dimensionless effective couplings, the so-called Wilson coefficients. Thus, SMEFT extends the SM Lagrangian by a complete set of higher‑dimensional operators, provides a systematic expansion in $1/\Lambda$, preserves gauge invariance and automatically enforces correlations between different observables.  As a result, SMEFT offers a unified framework for interpreting LHC measurements of multiple processes based on a common set of Wilson coefficients. This allows for global, model-independent constraints on new physics, while retaining sufficient flexibility to match the obtained results onto a large class of specific ultraviolet models if necessary. 

Among various processes measured at the LHC, the top-quark sector is an interesting place to study SMEFT.  The mass of the top quark  is close to the electroweak scale $m_t \sim v= (G_F \sqrt{2} )^{-1}$, whereas its coupling to the SM Higgs boson is $Y_t= \sqrt{2}m_t/v \approx 1$. Many BSM models, which address the electroweak symmetry breaking, naturalness, compositeness, etc., will modify the top-quark sector first. If there is heavy new physics coupled to the SM, it will certainly leave traces in the processes involving top quarks. Many effective operators in top-quark physics generate amplitudes that grow with energy like ${\cal M} \sim (Q^2/\Lambda^2) \,{\cal C}_i^{(6)}$, where $Q$ is the typical scale of the hard process. Consequently, high‑$p_T$ tails of various observables in the top-quark related processes, also in connection with the Higgs boson, are especially sensitive to SMEFT. However, this kinematic regime is extremely difficult to model, as off-shell effects, electroweak (EW)  Sudakov logarithms \cite{Frixione:2015zaa,Denner:2016wet} and non-factorisable effects make the $pp\to t\bar{t}H$ processes involving top-quark decays one of the most interesting and complex environments to search for new physics. Indeed, as has been shown in Ref. \cite{Stremmer:2021bnk}, these high‑$p_T$ tails that are sensitive to new physics effects \cite{Hermann:2022vit} are affected by full off-shell effects of top quarks  and massive gauge bosons \cite{Denner:2010jp,Bevilacqua:2010qb,Denner:2012yc,Denner:2015yca}.

Over the past decade, significant progress has been made in NLO QCD calculations in SMEFT for the processes involving top quarks. Specifically, $pp\to t\bar{t}$, $t$-channel single-top-quark production, $pp\to t\bar{t}+X$, where $X=\gamma,Z,H,t\bar{t}$ have been studied including dimension‑6 insertions, partly at NLO in QCD, for the key operators \cite{BessidskaiaBylund:2016jvp,Zhang:2016omx,Maltoni:2016yxb,deBeurs:2018pvs,Neumann:2019kvk,Hartland:2019bjb,Brivio:2019ius,Aoude:2022deh,Degrande:2024mbg}. A public implementation  in the Monte Carlo program \textsc{MadGraph5${}_{-}$aMC@NLO} \cite{Alwall:2014hca} exists, based on model files imported via the Universal Feynman Output (\textsc{UFO}) format \cite{Degrande:2011ua,Darme:2023jdn}, 
\textsc{SmeftSim} \cite{Brivio:2017btx,Brivio:2020onw} at LO and  \textsc{Smeft@NLO} \cite{Degrande:2020evl} at NLO.  A significant update to the one-loop generator \textsc{Gosam3.0} \cite{Braun:2025afl} has recently been released, which includes new features, in particular functionality that facilitates SMEFT calculations. Feynman rules \cite{Dedes:2017zog,Dedes:2023zws} as well as operator mixing effects \cite{Jenkins:2013zja,Jenkins:2013wua,Alonso:2013hga} have also been systematically examined \cite{Aoude:2022aro,DiNoi:2023onw,Bartocci:2024fmm,DiNoi:2025uhu,DiNoi:2025arz,DiNoi:2025tka,Duhr:2025zqw} and even implemented in several computer codes \cite{Celis:2017hod,Aebischer:2018bkb,Fuentes-Martin:2020zaz,DiNoi:2022ejg}. However, the main work has been done either only for stable top quarks or by considering only top quark decays \cite{Boughezal:2019xpp}. The interplay between SMEFT operators and higher-order QCD effects in top-quark decays combined with the production process has not been fully considered so far. Furthermore, the impact of phase-space cuts on the cross-section predictions in SMEFT has not yet been investigated in detail.

The purpose of this paper is to alleviate this situation by taking the first step towards simultaneously accounting for SMEFT in the production and top-quark decays for the $pp\to t\bar{t}H$ process. Using the narrow-width-approximation (NWA), we investigate the impact of effective operators and higher-order corrections on the cross-section predictions at the LHC for the following decay chain $pp \to t\bar{t}H\to W^+W^- \, b\bar{b} \, H \to e^+\nu_e \, \mu^- \bar{\nu}_\mu \, b\bar{b} \,H$. We focus on a minimal, realistic set of dimension-6 operators to investigate their potential impact if they appear simultaneously in the production of $t\bar{t}H$ and in top quark decays, and also to analyse the effects resulting from the presence of new operators only in decays. In addition, we will discuss the associated theoretical uncertainties arising from the missing higher-order terms, as well as operator mixing. To achieve these goals, we have developed a new version of the \textsc{Helac-NLO} Monte Carlo program called \textsc{Helac-Smeft}, which we will also describe in more detail in the paper. 

The paper is organised as follows. In Section \ref{sec:operators} we describe the dimension-6 SMEFT operators used in our study. In Section \ref{sec:framework} we outline changes and various cross-checks that need to be incorporated in the \textsc{Helac-NLO} framework. This section is divided into four parts: program structure, numerical benchmarks, checks of UV vertices and validation of the computational framework. Our computational setup is provided in Section \ref{sec:setup}.  Our LO and NLO QCD integrated cross-section results are discussed in Section \ref{sec:integrated_cross_sections}. Furthermore, in Section \ref{sec:integrated_cross_sections} we analyse the differences between the two case-studies considered, namely the $pp \to t\bar{t} H+X$ process with stable top quarks and the $pp  \to  t\bar{t}H +X \to W^+W^-\, b\bar{b}\, H +X  \to e^+\nu_e\,\mu^-\bar{\nu}_{\mu}\, b\bar{b} \,H+X$ process. Various differential cross-section distributions are provided and examined in Section \ref{sec:differential_cross_sections_fiducial}. To better understand the impact of fiducial cuts on the top quark kinematics in the presence of SMEFT effects, we present a comparison of the results for $pp\to t\bar{t}H+X$ and $pp \to e^+\nu_e\, \mu^-\bar{\nu}_{\mu}\, b\bar{b} \,H +X$ in Section \ref{sec:top_kin}. Finally, we summarise our results in Section \ref{sec:summary}, where we also provide our final conclusions.

%
\section{Effective dimension-6 operators in SMEFT}
\label{sec:operators}
%

We follow the \textsc{Smeft@NLO} convention \cite{Degrande:2020evl,Aguilar-Saavedra:2018ksv} that provides dimension-6 SMEFT operators in  the Warsaw basis after canonical normalisation \cite{Grzadkowski:2010es}. Furthermore, it implements five-flavour scheme and enforces $U(2)_q\times U(2)_u \times U(3)_d \times (U(1)_l \times U(1)_e)^3$ flavour symmetry in the fermion sector. The current implementation also imposes ${\cal CP}$ conservation. In our study, we focus only on a small number of dimension-6 operators. For the process $pp \to t\bar{t}H$ with stable top quarks, the dominant SMEFT effects come from the following set of effective operators $\left\{{\cal O}_{t\phi}, {\cal O}_{tG}, {\cal O}_{\phi G}, {\cal O}_G \right\}$ \cite{Maltoni:2016yxb,Aoude:2022aro,Maltoni:2024dpn}. The first one directly rescales the top-Higgs Yukawa coupling, the second one (the top quark chromomagnetic-dipole operator) modifies the $gt\bar{t}$ vertex and induces $gt\bar{t}H$ and $ggt\bar{t}H$ contact terms, the third one describes loop-induced interaction  between the gluon and Higgs generating $Hgg, Hggg$ contact terms, and the fourth one is the purely gluonic dimension‑6 operator that modifies the three- and four-gluon vertices and adds additional five- and six-gluon interactions. These effective operators have the following form,
\begin{equation}
\begin{split}
{\cal O}_{t \phi} &  =\left(\phi^\dagger\phi -\frac{v^2}{2}\right) \bar{Q} \,t\, \tilde{\phi} + h.c.  \,,\\[0.2cm]
{\cal O}_{t G}  & = i g_s  \left( \bar{Q} \,\sigma^{\mu\nu} \,T_a \,t\right) \tilde{\phi} \, G^a_{\mu\nu} + h.c. \,,\\[0.2cm]
{\cal O}_{\phi G} & = \left( \phi^\dagger \phi - \frac{v^2}{2}\right) G^{\mu\nu}_a \,G^a_{\mu\nu} \,, \\[0.2cm]
{\cal O}_G & = g_s \,f_{abc} \,G^{a}_{\mu \nu} \, G^{b \, \nu\rho} \,G^{\,c \, \mu}_\rho \,,  
\end{split}
\end{equation}
where $\phi$ is the Higgs doublet, $\tilde{\phi}=i\sigma^2\phi^\dagger$, $Q$ stands for the left-handed quark doublet of the third generation, $t$ for the right-handed top, $\sigma^{\mu\nu}=\frac{1}{2} (\gamma^\mu\gamma^\nu - \gamma^\nu\gamma^\mu)$, while $G^a_{\mu\nu}$ describes the field-strength tensor of the gluon. All field definitions and operator normalisations are specified in the \textsc{Definition} file that can be found on the \textsc{Smeft@NLO} webpage~\footnote{\texttt{https://cp3.irmp.ucl.ac.be/projects/feynrules/wiki/SMEFTatNLO}}. The ${\cal O}_G$ operator is rather special as it is already constrained by LHC multi-jet events \cite{Krauss:2016ely,Hirschi:2018etq}. In addition, it does not enter the renormalisation group flow of the other operator coefficients listed above and will therefore not be further investigated in this work. Many other operators can be considered as well, including four-fermion operators. The latter also contribute to the $pp\to t\bar{t}$ process and can be better constrained by 
LHC $t\bar{t}$ measurements \cite{Rosello:2015sck,Buckley:2015nca,Buckley:2015lku}.  

In the case of top-quark decays, the situation is similar, as different SMEFT operators can be considered \cite{Boughezal:2019xpp}. However, to study the $Wtb$ structures and spin correlations, we additionally consider only the ${\cal O}_{tW}$ operator, which is given by
\begin{figure}[t!]
    \centering
    \includegraphics[width=\linewidth]{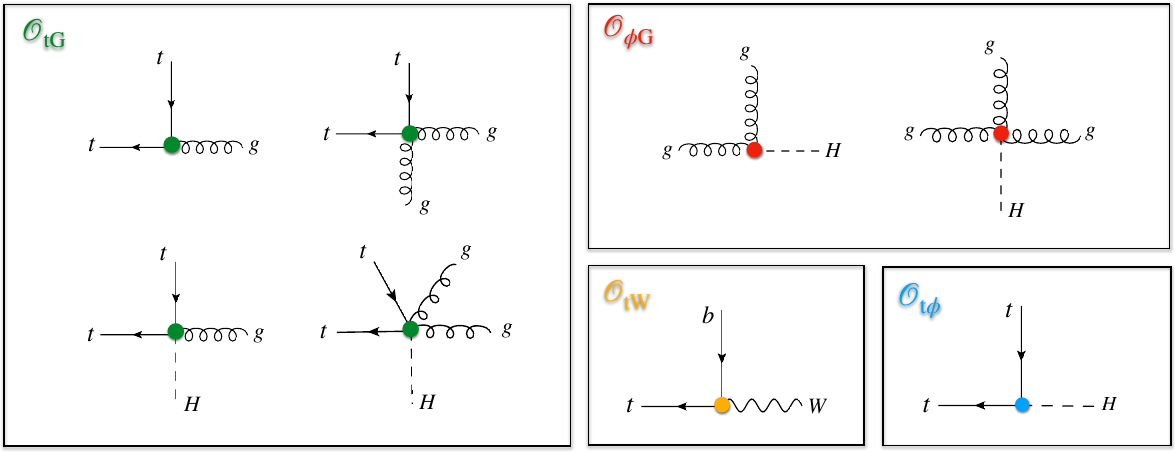}
\caption{Examples of effective vertices involved in the SMEFT calculation for the  $pp \to t\bar{t}H+X$ process in the di-lepton decay channel. The insertion operators are ${\cal O}_{t G}$, ${\cal O}_{\phi G}$, ${\cal O}_{t \phi}$ and  ${\cal O}_{tW}$.}
\label{fig:smeft_vertices}
\end{figure}
\begin{figure}[t!]
\centering
\includegraphics[width=0.9\linewidth]{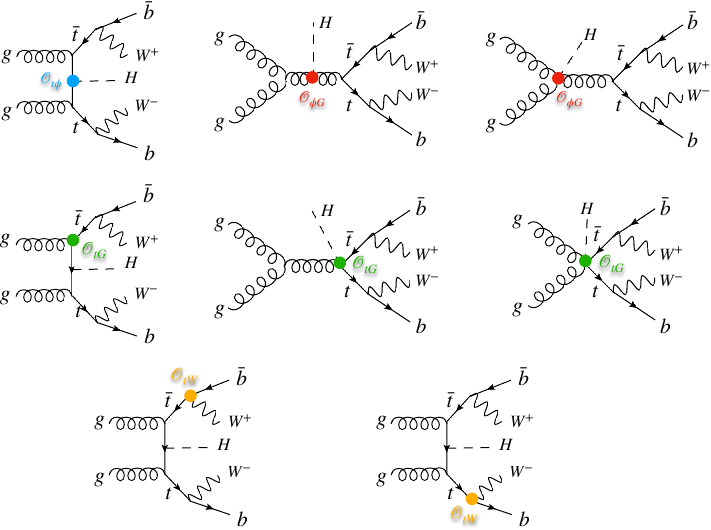}
\caption{Examples of effective-operator insertions of ${\cal O}_{t\phi}$, ${\cal O}_{\phi G}$,  ${\cal O}_{tG}$, ${\cal O}_{tW}$ entering $t\bar{t}H$ production and top-quark decays for the $pp \to t\bar{t}H+X$ process in the NWA. The leptonic decays of the $W$ gauge bosons considered in our study are omitted here for simplicity.}
\label{fig:ttH_tree}
\end{figure}
\begin{figure}[t!]
    \centering
    \includegraphics[width=0.75\textwidth]{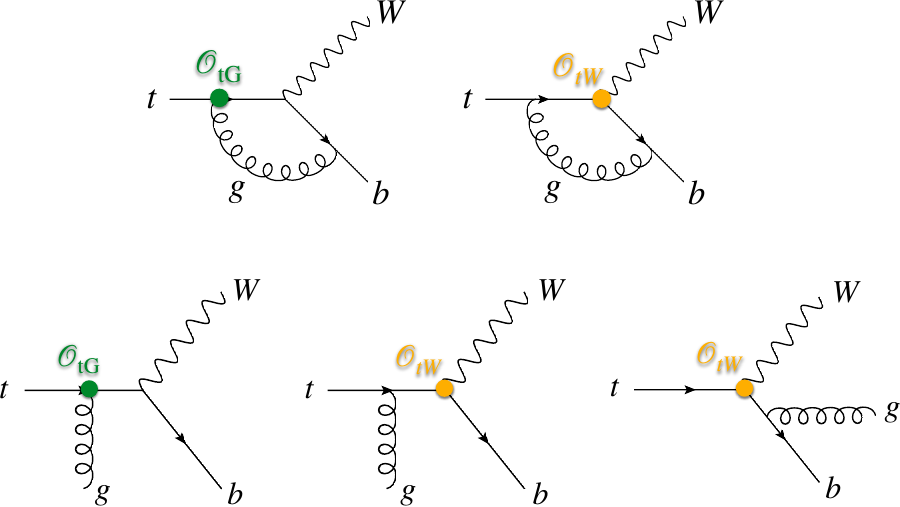}
    \label{fig:smeft_vertices_ttH_decay}
\caption{Examples of effective-operator insertions of ${\cal O}_{tG}$, ${\cal O}_{tW}$ entering top-quark decays for the $pp \to t\bar{t}H+X$ process in the NWA. The leptonic decays of the $W$ gauge bosons considered in our study are omitted here for simplicity.  }
\label{fig:ttH_decay_NLO}
\end{figure}
\begin{equation}
{\cal O}_{tW} = i\left(\bar{Q} \, \sigma^{\mu\nu} \,\sigma_I \, t\right) \tilde{\phi} \, W_{\mu\nu}^I + h.c.\,,
\end{equation}
where $\sigma_I$ are the Pauli matrices. Let us repeat at this stage that we are merely interested in understanding how including SMEFT effects in top-quark decays, and not just in production matrix elements, can affect the phenomenology of  the $pp\to t\bar{t}H$ process. In Figure \ref{fig:smeft_vertices} we show representative examples of the effective vertices that will appear in our study of the $pp\to t\bar{t}H$ process. Due to running and mixing with other dimension-6 operators at order $\alpha_s$ the following additional SMEFT operator, which enters the renormalisation of $\mathcal{O}_{tG}$, has to be in principle taken into account 
\begin{equation}
{\cal O}_{tB}=i\left(\bar{Q} \, \sigma^{\mu\nu} \, t\right)\tilde{\phi}\,B_{\mu\nu}+ h.c.\,.
\end{equation}
However, since we calculate NLO QCD corrections in the NWA, the ${\cal O}_{tB}$ operator will never enter into the calculations of LO matrix elements. Consequently, any counterterm generated by ${\cal O}_{tB}$ is also not needed in our case. If we instead wanted to consider full off-shell NLO QCD corrections to this process, the operator ${\cal O}_{tB}$ would have to be properly included. In this case, the Wilson coefficient of the $\mathcal{O}_{tB}$ operator, $C_{tB}$, could be replaced by $C_{tZ}$ with the help of the following relation 
\begin{equation}
C_{tZ}=-\sin\theta_W\,C_{tB}+\cos\theta_W\,C_{tW}\,,
\end{equation}
where $\theta_W$ is the weak mixing angle. In Figure \ref{fig:ttH_tree} we show examples of SMEFT effective-operator insertions entering the $pp\to t\bar{t}H$ process in the di-lepton channel when the NWA is used. For top-quark decays at the tree level we only have the ${\cal O}_{tW}$ operator. On the other hand, at NLO in QCD the decays also gain sensitivity to the $\mathcal{O}_{tG}$ operator, see Figure \ref{fig:ttH_decay_NLO}.

The evolution of the Wilson coefficients $C_i$ with the scale $\mu$ is driven by the renormalisation group equation (RGE) 
\begin{equation}
\label{eq:RGE_evolution_1}
\frac{dC_i}{d\log(\mu)}=\gamma_{ij}\,C_j \,,
\end{equation}
where $\gamma_{ij}$ is the anomalous dimension matrix. We consider the QCD running only and ignore terms in the anomalous dimension matrix that are not proportional to $\alpha_s$. At the level of QCD-induced running the latter can be expanded as
\begin{equation}
\gamma_{ij} = \sum_{k=1}^{\infty} \left(\frac{\alpha_s}{4\pi} \right)^k \gamma_{ij}^{{\rm QCD},\,k} \,.
\label{eq:anomalous_dimension_expasion}
\end{equation}
The solution to the RGE from Eq. \eqref{eq:RGE_evolution_1}, describing the evolution of the Wilson coefficients from a reference scale $\mu_0$ down to the scale $\mu$, is given by
\begin{equation}
    C_i(\mu) = \Gamma_{ij}(\mu,\mu_0)\,C_j(\mu_0)\,.
\end{equation}
The $\Gamma$ matrix can be evaluated by integrating the truncated RGE. For our case at one-loop order it is given by 
\begin{equation}
    \Gamma^{\rm QCD, \, 1}(\mu,\mu_0) = \exp\left(\int_{\mu_0}^{\mu}\frac{\alpha_s(\mu^\prime)}{4\pi\mu^\prime} \,d\mu^\prime\, \gamma^{\rm QCD, \, 1}\right)\,.
\end{equation}
It can be further rewritten using the one-loop beta function as follows 
\begin{equation}
    \Gamma^{\rm QCD, \, 1}(\mu,\mu_0) = \exp\left(\frac{1}{2\beta_0^{(5)}}\log \left(\frac{\alpha_s(\mu_0)}{\alpha_s(\mu)}\right)\,\gamma^{\rm QCD, \, 1}\right)\,,
\label{eq:RGE_matrix}
\end{equation}
where  $\beta^{(n_F)}_0=11 -2/3 \,n_F$ is the one-loop beta function coefficient and the number of light flavours is set to $n_F=5$. In practice, the numerical values of the Wilson coefficients are set at some reference scale $\mu_0$. Then the effective renormalisation scale $\mu=\mu_{EFT}$ can be chosen and the running to this scale setting is performed using Eq. \eqref{eq:RGE_matrix}. The effective scale $\mu_{EFT}$ can be defined either as a fixed or dynamic scale. 

Restricting to the effective operators considered in our analysis, the leading term of Eq. \eqref{eq:anomalous_dimension_expasion} reads \cite{Aoude:2022aro}
\begin{equation}
\gamma^{\textrm{QCD},1}=\frac{1}{3}\left(
\begin{matrix}
-24 & 96 \, Y_t          & 96 \, Y_t^2                        & 0 & 0\\[0.2cm]
 0 & -6 \,\beta_0^{(6)} & 12 \, Y_t                          & 0 & 0\\[0.2cm]
 0 & 0               & 4                              & 0 & 0\\[0.2cm]
 0 & 0               & 8\,g                           & 8 & 0\\[0.2cm]
 0 & 0               & 8\,g \cos\theta_W - \frac{40}{3} \,g^\prime \sin \theta_W & 0 & 8 \\[0.2cm]
\end{matrix}
\right)\,,
\label{eq:anomalous_dimension_matrix}
\end{equation}
where the Wilson coefficients are ordered as $C_i = (C_{t\phi}, \, C_{\phi G}, \, C_{tG}, \, C_{tW}, \, C_{tZ})$, $Y_t$ is the top-Higgs Yukawa coupling, while $g^\prime$ and $g$ are  the gauge coupling constants for the groups $U(1)_Y$ and $SU(2)_L$, respectively. Because  $\gamma^{\textrm{QCD},1}$ is not diagonal in general, different operators mix under renormalisation group (RG) evolution. We have included the $C_{tZ}$ coefficient, even if in practice it is not taken into account in our calculations, to show how it mixes with the relevant operators in our analysis. The mixing pattern of Wilson coefficients can be easily understood from Eq. \eqref{eq:anomalous_dimension_matrix}. For example, the Wilson coefficient $C_{tG}$ is the only coefficient that does not mix with the others during the RG evolution. On the other hand, $C_{tZ}$ mixes with $C_{tG}$, but not vice versa.

%
\section{Changes in the \textsc{Helac-NLO} framework}
\label{sec:framework}
%

The results presented in this paper are generated using the \textsc{Helac-Smeft} Monte Carlo program, which is an updated version of the \textsc{Helac-NLO} package \cite{Bevilacqua:2011xh}.  The \textsc{Helac-NLO} framework consists of \textsc{Helac-1loop} \cite{vanHameren:2009dr} and \textsc{Helac-Dipoles} \cite{Czakon:2009ss,Bevilacqua:2013iha}, but relies heavily on \textsc{CutTools} \cite{Ossola:2007ax}, \textsc{OneLOop} \cite{vanHameren:2010cp} and \textsc{Kaleu} \cite{vanHameren:2010gg} as cornerstones. To enable SMEFT in \textsc{Helac-NLO}, significant modifications had to be made to both \textsc{Helac-1loop} and \textsc{Helac-Dipoles}. Below we describe the main structural changes introduced compared to the \textsc{Helac-NLO} code and comment on the most important optimizations we had to implement.

When designing \textsc{Helac-Smeft}, the aim of which is to study the phenomenology of SMEFT with NLO QCD accuracy, we set ourselves the following goals:
\begin{itemize}
\item Create a program that is flexible in the number of effective operators that can be included.
\item Calculate matrix elements with effective operators for unstable particles either with full off-shell effects included or in the NWA, preserving  spin correlations.  
\item Consistently take into account the contributions of the appropriate effective operators and higher-order QCD corrections in the production part of the process and in the decays of unstable particles.
\item  Calculate in a single run the complete SMEFT cross-section result as well as all partial contributions from effective operators, i.e. all linear and quadratic terms in the operator coefficients.
\item Define the  effective scale $\mu_{EFT}$ either as a constant value, e.g.   $\mu_{EFT}=m_t$, or as some value that depends on the process kinematics, which changes  from one phase-space point to another.
\item Store and reuse the obtained results to provide theoretical predictions for different renormalisation, factorisation and effective scale settings and different PDF sets.
\end{itemize}

With these goals in mind, automation becomes a necessity. The original \textsc{Helac-NLO} framework is fully automated for recursive amplitude computations in the SM. In fact, the built-in vertex functions and generation algorithms for tree-level and one-loop topologies are in many respects tailored to the SM. Although extending existing functionalities by hand is technically possible once the Lorentz and colour structures of the desired effective vertices are understood \cite{Hermann:975160}, the practical implementation can be tedious and error-prone. To give the reader an idea of the complexity, note that for the process $pp\to t\bar{t}H$ considered in this analysis with four dimension-6 operators in addition to the SM contributions, the total number of vertex functions that need to be considered in the NLO QCD calculation is 1568 (split into 256 LO-type, 735 UV-type, and 577 $R_2$-type). These exact numbers are extracted from the
\textsc{Smeft@NLO UFO} model. Of course, it is possible to combine the latter based on common analytical structures, but this requires non-trivial (operator-dependent) classification work. Not to mention that some vertices (especially those of $R_2$-type) may have cumbersome analytical expressions that are difficult to read. A practical approach to solve this problem, used by several matrix-element generators to date, is to automate the entire vertex-generation chain for a given input model. We decided to adopt the same strategy. We further note that the \textsc{Helac-NLO} program can already handle the higher-order calculations with full off-shell effects included and the computations in the NWA in a fully automated manner \cite{Bevilacqua:2019quz}. The same applies to the dynamical $\mu_R$ and $\mu_F$ scale setting. Extending all these functionalities to the SMEFT case is rather straightforward, although time-consuming.

An important issue is the optimization of the procedure for obtaining the cross-section contributions from individual effective operators. In most cases, analysing the results of partial cross-section contributions is more interesting than analysing the full SMEFT results. For example, to assess which phase-space regions are most sensitive to different SMEFT operators, or to highlight any kinematic differences between them. Disentangling partial SMEFT contributions can be achieved either at the cost of CPU time or of increased structural complexity. Normally, a cross section integrator keeps track of a single weight during Monte Carlo integration. This weight corresponds to the full amplitude. Accessing partial SMEFT cross sections in this approach typically requires to run $N\,(N+1)/2$ samples (where $N$ is the number of effective operators considered, assuming that a different coupling is given to each operator). An alternative approach is possible if one  keeps track of weights from partial SMEFT amplitudes,  besides the full one, inside the program. The larger internal bookkeeping is compensated by the fact that all $N\,(N+1)/2$ partial cross sections are accessible within a single run. We decided to adopt the latter approach.

The last point on the list is primarily intended to facilitate the estimation of theoretical uncertainties. At the partonic level that we consider in our analysis, the dominant sources of theory uncertainty stem from renormalisation/factorisation scale dependence as well as from the PDF parametrisation. When working in the SMEFT framework, another important aspect to be considered is the evolution of the operator coefficients with the effective scale. The evolution itself, governed by the RGE, can be treated using different levels of approximation, either based on analytic solutions or numerical codes. Using a brute force method to address all of the above sources of uncertainties could potentially require hundreds of dedicated runs and would not be an optimal and practical solution for particularly complex processes. A more convenient alternative  is offered by event reweighting, which requires less CPU resources and scales to arbitrary processes with reasonable accuracy.  The \textsc{Helac-NLO} framework already has this option built-in for the SM case. In that case, we store the obtained results  in the modified \textsc{Les Houches Files} \cite{Alwall:2006yp}, which are subsequently converted into the \textsc{Root Ntuple Files} \cite{Antcheva:2009zz} based on the ideas introduced in Ref. \cite{Bern:2013zja}. This approach offers a significant advantage as it allows to efficiently reweight all the results. With this approach we can  generate  differential cross-section distributions for new observables, perform their rebinning or generate theoretical predictions for various scenarios, including different PDF sets, other scale settings, and various (more exclusive) fiducial cuts, all without the need to rerun the entire process from scratch. All the above tasks are carried out using the \textsc{HEPlot} program \footnote{In-house on-the-fly reweighting program to obtain results for another scale setting and PDF set.}. To extend this functionality also to the SMEFT case would be very beneficial. This is possible provided that the generated event samples contain all the necessary additional  matrix-element information.

%
\subsection{Program structure}
\label{sec:program_structure}
%

\begin{figure}[h!]   
        \centering
        \begin{tikzpicture}
            \fill[ gray!15, rounded corners = 10pt ] (-1.75,-3) rectangle (8.2,3);
            \fill[ gray!15, rounded corners = 10pt ] (8.3,-3) rectangle (11.7,3);
            \fill[ gray!15, rounded corners = 10pt ] (11.8,-3) rectangle (15.5,3);
            \node at ( 3,2.65) {\textsc{python}};
            \node at (10,2.65) {\textsc{Fortran}};
            \node at (13.75,2.65) {\textsc{C++}};    
            \node[draw, rectangle, rounded corners, align=center       ] (ufo)  at (0, +1.5) {UFO model};
            \node[draw, rectangle, rounded corners, blue, align=center ] (frg)  at (0, -1.5) {in-house\\ Feynman rules \\ generator};            
            \node[draw, rectangle, rounded corners, blue ] (ecru) at (3, 0) {\textsc{ECRU}};
            \node[draw, rectangle, rounded corners, black] (form) at (3, 2) {\textsc{Form}};
            \node[draw, rectangle, rounded corners, black] (sympy)at (3,-2) {\textsc{SymPy}};
            \node[draw, rectangle, rounded corners, blue ] (hcg)  at (6, 0) {\textsc{HModelGenerator}};
            \node[draw, rectangle, rounded corners, blue ] (dipl) at (10, 2) {\textsc{Helac-Dipoles}};
            \node[draw, rectangle, rounded corners, blue ] (loop) at (10,-2) {\textsc{Helac-1Loop}};
            \node[draw, rectangle, rounded corners, blue ] (hepl) at (14,0) {\textsc{HEPlot}}; 
            \node[draw, rectangle, rounded corners, black] (root) at (14,-2) {\textsc{Root}};   
            \path[line width = 0.35mm, ->, shorten >= 3pt ] (ufo)   edge node[left, align=center, pos=0.8, xshift=-20pt] {\small Feynman\\ rules} (ecru);
            \path[line width = 0.35mm, <->, shorten >=2pt, shorten <=2pt] (form)  edge (ecru);
            \path[line width = 0.35mm, <->, shorten >=2pt, shorten <=2pt] (sympy) edge (ecru);
            \path[line width = 0.35mm, ->, shorten >= 3pt ] (frg)   edge (ecru);
            \path[line width = 0.35mm, ->, shorten >= 2pt ] (ecru)  edge (hcg);
            \path[line width = 0.35mm, ->, shorten >= 4pt ] (hcg)   edge node[ left, pos=0.55, xshift = -10pt ] {\small vertices} (dipl);
            \path[line width = 0.35mm, ->, shorten >= 4pt ] (hcg)   edge node[ left, pos=0.55, xshift = -10pt ] {\small vertices} (loop);
            \path[line width = 0.35mm, ->, shorten >= 4pt ] (dipl)  edge node[ right, pos = 0.45, xshift = +10pt ] {\small events} (hepl);
            \path[line width = 0.35mm, ->, shorten >= 4pt ] (loop)  edge node[ right, pos = 0.45, xshift = +10pt ] {\small events} (hepl);
            \path[line width = 0.35mm, <->, shorten >=2pt, shorten <=2pt] (hepl) edge (root);           
        \end{tikzpicture}
        \caption{Schematic representation of the \textsc{Helac-Smeft} framework and its  workflow. Newly developed modules and updated parts of the \textsc{Helac-NLO} program are highlighted in blue.}
        \label{fig:ufo_to_helac_interface}
\end{figure}
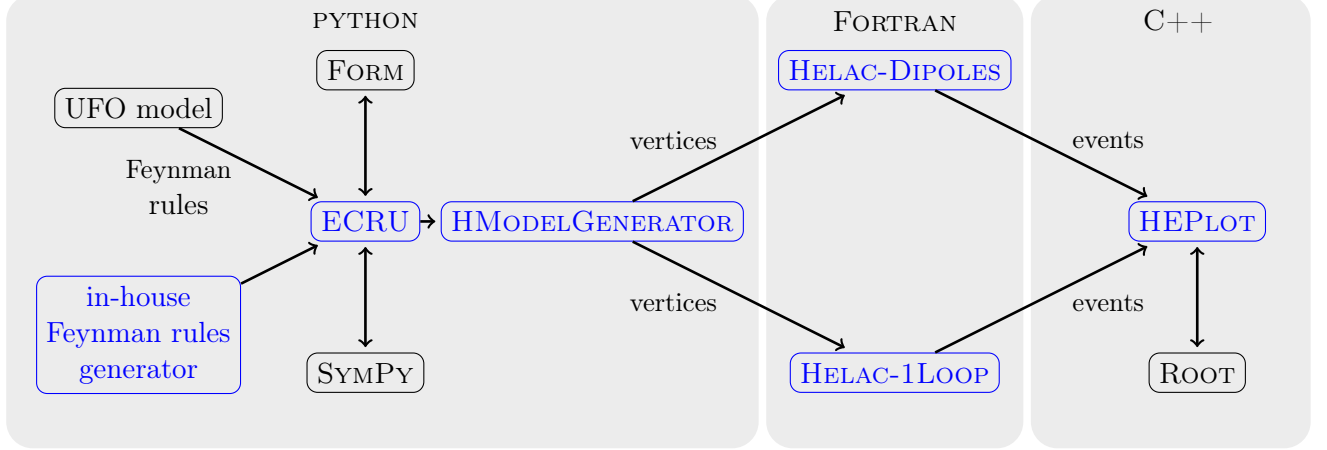

The structure of the \textsc{Helac-Smeft} framework is reported schematically in Figure \ref{fig:ufo_to_helac_interface}. We use three main blocks to help the reader identify the different tasks the various components are designed for. The block on the left concerns the automated generation of \textit{vertex functions} for a given input model. Vertex functions, $V_X$, form the basis for recursive amplitude calculations. They take a set of input currents, $J_1,\dots,J_N$, and return a new current as output, $J_0$. Schematically we can write
\begin{equation}
J_0 = V_X(J_1,J_2,\dots,J_N) \,.
\end{equation}
The output current represents a sub-amplitude and, in turn, serves as input data for the next stage of the recursion. The recursion is repeated until the full amplitude is constructed, see e.g. Refs. \cite{Kanaki:2000ey,Cafarella:2007pc} for more details. 

Particular attention should be paid to the treatment of colour degrees of freedom in QCD. The \textsc{Helac-Smeft} framework uses the so-called colour-flow representation \cite{tHooft:1973alw}. This concept can be better illustrated using the example of a four-gluon vertex in QCD. The starting point is the Feynman rule,
\begin{align}
\label{eq:4gvertex}
F_{(4g)}^{\mu,\,\nu,\,\rho,\,\sigma; \,a, \,b, \, c,\, d}  =  
-i g_s^2 & \: [ \:   f^{eab}f^{ecd} \, \underbrace{ \left( g^{\mu\rho}g^{\nu\sigma} - g^{\mu\sigma}g^{\nu\rho} \right)}_{ \equiv \: T_1^{\mu\nu\rho\sigma}} \nonumber \\[0.2cm]
& +  f^{eac}f^{edb} \underbrace{ \left( g^{\mu\sigma}g^{\rho\nu} - g^{\mu\nu}g^{\rho\sigma}  \right)}_{ \equiv \: T_2^{\mu\nu\rho\sigma}}  \nonumber \\[0.2cm] 
& +  f^{ead}f^{ebc}  \underbrace{ \left( g^{\mu\nu}g^{\rho\sigma} - g^{\mu\rho}g^{\nu\sigma}  \right)}_{ \equiv \: T_3^{\mu\nu\rho\sigma}}  \: ] \,.
\end{align}
Each free index $a$ can be contracted with a matrix $\tau^{a}_{ij} \equiv \sqrt{2} \, T^{a}_{ij}$, where $T^{a}_{ij}$ are the SU(3) generators. By repeated use of SU(3) Fierz identities
\begin{equation}
\tau^{a}_{ij}\tau^{a}_{kl} = \delta_{il}\delta_{kj} - \frac{1}{3}  \delta_{ij}\delta_{kl}  \,,
\end{equation}
we can decompose  Eq. \eqref{eq:4gvertex} into the so-called colour-flow basis
\begin{align}
\label{eq:4gvertex_reduced}
F_{(4g)}^{\mu,\nu,\rho,\sigma; a,b,c,d}  \, \tau^{a}_{i_1,j_1} \, \tau^{b}_{i_2,j_2} \, \tau^{c}_{i_3,j_3} \, \tau^{d}_{i_4,j_4} & =  \; 
 \delta_{i_1j_2}\delta_{i_2j_4}\delta_{i_3j_1}\delta_{i_4j_3}  \, L_1^{\mu\nu\rho\sigma} 
    + \delta_{i_1j_3}\delta_{i_2j_1}\delta_{i_3j_4}\delta_{i_4j_2}  \, L_2^{\mu\nu\rho\sigma} \nonumber \\
 & +  \;  \delta_{i_1j_3}\delta_{i_2j_4}\delta_{i_3j_2}\delta_{i_4j_1} \, L_3^{\mu\nu\rho\sigma} 
    + \delta_{i_1j_4}\delta_{i_2j_3}\delta_{i_3j_1}\delta_{i_4j_2} \, L_4^{\mu\nu\rho\sigma} \nonumber \\
 & +  \;  \delta_{i_1j_2}\delta_{i_2j_3}\delta_{i_3j_4}\delta_{i_4j_1}  \, L_5^{\mu\nu\rho\sigma} 
    + \delta_{i_1j_4}\delta_{i_2j_1}\delta_{i_3j_2}\delta_{i_4j_3}  \, L_6^{\mu\nu\rho\sigma}  \,,
\end{align}
where we have defined
\begin{align}
 L_1^{\mu\nu\rho\sigma}  & =  L_2^{\mu\nu\rho\sigma}  \equiv -ig_s^2 \: \frac{T_1^{\mu\nu\rho\sigma}  - T_2^{\mu\nu\rho\sigma}  }{2} \,, \nonumber \\
 L_3^{\mu\nu\rho\sigma}  & =  L_4^{\mu\nu\rho\sigma}  \equiv -ig_s^2 \: \frac{ T_2^{\mu\nu\rho\sigma}  - T_3^{\mu\nu\rho\sigma}  }{2} \,, \nonumber \\
 L_5^{\mu\nu\rho\sigma}  & =  L_6^{\mu\nu\rho\sigma}  \equiv -ig_s^2 \: \frac{ T_3^{\mu\nu\rho\sigma} -T_1^{\mu\nu\rho\sigma} }{2} \,.
\end{align}
We note here that Eq. \eqref{eq:4gvertex_reduced} matches the internal colour representation used in the \textsc{Helac-Smeft} program. The various terms in Eq. \eqref{eq:4gvertex_reduced} can be  interpreted as different ways of connecting particles in the vertex through colour-flow lines, see e.g. Ref. \cite{Papadopoulos:2005ky}. In the end, analytic expressions for colour-stripped vertex functions are obtained contracting each term in Eq. \eqref{eq:4gvertex_reduced} with the input currents ($J_1^{\nu},J_2^{\rho},J_3^{\sigma}$)
\begin{equation}
V_{(4g; \,i)}^{\mu} (J_1, J_2, J_3) \equiv  L_i^{\mu\nu\rho\sigma} J_{1\nu} \, J_{2\rho} \,J_{3\sigma}  \,.
\end{equation}

To put automation in practice, we have developed two python modules: \textsc{Ecru} and \textsc{HModelGenerator}. The first module (\textsc{Ecru}=Explicit Currents fRom UFO), reads and assembles the information stored in the input model and generates analytic formulae for all vertex functions. It uses FORM \cite{Ruijl:2017dtg} to handle colour-flow decomposition and core python libraries (in particular \textsc{SymPy}) to generate fully simplified analytic expressions. The UFO model format has become the standard adopted by several high-energy physics programs to date. For the case study presented in this paper, we have used the public UFO model \textsc{Smeft@NLO}. The second module, \textsc{HModelGenerator}, reads the analytic expressions produced by \textsc{Ecru} and generates Fortran code ready to be integrated into the \textsc{Helac-Smeft} framework. For each vertex, two  routines are generated. The first one is used to handle all checks (flavours, perturbative couplings, colour conservation) that are necessary in the first phase of the run, when the workflow of the recursive calculation is being set. The second routine contains the actual numerical implementation of the vertex function. The \textsc{HModelGenerator} module creates also additional files necessary to embed the above routines in the \textsc{Helac} code. Besides the two modules mentioned above, a in-house python module for Feynman-rule generation has also been developed. The latter is useful to perform independent checks against the supplied UFO models. It has been used to generate the Feynman rules for  UV-type vertices for the $pp\to t\bar{t}H$ process that we study. We note at this point that we have found a problem with the $\mathcal{O}(\varepsilon^0)$ part of the contribution induced by $C_{\phi G}$ coming from the UV vertex $gggH$ in the public  \textsc{Smeft@NLO} UFO model (version 1.0.3), which leads to a violation of the QCD Ward identity.  We will return to this part later.

The central block in Figure \ref{fig:ufo_to_helac_interface} is the core of the \textsc{Helac-Smeft} system. It contains the actual matrix-element calculations, cross-section integration, and event generation. It also constructs the recursive-computation workflow in the initial phase, a process also known as \textit{skeleton generation} in standard \textsc{Helac-NLO} terminology. It should be clear that the presence of effective interactions complicates the skeleton generation phase. Not only do we have to deal with a larger number of vertex functions compared to the SM case, but we must also track the partial contributions from SMEFT operators. This constitutes an additional combinatorial burden, contributing to an increase in the number of possible ways to combine the available currents at each stage of the recursion. A key requirement in developing recursive computations is to count the powers of perturbative couplings (e.g., powers of $\alpha, \alpha_s$ for SM) step by step in the recursion. In the SMEFT framework, the power counting must be extended to include operator coefficients. All these aspects contribute to an increase in the number of intermediate steps that must be performed in the recursive calculations. To limit the increasing complexity, important optimizations have been introduced in \textsc{Helac-Smeft}:
\begin{itemize}
\item  A simultaneous calculation of the Dyson-Schwinger recursion for all SMEFT contributions, which allows to efficiently reuse shared sub-currents between all partial amplitudes. For example, an intermediate, pure SM-like current can be reused as an input to construct both SM- and SMEFT-like  currents in subsequent steps in the recursion. 
\item A caching system which avoids re-computation of loop-momentum independent currents during the process of the 1-loop amplitude reduction that is performed with the \textsc{CutTools} program. 
\item An optimised flavour-dressing algorithm. The original \textsc{Helac-Phegas} and \textsc{Helac-NLO} algorithms for dressing tree-order and one-loop topologies with flavours have been improved with the help of the \textit{vertex-flavour maps}. In short, instead of selecting all possible flavour combinations by checking the conservation of quantum numbers at each vertex, we directly specify the proper flavour combinations as extracted from Feynman's rules in the UFO model. The approach allows for a more efficient identification of the  subset of vertices that can be used at any step of the recursion.
\end{itemize}
The above optimizations contribute to faster skeleton and event generation. As a bonus, we note that the new \textsc{Helac-Smeft} program outperforms the \textsc{Helac-NLO} code for SM process generation.

We would like to add here that, as we mentioned earlier, our system keeps track of several partial weights during Monte Carlo integration for SMEFT computations.  In the current implementation, phase-space adaptation is tailored to the full SMEFT amplitude. We are aware that this choice may prove suboptimal from the point of view of the partial SMEFT cross-section results, especially when they span different orders of magnitude. Indeed, we observe some degradation in the statistical accuracy of the partial SMEFT results and  plan to explore alternative ways of the phase-space optimization in the future.  

\begin{figure}[t!]
\centering 
\includegraphics[width=0.35\textwidth]{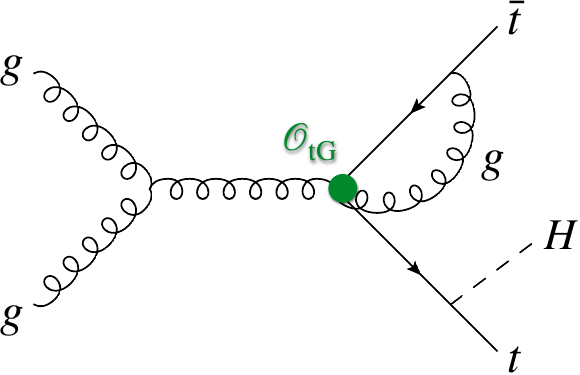}
\caption{Example of a one-loop topology relevant for the $pp\to t\bar{t}H+X$ process in SMEFT, not included in the original \textsc{Helac-1Loop} Monte Carlo program. The vertex marked in green is associated with the $\mathcal{O}_{tG}$ operator.}
\label{fig:ttH_1loop}
\end{figure}

\begin{figure}[t!]
\centering 
\includegraphics[width=0.7\textwidth]{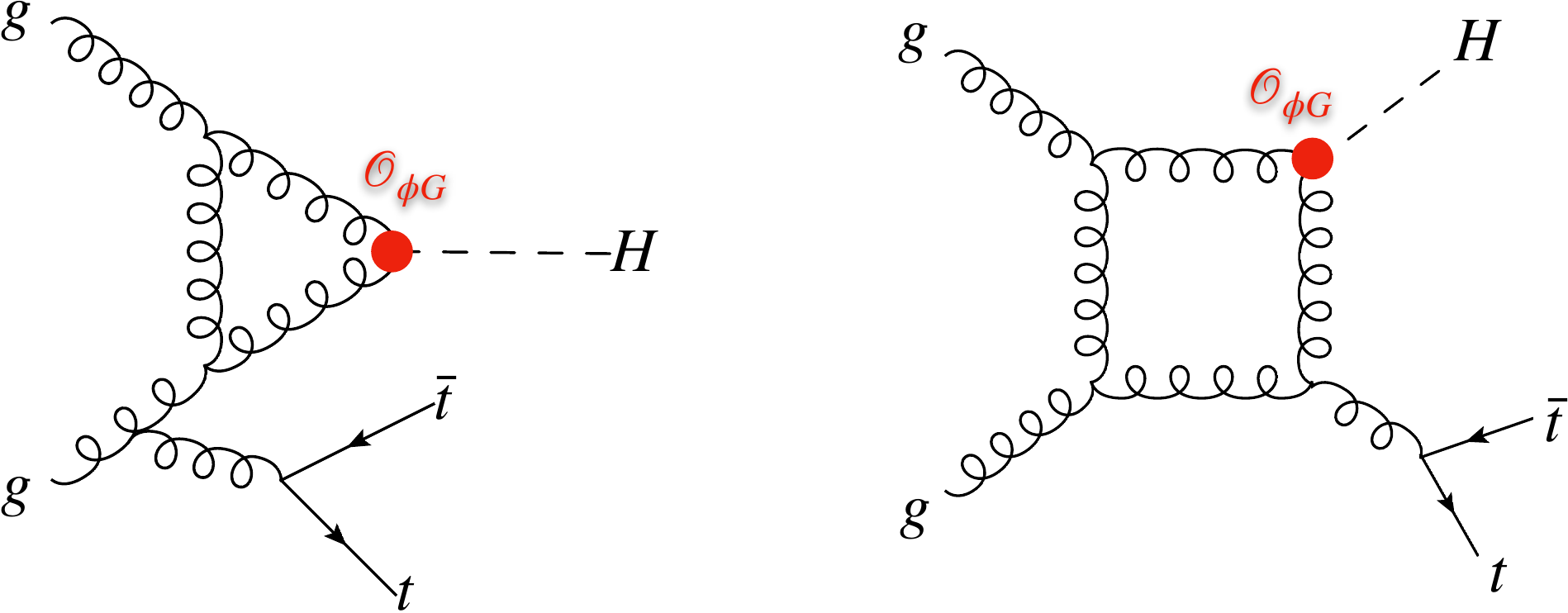}
\caption{Example of one-loop topologies relevant for the $pp\to t\bar{t}H+X$ process in SMEFT, for which  the rank of one-loop integrals can exceed the number of propagators. The vertices marked in red are associated with the $\mathcal{O}_{\phi G}$ operator.}
\label{fig:ttH_1loop_rankbiggerpropagators}
\end{figure}

Furthermore, various additional changes to the algorithms used for the one-loop amplitude generation have been incorporated into the  \textsc{Helac-Smeft} framework. The generation of one-loop topologies in the original implementation of \textsc{Helac-1Loop} is tailored to the SM at NLO QCD. In QCD when considering the on-shell scheme, bubble diagrams where one of the two legs relates to an external particle do not have to be computed, because they either are absorbed by the renormalisation procedure, or lead to vanishing scaleless integrals. Thus, the original implementation simply discards all such topologies during the skeleton-generation phase.  In the case of SMEFT, or even when considering electroweak corrections within the  SM, this approach is not generally correct.  Figure \ref{fig:ttH_1loop} shows one such contribution in our calculation, which would be erroneously removed. Consequently, the algorithm in \textsc{Helac-Smeft} has been modified to properly account for such one-loop topologies. In addition, in four‑dimensional renormalizable gauge theories, such as the Standard Model, any one‑loop $N$‑point diagram can be represented in terms of tensor integrals whose rank $(R)$ does not exceed the number of propagators, so we always have $R\le N$. This condition is no longer true in the presence of higher‑dimensional (non‑renormalizable) operators. Thus, for SMEFT \footnote{SMEFT is a non-renormalizable theory in the strict sense, because it contains higher-dimensional operators that require an infinite number of counterterms to absorb UV divergences. However, SMEFT is renormalizable as an effective field theory, meaning one can renormalize it systematically order by order in $1/\Lambda$, and run the Wilson coefficients with the renormalisation group equation.} tensor integrals with  $R>N$ can genuinely appear. For example, if we have a dimension‑6 operator with two extra derivatives, each insertion can bring extra powers of momentum into the numerator. This is the case for the $\mathcal{O}_{\phi G}$ operator. The $ggH$ vertex induced by this operator carries a loop momentum on each gluon leg. Example of one-loop topologies for which  the rank of one-loop tensor integrals exceed the number of propagators are presented in Figure \ref{fig:ttH_1loop_rankbiggerpropagators}. In each case we have $R=N+1$. The calculation of such contributions is supported in the modern version of \textsc{CutTools}, which we have interfaced to \textsc{Helac-Smeft}. It should be noted, however, that the version  v1.9.3 of \textsc{CutTools}  comes interfaced with the somewhat outdated version of \textsc{OneLOop} (v3.4). The newest available version of \textsc{OneLOop} (v3.7.2) comes with several bugfixes and stability improvements.  Therefore, we have also included the latest version of \textsc{OneLOop} in the \textsc{Helac-Smeft} program.

Finally, the block on the right-hand side of Figure \ref{fig:ufo_to_helac_interface} is dedicated to the processing of the obtained results.  As with SM calculations, we store our theoretical predictions in modified \textsc{Les Houches Files} and then convert them to \textsc{Root Ntuple Files}.  Of course, this required significant modifications to the file format for the SMEFT project and the retention of much more information for later reuse. However, these modified files now allow us to efficiently change the effective, renormalisation and factorisation scale settings, use different PDF sets, calculate different infrared-safe observables, all without having to recalculate the entire process from scratch. All of these tasks are accomplished within the \textsc{HEPlot} framework.

%
\subsection{Numerical benchmarks}
\label{sec:numerical_benchmarks}
%
%

As already mentioned, we have implemented several optimizations to limit the increasing computational complexity of NLO QCD calculations in SMEFT. These optimizations also work for the SM calculation. Therefore, it is interesting to compare the performance of the \textsc{Helac-NLO} and \textsc{Helac-Smeft} codes for several SM processes. In the following, we provide such a comparison for a few benchmarks processes, where we analyse separately the timing of skeleton generation and of numerical amplitude evaluation. The obtained results are generated on the Intel(R) Core(TM) i5-14600 processor, while the software is compiled with \texttt{gcc-7.5 -O3}. In all cases, full off-shell matrix elements have been calculated. 
\begin{table}[t!]
    \centering
    \begin{tabular}{l|c|c|c}
    \hline
    \multicolumn{4}{c}{\textsc{Tree-Level Skeleton - CPU Time} (s)}\\
    \hline 
      \multicolumn{1}{c}{\textsc{Process (SM)}} & \multicolumn{1}{|c|}{\textsc{Helac-Smeft}} & \multicolumn{1}{c|}{\textsc{Helac-NLO}} & \multicolumn{1}{c}{\textsc{Ratio}}\\
      \hline
       $ gg \rightarrow ggg $                                         & 0.011 &    0.173 & 15.73 \\
       $ gg \rightarrow gggg $                                        & 0.070 &    5.760 & 82.3 \\ 
       $ gg \rightarrow ggggg $                                       & 2.574 &  222.410 & 86.4 \\
       $ gg \rightarrow e^+ \nu_e\, \mu^- \bar{\nu}_\mu \,b\bar{b}$   & 0.016 &    0.069 & 4.3 \\
       $ gg \rightarrow e^+ \nu_e\, \mu^- \bar{\nu}_\mu \,b\bar{b}\, H$   & 0.030 &    0.286 & 9.5 \\ 
       $ gg \rightarrow e^+ \nu_e\, \mu^- \bar{\nu}_\mu \,b\bar{b}\, g$   & 0.064 &    0.786 & 12.3 \\ 
       $ gg \rightarrow e^+ \nu_e\, \mu^- \bar{\nu}_\mu \,b\bar{b}\, gg$  & 1.016 &   24.094 & 23.7 \\
    \end{tabular}
    \caption{CPU time required to generate skeletons for various SM processes at leading-order. Comparison between the \textsc{Helac-Smeft} framework and the \textsc{Helac-NLO} code. Results are generated on the Intel(R) Core(TM) i5-14600 processor, while the software is compiled with \texttt{gcc-7.5 -O3}. In all cases, full off-shell matrix elements have been calculated.}
    \label{tab:performance_skeleton_LO}
\end{table}
\begin{table}[t!]
    \centering
    \begin{tabular}{l|c|c|c}
    \hline
    \multicolumn{4}{c}{\textsc{One-Loop Skeleton - CPU Time} (s)}\\
    \hline
      \multicolumn{1}{c}{\textsc{Process (SM)}} & \multicolumn{1}{|c|}{\textsc{Helac-Smeft}} & \multicolumn{1}{c|}{\textsc{Helac-NLO}} & \multicolumn{1}{c}{\textsc{Ratio}}\\
      \hline
       $ gg \rightarrow gg $  & 1.6 & 2.5 & 1.6 \\ 
       $ gg \rightarrow ggg $ & 64.5 & 279.1 & 4.3 \\  
       $ gg \rightarrow e^+ \nu_e\, \mu^- \bar{\nu}_\mu \,b\bar{b}$  & 122.7 & 261.8 & 2.1  \\   
       $ gg \rightarrow e^+ \nu_e\, \mu^- \bar{\nu}_\mu \,b\bar{b}\, H $  & 1462.4 & 5198.8 & 3.6  \\ 
    \end{tabular}
    \caption{CPU time required to generate skeletons for various SM processes at one-loop. Comparison between the \textsc{Helac-Smeft} framework and the \textsc{Helac-NLO} code. Results are generated on the Intel(R) Core(TM) i5-14600 processor, while the software is compiled with \texttt{gcc-7.5 -O3}. In all cases, full off-shell matrix elements have been calculated.}
    \label{tab:performance_skeleton_1L}
\end{table}
\begin{table}[t!]
    \centering
    \begin{tabular}{l|c|c|c}
    \hline
    \multicolumn{4}{c}{\textsc{One-Loop Amplitude - CPU Time} (s)}\\
    \hline
      \multicolumn{1}{c}{\textsc{Process (SM)}} & \multicolumn{1}{|c|}{\textsc{Helac-Smeft}} & \multicolumn{1}{c|}{\textsc{Helac-NLO}} & \multicolumn{1}{c}{\textsc{Ratio}}\\
      \hline
       $ gg \rightarrow gg $  & 2.13 & 2.42 & 1.1 \\ 
       $ gg \rightarrow ggg $ & 105.4 & 114.1 & 1.1 \\  
       $ gg \rightarrow e^+ \nu_e\, \mu^- \bar{\nu}_\mu \,b\bar{b} $ & 5.7 & 14.4 & 2.5 \\   
       $ gg \rightarrow e^+ \nu_e\, \mu^- \bar{\nu}_\mu \,b\bar{b}\, H $ & 41.6 & 61.1 & 1.5 \\ 
    \end{tabular}
    \caption{CPU time required to compute one-loop amplitudes for various SM processes for one phase-space point (fully summed over colour and helicity configurations). Comparison between the \textsc{Helac-Smeft} framework and the \textsc{Helac-NLO} code. Results are generated on the Intel(R) Core(TM) i5-14600 processor, while the software is compiled with \texttt{gcc-7.5 -O3}. In all cases, full off-shell matrix elements have been calculated.}
    \label{tab:performance_amplitude_1L}
\end{table}

Table \ref{tab:performance_skeleton_LO} shows the time required to generate the skeleton for several tree-level processes. The selected processes are chosen to cover a variety of final states and different colour structures. In all cases, we observe a speed-up in skeleton generation, which is primarily attributed to the optimization of the flavour assignment. The increase in speed is more pronounced for processes characterized by a more complex colour structure, as can be observed when analysing processes with many gluons. This can be easily understood by considering that in the \textsc{Helac-Smeft} framework each colour configuration has its own recursion workflow.

In Table \ref{tab:performance_skeleton_1L} we report the time needed to generate the skeleton for several one-loop processes. In this case, we observe that the speed-up factor varies from several tens of percent up to a factor $4$, depending on the process. The speed gain is still noticeable, although less significant compared to the tree-level results. This is a consequence of the fact that other aspects, particularly colour and flavour assignment inside the loops,  are currently the bottleneck of the one-loop skeleton generation.

Finally, in Table \ref{tab:performance_amplitude_1L} we report the time needed to compute the one-loop amplitude, fully summed over colour and helicity configurations, for one phase-space point. In this case, we only see a slight improvement in speed. However, in this case there is still room for further optimizations.

%
\subsection{Checks of UV vertices}
\label{sec:cross_checks_UV}
%
%

The Ward identity offers robust consistency checks for matrix element calculations. It can also be used to certify precision in numerical evaluations. In the \textsc{Helac-NLO} framework, we perform checks of the Ward identity separately for tree-order amplitudes and one-loop amplitudes. This helps us to most effectively capture numerical instabilities and isolate events that require re-computation with higher (quadruple) precision.  While cross-checking our implementation in \textsc{Helac-Smeft}, we found a problem with the UV $gggH$ vertex available in the public version of \textsc{Smeft@NLO}, leading to a violation of the Ward identity. This vertex is relevant for the  $gg \to t\bar{t}H$ partonic subprocess. More specifically, the violation occurs in partial one-loop amplitudes proportional to $C_{\phi G}$. We pinned down the problem to the UV $gggH$ vertex (vertex \texttt{V\_622} in the \textsc{Smeft@NLO} UFO model).  Since pole cancellation worked correctly, this narrowed the problem down to $C_{\phi G}$-induced terms at  $\mathcal{O}(\varepsilon^0)$. We also noticed that the problem disappeared after putting $\mu_R = \mu_{EFT}$, which is actually the setting recommended by the authors of the \textsc{Smeft@NLO} package. This indicated to us that there are possible problems with some terms proportional to $\log(\mu_{EFT}^2/\mu_R^2)$. We have verified that this problem also appears in other processes, e.g. in the case of the partonic subprocess $gg \to gH$ in which the considered vertex is also included. It should be emphasized that this problem only concerns the one-loop virtual amplitude, $\mathcal{M}^{(1)}$, while it does not occur for the tree-order amplitude, $\mathcal{M}^{(0)}$. Thus, it may not be detected if the numerical check of the Ward identity is performed at the interference level, i.e. for the  $2 \Re \left(\mathcal{M}^{(0)}\,\mathcal{M}^{(1)\, *}\right)$ term.

We opted for an independent cross-check of all UV-type vertex contributions stemming from the $\mathcal{O}_{\phi G}$ operator. To this end, we derived  Feynman rules afresh using the anomalous dimension matrix presented in Ref. \cite{Aoude:2022aro} and the expression of the SMEFT Lagrangian. More details can be found in Appendix \ref{appendix:a}. As a first consistency check, we derived all LO-type Feynman rules for the operators ${\cal O}_{t\phi}$, ${\cal O}_{\phi G}$, ${\cal O}_{tG}$, $ {\cal O}_{tW}$ as well as ${\cal O}_{tB}$ and achieved perfect agreement with the content of the \textsc{Smeft@NLO} UFO model.  Using the UV vertices we derived, we observe that the Ward identity test is restored for this operator. We add that for our fixed scale settings, i.e., for $\mu_R=m_t+m_H/2$ and $\mu_{EFT}=m_t$, the impact of this logarithm is not particularly large. However, when various dynamic scale settings will be considered instead, such contributions may begin to play a more significant phenomenological role. We leave such investigations for future research.

%
\subsection{Validation of the computational  framework}
\label{sec:validation}
%
%

\begin{table}[t!]
    \centering
    \renewcommand{\arraystretch}{1.3}
    \begin{tabular}{|l|c|c|c|}
        \hline\hline
                                  & \textsc{Helac-Smeft}   [pb]               & \textsc{Ref.}~\cite{Maltoni:2016yxb}  [pb]     & \textsc{Ratio} \\
        \hline
        $\sigma_{\rm SM}$         & $ ( +5.060 \pm 0.025 )\cdot 10^{-1} $ & $+5.07\cdot 10^{-1}$ & 1.00 \\
        \rowcolor{gray!15}
        $\sigma_{t\phi} $         & $ ( -6.231 \pm 0.030 )\cdot 10^{-2} $ & $-6.20\cdot 10^{-2}$ & 1.01\\
        $\sigma_{\phi G} $        & $ ( +8.732 \pm 0.040 )\cdot 10^{-1} $ & $+8.72\cdot 10^{-1}$ & 1.00\\
        \rowcolor{gray!15}
        $\sigma_{tG} $            & $ ( +5.025 \pm 0.035 )\cdot 10^{-1} $ & $+5.03\cdot 10^{-1}$ & 1.00\\
        $\sigma_{t\phi,\,t\phi} $   & $ ( +1.918 \pm 0.009 )\cdot 10^{-3} $ & $+1.90\cdot 10^{-3}$ & 1.01\\
        \rowcolor{gray!15}
        $\sigma_{\phi G,\,\phi G} $ & $ ( +1.017 \pm 0.006 )\cdot 10^{+0} $ & $+1.02\cdot 10^{+0}$ & 1.00\\
        $\sigma_{tG,\,tG} $         & $ ( +6.739 \pm 0.036 )\cdot 10^{-1} $ & $+6.74\cdot 10^{-1}$ & 1.00\\
        \rowcolor{gray!15}
        $\sigma_{t\phi, \,\phi G} $  & $ ( -5.341 \pm 0.024 )\cdot 10^{-2} $ & $-5.30\cdot 10^{-2}$ & 1.01\\
        $\sigma_{t\phi, \,tG} $      & $ ( -3.099 \pm 0.022 )\cdot 10^{-2} $ & $-3.10\cdot 10^{-2}$ & 1.00\\
        \rowcolor{gray!15}
        $\sigma_{\phi G, \,tG} $     & $ ( +8.496 \pm 0.059 )\cdot 10^{-1} $ & $+8.59\cdot 10^{-1}$ & 0.99\\
        \hline\hline
    \end{tabular}
    \caption{NLO integrated cross-section results for the $pp\to t\bar{t}H+X$ process at the LHC with $\sqrt{s}=13$ TeV. Comparison with the results from Ref. \cite{Maltoni:2016yxb} is provided. The results from  Ref. \cite{Maltoni:2016yxb} have been rescaled in order to match the operator definitions used in  the \textsc{Smeft@NLO} UFO model.}
    \label{tab:maltoni_validation}
\end{table}

We performed a series of checks to validate our computational framework. These checks included both the internal consistency checks of the \textsc{Helac-Smeft} framework and comparisons with external public codes and results in the literature. These can be summarized as follows:
\begin{itemize}
\item The colour-flow decomposition, fully automated in the package \textsc{Ecru} via an interface to \textsc{Form}, has been cross-checked with an independent implementation in python fully based on the \textsc{Sympy} core library. We  have observed that the interface to \textsc{Form} significantly improves the performance. 
\item The full chain of vertex generation, handled by \textsc{Ecru} and \textsc{HModelGenerator}, has been checked against the implementation of the \textsc{Helac-NLO} code for what concerns the SM part. Other genuine SMEFT contributions have been checked against a dedicated LO implementation of the SMEFT operators $\mathcal{O}_{t\phi}$, $\mathcal{O}_{\phi G}$, $\mathcal{O}_{tG}$, $\mathcal{O}_{tW}$ in the  \textsc{Helac-Dipoles} code \cite{Hermann:975160}.
\item The QCD Ward identity has been verified numerically for single phase-space points for a wide range of LHC processes to test the correctness of our implementation in all its parts. In our framework, the Ward identity is also checked at runtime for all one-loop calculations as a probe of numerical stability.
\item We have verified the numerical cancellation of infrared singularities between virtual and real corrections for different processes, both within the SM and SMEFT frameworks, with respect to the operators studied in this paper. The cancellation of IR poles in the regularization parameter $\varepsilon$ ($d=4-2\varepsilon$) has been verified at the level of single phase-space points using built-in integrated subtraction term formulas (also known as the ${\cal I}$-operator) based on the Catani-Seymour and Nagy-Soper subtraction schemes.
\item We have performed extensive checks of Born-level and one-loop amplitudes against the public codes \textsc{MadGraph5\_aMC@NLO} and \textsc{Gosam3.0} for processes with stable top quarks to test the SM and SMEFT contributions, always obtaining excellent agreement between the results.
\item We have performed dedicated calculations of NLO QCD corrections to many processes with and without the SMEFT contributions using two independent subtraction schemes and verified that the numerical results are scheme-independent. This allows us to cross check the correctness of the real emission part of the calculation in a very robust way.
\item  We have also considered a phase-space restriction on the contribution of the subtraction
terms for the Catani-Seymour and Nagy-Soper subtraction schemes, also known as the so-called $\alpha_{max}$. We considered two extreme choices to check the independence of the results from this parameter, obtaining in each case perfect agreement of the results.
\item We have reproduced the results of Ref. \cite{Maltoni:2016yxb} for what concerns the LO and NLO cross sections for the $pp \to t\bar{t}H+X$ process with stable top quarks at the LHC with  $\sqrt{s}=13$ TeV, finding an agreement at the per mille level. In Table \ref{tab:maltoni_validation} we present an explicit comparison of the integrated cross-section results at the NLO QCD level.
\item The reweighting approach implemented within the \textsc{HEPlot} framework has been validated using integrated and differential level comparisons against dedicated \textsc{Helac-Smeft} runs based on different scale settings for $\mu_R, \,\mu_F$ and $\mu_{EFT}$, as well as for different PDF sets.
\end{itemize}

%
\section{Computational setup}
\label{sec:setup}
%

We calculate the NLO QCD predictions for the $pp \to t\bar{t}H$ process for LHC Run III center-of-mass energy of  $\sqrt{s} = 13.6$ TeV.  Two different calculation approaches are considered and the differences between them are discussed. In the first approach, top quarks are considered as stable particles
\begin{equation}
pp  \to t\bar{t}H  +X\,.
\end{equation}
This approach is suitable to study observables such as inclusive production rates, but clearly provides no information on the kinematics of decay products. The second approach incorporates decays of on-shell top quarks using the NWA and is suitable to perform  fiducial-level studies. In the latter approach, QCD corrections and SMEFT effects are consistently taken into account in production and decay matrix elements. Thus, in practice we consider the following decay chain:
\begin{equation}
pp   \to  t\bar{t}H +X \to W^+ W^-\, b\bar{b}\,H  +X \to e^+\nu_e\,\mu^-\bar{\nu}_{\mu}\, 
b\bar{b}\, H +X\,.
\end{equation}
The results for arbitrary leptonic configurations can be obtained by rescaling the cross-section results reported in this paper by a lepton-factor of  $4$ to account for the $\ell = e^\pm, \mu^\pm$ configurations. In the following, we will also refer to the two approaches  as \textit{Stable}  for $pp  \to t\bar{t}H  +X$  and \textit{NWA} for $pp \to e^+\nu_e\,\mu^-\bar{\nu}_{\mu}\, b\bar{b}\, H +X$. 

The LO cross section in the NWA can be written as follows
\begin{equation}
d\sigma^{\rm LO}_{\rm NWA} =  d\sigma^{\rm LO}_{t\bar{t}H} \, \frac{ d\Gamma^{0}_{t \to W^+ b} \,\, d\Gamma^{0}_{ \bar{t} \to W^- \bar{b} }  }{ \left( \Gamma^{\rm LO}_{t} \right)^2} \,,
\end{equation}
where the subsequent decay of the $W$ gauge boson is not included to keep the notation clearer, $d\sigma^{\rm LO}_{t\bar{t}H}$ describes the LO cross section for on-shell ${t\bar{t}H}$ production, while  $d\Gamma^{0}_{t}$ and $d\Gamma^{0}_{\bar{t}}$ denote the LO differential decay rates. In this way, spin correlations are fully taken into account. Similarly, the NLO cross section in the NWA reads
\begin{eqnarray}
d\sigma^{\rm NLO}_{\rm NWA} & = &  d\sigma^{\rm NLO}_{t\bar{t}H} \, \frac{ d\Gamma^{0}_{t \to W^+ b} \,\, d\Gamma^{0}_{ \bar{t} \to W^- \bar{b} }  }{ \left( \Gamma^{\rm NLO}_{t} \right)^2}  \nonumber 
\\[0.2cm]
& + & d\sigma^{\rm LO}_{t\bar{t}H} \, \left(  \frac{ d\Gamma^{1}_{t \to W^+ b} \,\, d\Gamma^{0}_{ \bar{t} \to W^- \bar{b} }  }{ \left(  \Gamma^{\rm NLO}_{t} \right)^2}  +  \frac{ d\Gamma^{0}_{t \to W^+ b} \,\, d\Gamma^{1}_{ \bar{t} \to W^- \bar{b} }  }{ \left( \Gamma^{\rm NLO}_{t} \right)^2}  \right) \,,
\label{eq:sigma_NLO_unexpanded}
\end{eqnarray}
where $d\Gamma^{1}_{t}$ and $d\Gamma^{1}_{\bar{t}}$ are NLO differential decay rates. Moreover, $\Gamma^{\rm LO}_{t}$ and $\Gamma^{\rm NLO}_{t}$ are the  top-quark widths computed respectively at LO and NLO QCD accuracy. Taking a closer look at Eq. \eqref{eq:sigma_NLO_unexpanded}, we note that the rigorous expansion in powers of $\alpha_s$ is spoiled by the fact that the top-quark width, $\Gamma^{\rm NLO}_{t}$, depends itself on $\alpha_s$. In other words, $d\sigma^{\rm NLO}_{\rm NWA}$ contains spurious contributions which are formally $\mathcal{O}(\alpha_s^2)$ with respect to the Born-level result. They can be systematically removed by considering the following expansion 
\begin{equation}
d\sigma^{\rm NLO}_{\rm NWA,exp} =  d\sigma^{\textrm{NLO}}_{\rm NWA} \times\left(\frac{\Gamma_t^{\textrm{NLO}}}{\Gamma_t^{\textrm{LO}}}\right)^2  - d\sigma^{\textrm{LO}}_{\rm NWA} \times \frac{2\,(\Gamma_t^{\textrm{NLO}}-\Gamma_t^{\textrm{LO}})}{\Gamma_t^{\textrm{LO}}}\,.
\label{eq:sigma_NLO_expanded}
\end{equation}
All the NLO cross-section predictions in  the  NWA reported in this paper are computed according to Eq. \eqref{eq:sigma_NLO_expanded}.

For our LO and NLO results we use the NNPDF3.1 \cite{NNPDF:2017mvq} PDF set with $\alpha_s(m_Z) = 0.118$.  The running of $\alpha_s$ is performed with a two-loop (one-loop) accuracy at NLO (at LO) and provided by the LHAPDF library \cite{Buckley:2014ana} involving five active flavours. We consider the following SM input parameters
\begin{equation}
\begin{array}{lll}
 G_{ \mu}=1.166378 \cdot 10^{-5} ~{\rm GeV}^{-2}\,, &\quad \quad \quad
&   m_{t}=172.5 ~{\rm GeV} \,, \\[0.2cm]
 m_{W}=80.379 ~{\rm GeV} \,, &
&\Gamma_{W} = 2.0972 ~{\rm GeV}\,, \\[0.2cm]
  m_{Z}=91.1876  ~{\rm GeV} \,, & 
& m_{H} = 125.0  ~{\rm GeV}\,.
\end{array}
\end{equation}
In the $G_\mu$-scheme that we are using, the remaining EW parameters, $\alpha$ as well as $\sin^2\left(\theta_W\right)$, are derived from the above inputs, and are not modified by the set of SMEFT operators that we are considering. This is generally not true for all operators. We keep the  Cabibbo-Kobayashi-Maskawa (CKM) matrix diagonal. With the parameters above, the numerical value of top-quark widths at LO and NLO, which are used to compute fiducial cross-section predictions, is computed using the formulae from Ref. \cite{Jezabek:1988iv}:
\begin{equation}
\Gamma^{\rm LO}_{t} = 1.4806842 \; \mbox{GeV}\,,  \qquad \qquad \Gamma^{\rm NLO}_{t} = 1.3535983 \; \mbox{GeV} \,.
\end{equation}
A comment is in order here. The total top-quark widths used  as input parameters in our study  do not include SMEFT contributions. However, in general $\Gamma_t$ also receives SMEFT corrections. Therefore,  one should incorporate the relevant SMEFT operators to $\Gamma_t$ and perform a consistent expansion in 
$C_i^{(6)}/\Lambda^2$. Schematically, for the SM cross-section result, its dependence on $\Gamma_t$, where $\Gamma_t= {\Gamma_t^{\text{SM}} + \delta\Gamma_t^{(6)}}$, can be written as follows 
\begin{equation}
\sigma^{\rm SM} ({\Gamma_t^{\text{SM}} + \delta\Gamma_t^{(6)}}) = 
\sigma^{\rm SM}(\Gamma_t^{\rm SM})\left( 1 - 2 \frac{\delta \Gamma^{(6)}_t}{\Gamma_t^{\rm SM}} + 3 \frac{\delta \Gamma^{(6)}_t}{\Gamma_t^{\rm SM}}\frac{\delta \Gamma^{(6)}_t}{\Gamma_t^{\rm SM}}  + {\cal O} \left(\frac{1}{\Lambda^6} \right)\right)\,,
\end{equation}
where $\delta\Gamma_t^{(6)}$ is the LO contribution in $C_i^{(6)}/\Lambda^2$. In our study this contribution can be given by the ${\cal O}_{tW}$ operator, schematically written as $\delta \Gamma_t^{(tW)} \sim C_{tW}^{(6)}/\Lambda^2$, or by the ${\cal O}_{tG}$ operator where we would have $\delta\Gamma_t^{(tG)}\sim \alpha_s C_{tG}^{(6)}/\Lambda^2$.  However,  expanding the insertions of SMEFT operators in the top-quark decay widths generates higher-order terms in $C_i^{(6)}/\Lambda^2$, which might have a non-trivial interplay with dimension-8 operators \cite{Zhang:2014rja,Boughezal:2019xpp}.  In addition,  different ways of performing and truncating this expansion can lead to significantly different results \cite{Hermann:975160}. On the other hand, experimental measurements of the top-quark width indicate that the impact of complete SMEFT effects on $\Gamma_t$ cannot be large. The most precise (indirect) measurement of the top-quark width is $\Gamma_t = 1.36 \pm 0.02 \,({\rm stat}) {}^{+0.14}_{-0.11} \,({\rm syst})$  GeV \cite{CMS:2014mxl}, which is in excellent  agreement with the SM result. It was obtained by the CMS collaboration from a measurement of the branching fraction ${\cal B}(t\to Wb)$ assuming the validity of the SM and combined with a dedicated single-top-quark cross-section measurement in the
$t$-channel \footnote{All direct measurements of $\Gamma_t$ are less precise than the indirect measurements but are less model-dependent. For example, the measured values as obtained by the ATLAS collaboration using direct approaches are $\Gamma_t = 1.76 \pm 0.33 \,({\rm stat}) {}^{+0.79}_{-0.68} \,({\rm syst})$  GeV \cite{ATLAS:2017vgz} and $\Gamma_t = 1.28 \pm 0.30$ GeV \cite{Herwig:2019obz}.}. Therefore, in this study, we decided to adopt the SM value for $\Gamma_t$. Aspects related to SMEFT corrections in $\Gamma_t$ are beyond the scope of this paper and will be investigated in detail in a separate publication.

Our definition of  fiducial phase-space regions is based on the following requirements. The final states must contain exactly two charged leptons and at least two $b$-jets. Jets are defined according to the anti-$k_T$ algorithm \cite{Cacciari:2008gp} using the resolution parameter $R=0.4$. We impose the following kinematical cuts on  charged leptons and the two hardest $b$-jets
\begin{equation}
\setlength{\arraycolsep}{12pt}
\begin{array}{ccccc}
    p_{T, \,b} > 25\,\textrm{GeV}\,, && |y_b| < 2.5\,, && \Delta R_{bb} > 0.4\,,  \\[0.2cm]
    p_{T, \, \ell} > 25\,\textrm{GeV}\,, && |y_\ell| < 2.5\,, && \Delta R_{\ell\ell} > 0.4\,,
\end{array}
\end{equation}
where $\Delta R_{ij}=\sqrt{(\Delta y_{ij})^2+(\Delta\phi_{ij})^2}$ is the distance in the rapidity-azimuthal angle plane. We do not impose any kinematic constraints on the missing transverse momentum and the additional jet from the real-emission part of NLO QCD calculations.

The cross section, split into the SM, linear and quadratic SMEFT partial contributions, can be parametrised as follows
\begin{equation}
\sigma(\mu_R,\mu_F,\mu_{EFT})  =  \sigma_{\rm SM}(\mu_R,\mu_F)  + \sum_i  \Sigma^{\rm SMEFT}_{i}(\mu_R,\mu_F,\mu_{EFT})  + \sum_{ij} \, \Sigma^{\rm SMEFT}_{ij}(\mu_R,\mu_F,\mu_{EFT}) \,,
\label{eq:parametrization_cross-section}
\end{equation}
where we have explicitly provided the dependence on the various scales entering each term. We consider the renormalisation and factorisation scales, $\mu_R$ and $\mu_F$ respectively, as well as the effective scale $\mu_{EFT}$. When fixed-scale settings for $\mu_{EFT}$ are considered, Wilson coefficients become phase-space independent and can be factored out
\begin{equation}
\begin{aligned}
\Sigma^{\rm SMEFT}_{i}(\mu_R,\mu_F,\mu_{EFT}) & = \frac{1 \,{\rm TeV}^{2}}{\Lambda^2}  \, C_i(\mu_{EFT}) \, \sigma_{i}(\mu_R,\mu_F,\mu_{EFT})  \,, \\[0.2cm]
\Sigma^{\rm SMEFT}_{ij}(\mu_R,\mu_F,\mu_{EFT}) & = \frac{1 \,{\rm TeV}^{4}}{\Lambda^4} \, C_i(\mu_{EFT}) \, C_j(\mu_{EFT})\, \sigma_{ij}(\mu_R,\mu_F,\mu_{EFT}) \,,
\end{aligned}
\end{equation}
where a residual dependence on $\mu_{EFT}$ in the $\sigma_i$ and $\sigma_{ij}$ term stems from the UV counterterms.  Following Ref. \cite{Maltoni:2016yxb}, we will present the results for each partial contribution ($\sigma_{\rm SM}$, $\sigma_{i}$ and $\sigma_{ij}$) at the integrated and differential cross-section level.  Central values for $\mu_R$, $\mu_F$ and $\mu_{EFT}$ are defined as follows: 
\begin{equation}
\mu_0 = \mu_{R} = \mu_{F} = m_t + \frac{m_H}{2} \,, \qquad \qquad  \mu_0=\mu_{EFT} = m_t \,.
\label{eq:scale_setup}
\end{equation}
Theoretical uncertainties due to missing higher-order effects in the $\alpha_s$ expansion  are estimated using a standard  $7$-point scale variation
\begin{equation}
\left(\frac{\mu_R}{\mu_{0}}\,,\frac{\mu_F}{\mu_{0}}\right) = \Big\{ \left(2,1\right),\left(0.5,1\right),\left(1,2\right), 
(1,1), (1,0.5), (2,2),(0.5,0.5) \Big\} \,,
\end{equation}
where $\mu_{EFT}$ is kept fixed at its central value. By searching for the minimum and maximum of the resulting cross sections one obtains an uncertainty
band. On the other hand, the uncertainty related to the $\mu_{EFT}$ scale, at which the operators are defined,  are calculated based on a 3-point scale variation 
\begin{equation}
\mu_{EFT}=\mu_0\,,  \quad \quad \quad \mu_{EFT}=2 \mu_0\,,  \quad \quad \quad \mu_{EFT}=0.5\mu_0\,, 
\end{equation}
which in practice means that the scale dependence of $\mu_{EFT}$ and $\mu_R/\mu_F$ are disentangled from each other. Indeed,  the former one characterises the renormalisation group running of the operators and probes the $\log(\mu_{EFT}^2/\mu_R^2)$ terms. This uncertainty is related to the renormalisation group running and mixing of the operators. For the prediction obtained at  $\mu_0=\mu_{EFT}$ that are  given by 
\begin{equation}
\sigma(\mu_0) = \sigma_{\rm SM} + \sum_i \frac{1\, {\rm TeV}^2}{\Lambda^2} \, C_i(\mu_0) \, \sigma_i(\mu_0)
+ \sum_{i j} \frac{1\, {\rm TeV}^4}{\Lambda^4} \, C_i(\mu_0) \, C_j(\mu_0) \, \sigma_{ij}(\mu_0)\,,
\end{equation}
a result at a different scale setting $\mu_{EFT}=\mu$ 
\begin{equation}
\sigma(\mu)  =   \sigma_{\rm SM} + \sum_{i} \frac{1 \, {\rm TeV}^2}{\Lambda^2}  \,C_i(\mu)\,\sigma_{i}(\mu) + \sum_{ij} \frac{1\, {\rm TeV}^4}{\Lambda^4}  \,C_i(\mu)\,C_j(\mu)\,\sigma_{ij}(\mu) \,,
\end{equation}
is calculated by inserting 
\begin{equation}
C_i(\mu) = \Gamma_{ij}(\mu,\mu_0)  \, C_j(\mu_0)\,,
\end{equation}
where $\Gamma_{ij}$ describes the running of operator coefficients that is given by Eq. \eqref{eq:RGE_matrix}. In the end, we obtain 
\begin{equation}
\sigma(\mu)  =   \sigma_{\rm SM} + \sum_{i} \frac{1\, {\rm TeV}^2}{\Lambda^2} \, C_i(\mu_0)\,\sigma_{i}(\mu_0,\mu) + \sum_{ij} \frac{1\, {\rm TeV}^4}{\Lambda^4} \, C_i(\mu_0)\,C_j(\mu_0)\,\sigma_{ij}(\mu_0,\mu) \,,
\label{eq:SMEFT_parametrisation}
\end{equation}
where we have defined   
\begin{align}
\begin{split}
    \label{eq:mueft_uncertainty}                                           \sigma_i(\mu_0,\mu)    
    &= \Gamma_{ji}(\mu,\mu_0)\,\sigma_j(\mu) \,,\\[0.2cm]
    \sigma_{ij}(\mu_0,\mu) &= \Gamma_{ki}(\mu,\mu_0)\,\Gamma_{lj}(\mu,\mu_0)\,\sigma_{kl}(\mu)\,.
\end{split}
\end{align}
 By default we use $\mu_{EFT}= \mu_0= m_t$  as a central scale, but we also provide results for the $\mu_{EFT}=m_t+m_H/2$ case. In both cases the 3-point scale variation is performed. Thus, $\sigma_i(\mu_0,\,\mu)$,  $\sigma_{ij}(\mu_0,\,\mu)$ and $\sigma_{ii}(\mu_0,\,\mu)$ can be understood  as the cross-section results corresponding to the Wilson coefficients that are evaluated at $\mu$ and then evolved back to $\mu_0$ including all the mixing and running effects. 
\begin{table}
\centering
\renewcommand{\arraystretch}{1.5}
\begin{subtable}[t]{0.45\textwidth}
\vspace{0pt}
\centering
\begin{tabular}{|lccccc|}
     \hline\hline
     Contribution & SM & $C_{t\phi}$ & $C_{\phi G}$ & $C_{tG}$ & $C_{tW}$\\
     \hline\hline
     \rowcolor{gray!15}
     \multicolumn{6}{|c|}{ $ gg \to \WWbb H $ } \\
     \hline 
     Born   &   8 &   8 &   8 &  22 &  16 \\
     one-loop & 178 & 178 & 289 & 774 & 356 \\
     \hline
     \rowcolor{gray!15}
     \multicolumn{6}{|c|}{ $ q\bar{q}/\bar{q}q \to \WWbb H$ } \\
     \hline
     Born   &   2 &   2 &   1 &   3 &   4 \\
     one-loop &  43 &  43 &  42 & 122 &  86 \\
     \hline
     \rowcolor{gray!15}
     \multicolumn{6}{|c|}{ $ gg \to \WWbb Hg$ } \\
     \hline
     Real prod. &  50 &  50 &  74 & 183 & 100 \\ 
     Real decay &  16 &  16 &  16 &  52 &  32 \\ 
     \hline
     \rowcolor{gray!15}
     \multicolumn{6}{|c|}{ $ q\bar{q}/\bar{q}q \to \WWbb Hg$ } \\
     \hline
     Real prod. &  12 &  12 &  12 &  28 &  24 \\ 
     Real decay &   4 &   4 &   2 &   8 &   8 \\ 
     \hline
     \rowcolor{gray!15}
     \multicolumn{6}{|c|}{ $ qg/gq\to \WWbb Hq$ } \\
     \hline
     Real prod. &  12 &  12 &  12 &  28 &  24 \\ 
     \hline
\end{tabular}
\end{subtable}
\begin{subtable}[t]{0.45\textwidth}
\vspace{0pt}
\centering
\begin{tabular}{|lcccc|}
     \hline\hline
     Contribution & SM & $C_{t\phi}$ & $C_{\phi G}$ & $C_{tG}$\\
     \hline\hline
     \rowcolor{gray!15}
     \multicolumn{5}{|c|}{ $ gg \to t\bar{t}H $ } \\
     \hline 
     Born   &   8 &   8 &   8 &  22 \\
     one-loop & 162 & 162 & 273 & 714 \\
     \hline
     \rowcolor{gray!15}
     \multicolumn{5}{|c|}{ $ q\bar{q}/\bar{q}q \to t\bar{t}H$ } \\
     \hline
     Born   &   2 &   2 &   1 &   3 \\
     one-loop &  39 &  39 &  40 & 112 \\
     \hline
     \rowcolor{gray!15}
     \multicolumn{5}{|c|}{ $ gg \to t\bar{t}Hg$ } \\
     \hline
     Real emission &  50 &  50 &  74 & 183 \\ 
     \hline
     \rowcolor{gray!15}
     \multicolumn{5}{|c|}{ $ q\bar{q}/\bar{q}q \to t\bar{t}Hg$ } \\
     \hline
     Real emission &  12 &  12 &  12 &  28 \\ 
     \hline
     \rowcolor{gray!15}
     \multicolumn{5}{|c|}{ $ qg/gq\to t\bar{t}Hq$ } \\
     \hline
     Real emission &  12 &  12 &  12 &  28 \\ 
     \hline
\end{tabular}
\end{subtable}
\caption{\it Number of Feynman diagrams required for  NLO calculations for  the  $pp \to e^+\nu_e \, \mu^- \bar{\nu}_\mu \, b \bar{b} \, H+X$ process  in the  NWA (left) and for the $pp\to t\bar{t}H$ process with the stable top quarks (right). Genuine SM contributions are reported for various partonic subprocesses along with contribution from the four dimension-6   operators considered. The labels "Real prod." and "Real decay" denote the real-radiation contribution where the extra parton is emitted in production and decay matrix elements, respectively.}
\label{table:number_feynman_diagrams}
\end{table}

The theoretical uncertainties arising from the PDF parametrization have already been calculated in Ref. \cite{Maltoni:2016yxb} for the stable top-quark case. Therefore, we do not repeat that work here. Furthermore, we do not provide PDF uncertainties for the $pp\to t\bar{t}H$ process including top-quark decays in the di-lepton decay channel, because SMEFT effects do not directly affect their size. For the SM  $pp\to t\bar{t}H$ process, including top quark decays, they have been studied in detail in Ref.  \cite{Stremmer:2021bnk}. 

Before we conclude this section, we would like to briefly discuss the complexity of our calculations. To this end, in Table \ref{table:number_feynman_diagrams} we provide the number of Feynman diagrams entering matrix-element calculations at LO and NLO in QCD for both cases, i.e.  for the $pp \to t\bar{t}H +X$ process  with the stable top quarks at ${\cal O}(\alpha\,\alpha_s^3)$ and for the $pp \to e^+ \nu_e  \, \mu^- \bar{\nu}_\mu \, b\bar{b} \, H +X$ process in the NWA at ${\cal O}(\alpha^5\alpha_s^3)$. The results are divided according to the partonic subprocess. Real-emission diagrams are additionally split into contributions where the extra parton originates at the level of production or decay matrix elements.  Furthermore, we show the number of Feynman diagrams separately for each SMEFT operator included in the study. Although we do not use Feynman diagrams to compute the scattering amplitudes but rely on a recursive approach, we provide these numbers to better illustrate the computational cost of the analysis.

%
\section{Integrated cross-section results}
\label{sec:integrated_cross_sections}
%
%

In the following section we will analyse the phenomenological differences between the two case-studies considered, namely the $pp \to t\bar{t} H+X$ process with stable top quarks and the $pp  \to  t\bar{t}H +X \to W^+W^-\, b\bar{b}\, H +X  \to e^+\nu_e\,\mu^-\bar{\nu}_{\mu}\, b\bar{b} \,H+X$ process in the fiducial phase-space regions. As it is customary for such studies, we start presenting our findings at the integrated cross-section level. 

In Table \ref{tab:ttH_xsec} we give the integrated cross-section predictions  at LO and NLO in QCD. The provided results are calculated for the  NNPDF3.1 PDF set employing $\mu_R = \mu_F = m_t + m_H/2$ and $\mu_{EFT} = m_t$. We display the SM results as well as the SMEFT predictions for the following contributions: 
\begin{itemize}
\item  The linear $\sigma_i$ terms that come from the interference between the SM and dimension‑6 operators at $C_i^{(6)}/\Lambda^2$.
\item The cross $\sigma_{ij}$  terms with $i\ne j$ that stem from the interference between different dimension‑6 operators at $C_i^{(6)} C_j^{(6)}/\Lambda^4$. 
\item The quadratic $\sigma_{ii}$ terms that result from the interference between the same dimension‑6 operators at $C_i^{(6)} C_i^{(6)}/\Lambda^4$.
\end{itemize}
For the  $\sigma_{ij}$ terms factor of $2$ is included to account for the $i \leftrightarrow j$ transitions. Furthermore, we show the ratios to the respective SM cross-section results and the ${\cal K}$-factors, which quantify the impact of higher-order corrections. Two kinds of uncertainties are provided for each case. The first reported uncertainty refers to the standard 7-point scale variation while the second one to the $\mu_{EFT}$ scale variation. Theoretical predictions obtained in the NWA and for stable top quarks are presented together to facilitate comparisons. This is particularly instructive to examine the similarities and differences between the two approaches.
\begin{table}[h!]
    \centering
    \renewcommand{\arraystretch}{1.5}
    \small
    \begin{subtable}[t]{\linewidth}        
\begin{tabular}{|c|rrrrc|}
 \hline
 \hline
 \multicolumn{6}{|c|}{$pp \to t\bar{t}H+X$ \quad \textsc{(Stable)} } \\
 \hline
 \hline
 & \multicolumn{1}{c}{$\sigma^{\textrm{LO}}$[pb]} & \multicolumn{1}{c}{$\sigma^{\textrm{LO}}/\sigma^{\textrm{LO}}_{\textrm{SM}}$} & \multicolumn{1}{c}{$\sigma^{\textrm{NLO}}$[pb]} & \multicolumn{1}{c}{$\sigma^{\textrm{NLO}}/\sigma^{\textrm{NLO}}_{\textrm{SM}}$} & $\mathcal{K}$\\
 \hline
 $\sigma_{SM}$              &    $  0.4424^{+0.1337\,+0.0000}_{-0.0956\,-0.0000} $    &    $  1.000^{+0.000\,+0.000}_{-0.000\,-0.000} $    &    $  0.5623^{+0.0327\,+0.0000}_{-0.0519\,-0.0000} $    &    $  1.000^{+0.000\,+0.000}_{-0.000\,-0.000} $    &   1.27\\
\rowcolor{gray!15}
 $\sigma_{t\phi}$           &    $ -0.0541^{+0.0117\,+0.0024}_{-0.0164\,-0.0028} $    &    $ -0.122^{+0.000\,+0.005}_{-0.000\,-0.006} $    &    $ -0.0713^{+0.0071\,+0.0009}_{-0.0051\,-0.0012} $    &    $ -0.127^{+0.001\,+0.002}_{-0.002\,-0.002} $    &   1.32\\
 $\sigma_{\phi G}$          &    $  0.3023^{+0.0945\,+0.0392}_{-0.0671\,-0.0324} $    &    $  0.683^{+0.005\,+0.088}_{-0.005\,-0.073} $    &    $  0.4941^{+0.0724\,+0.0247}_{-0.0683\,-0.0204} $    &    $  0.879^{+0.073\,+0.044}_{-0.044\,-0.036} $    &   1.63\\
\rowcolor{gray!15}
 $\sigma_{tG}$              &    $  0.4501^{+0.1393\,+0.0004}_{-0.0992\,-0.0008} $    &    $  1.017^{+0.006\,+0.001}_{-0.005\,-0.002} $    &    $  0.5641^{+0.0268\,+0.0022}_{-0.0500\,-0.0024} $    &    $  1.003^{+0.004\,+0.004}_{-0.010\,-0.004} $    &   1.25\\
 $\sigma_{t\phi,t\phi}$     &    $  0.0017^{+0.0005\,+0.0002}_{-0.0004\,-0.0001} $    &    $  0.004^{+0.000\,+0.000}_{-0.000\,-0.000} $    &    $  0.0023^{+0.0002\,+0.0001}_{-0.0002\,-0.0001} $    &    $  0.004^{+0.000\,+0.000}_{-0.000\,-0.000} $    &   1.36\\
\rowcolor{gray!15}
 $\sigma_{t\phi,\,\phi G}$    &    $ -0.0185^{+0.0041\,+0.0027}_{-0.0058\,-0.0035} $    &    $ -0.042^{+0.000\,+0.006}_{-0.000\,-0.008} $    &    $ -0.0311^{+0.0044\,+0.0020}_{-0.0048\,-0.0023} $    &    $ -0.055^{+0.003\,+0.004}_{-0.005\,-0.004} $    &   1.68\\
 $\sigma_{t\phi,\,tG}$        &    $ -0.0275^{+0.0061\,+0.0012}_{-0.0085\,-0.0014} $    &    $ -0.062^{+0.000\,+0.003}_{-0.000\,-0.003} $    &    $ -0.0357^{+0.0034\,+0.0003}_{-0.0022\,-0.0004} $    &    $ -0.064^{+0.000\,+0.001}_{-0.000\,-0.001} $    &   1.30\\
\rowcolor{gray!15}
 $\sigma_{\phi G,\,\phi G}$   &    $  0.1506^{+0.0548\,+0.0332}_{-0.0373\,-0.0251} $    &    $  0.340^{+0.016\,+0.075}_{-0.014\,-0.057} $    &    $  0.2991^{+0.0581\,+0.0318}_{-0.0511\,-0.0263} $    &    $  0.532^{+0.068\,+0.057}_{-0.046\,-0.047} $    &   1.99\\
 $\sigma_{\phi G,\,tG}$       &    $  0.2958^{+0.1027\,+0.0257}_{-0.0709\,-0.0223} $    &    $  0.669^{+0.023\,+0.058}_{-0.020\,-0.050} $    &    $  0.4965^{+0.0704\,+0.0151}_{-0.0712\,-0.0125} $    &    $  0.883^{+0.070\,+0.027}_{-0.050\,-0.022} $    &   1.68\\
\rowcolor{gray!15}
 $\sigma_{tG,\,tG}$           &    $  0.5852^{+0.2145\,+0.0106}_{-0.1459\,-0.0118} $    &    $  1.323^{+0.065\,+0.024}_{-0.056\,-0.027} $    &    $  0.7798^{+0.0337\,+0.0056}_{-0.0749\,-0.0074} $    &    $  1.387^{+0.005\,+0.010}_{-0.040\,-0.013} $    &   1.33\\
\hline
\end{tabular}
    \end{subtable} \\
\vspace{0.1cm}
    \begin{subtable}[t]{\linewidth}        
\begin{tabular}{|c|rrrrc|}
 \hline
 \hline
 \multicolumn{6}{|c|}{$pp \to e^+\nu_e\, \mu^-\bar{\nu}_{\mu}\, b\bar{b} \, H +X$  \quad \textsc{(NWA)}} \\
 \hline
 \hline
 & \multicolumn{1}{c}{$\sigma^{\textrm{LO}}$[fb]} & \multicolumn{1}{c}{$\sigma^{\textrm{LO}}/\sigma^{\textrm{LO}}_{\textrm{SM}}$} & \multicolumn{1}{c}{$\sigma^{\textrm{NLO}}$[fb]} & \multicolumn{1}{c}{$\sigma^{\textrm{NLO}}/\sigma^{\textrm{NLO}}_{\textrm{SM}}$} & $\mathcal{K}$\\
 \hline
 $\sigma_{SM}$              &    $  2.3166^{+0.7068\,+0.0000}_{-0.5044\,-0.0000} $    &    $  1.000^{+0.000\,+0.000}_{-0.000\,-0.000} $    &    $  2.7680^{+0.0560\,+0.0000}_{-0.1898\,-0.0000} $    &    $  1.000^{+0.000\,+0.000}_{-0.000\,-0.000} $    &   1.19\\
\rowcolor{gray!15}
 $\sigma_{t\phi}$           &    $ -0.2835^{+0.0617\,+0.0126}_{-0.0865\,-0.0145} $    &    $ -0.122^{+0.000\,+0.005}_{-0.000\,-0.006} $    &    $ -0.3519^{+0.0270\,+0.0037}_{-0.0091\,-0.0050} $    &    $ -0.127^{+0.001\,+0.001}_{-0.002\,-0.002} $    &   1.24\\
 $\sigma_{\phi G}$          &    $  1.6262^{+0.5137\,+0.2089}_{-0.3638\,-0.1730} $    &    $  0.702^{+0.006\,+0.090}_{-0.005\,-0.075} $    &    $  2.5292^{+0.2865\,+0.1154}_{-0.3098\,-0.0954} $    &    $  0.914^{+0.094\,+0.042}_{-0.053\,-0.034} $    &   1.56\\
\rowcolor{gray!15}
 $\sigma_{tG}$              &    $  2.4217^{+0.7568\,+0.0108}_{-0.5374\,-0.0137} $    &    $  1.045^{+0.006\,+0.005}_{-0.006\,-0.006} $    &    $  2.8249^{+0.0460\,+0.0133}_{-0.1774\,-0.0157} $    &    $  1.021^{+0.006\,+0.005}_{-0.016\,-0.006} $    &   1.17\\
 $\sigma_{tW}$              &    $  0.7238^{+0.2214\,+0.0111}_{-0.1579\,-0.0120} $    &    $  0.312^{+0.000\,+0.005}_{-0.000\,-0.005} $    &    $  0.8719^{+0.0182\,+0.0025}_{-0.0615\,-0.0041} $    &    $  0.315^{+0.001\,+0.001}_{-0.001\,-0.001} $    &   1.20\\
\rowcolor{gray!15}
 $\sigma_{t\phi,\,t\phi}$     &    $  0.0087^{+0.0026\,+0.0009}_{-0.0019\,-0.0008} $    &    $  0.004^{+0.000\,+0.000}_{-0.000\,-0.000} $    &    $  0.0112^{+0.0005\,+0.0003}_{-0.0009\,-0.0003} $    &    $  0.004^{+0.000\,+0.000}_{-0.000\,-0.000} $    &   1.29\\
 $\sigma_{t\phi,\,\phi G}$    &    $ -0.0995^{+0.0223\,+0.0146}_{-0.0314\,-0.0185} $    &    $ -0.043^{+0.000\,+0.006}_{-0.000\,-0.008} $    &    $ -0.1592^{+0.0203\,+0.0095}_{-0.0196\,-0.0111} $    &    $ -0.058^{+0.004\,+0.003}_{-0.006\,-0.004} $    &   1.60\\
\rowcolor{gray!15}
 $\sigma_{t\phi,\,tG}$        &    $ -0.1482^{+0.0329\,+0.0060}_{-0.0463\,-0.0067} $    &    $ -0.064^{+0.000\,+0.003}_{-0.000\,-0.003} $    &    $ -0.1794^{+0.0128\,+0.0009}_{-0.0034\,-0.0014} $    &    $ -0.065^{+0.000\,+0.000}_{-0.000\,-0.000} $    &   1.21\\
 $\sigma_{t\phi,\,tW}$        &    $ -0.0443^{+0.0097\,+0.0013}_{-0.0135\,-0.0015} $    &    $ -0.019^{+0.000\,+0.001}_{-0.000\,-0.001} $    &    $ -0.0553^{+0.0043\,+0.0004}_{-0.0016\,-0.0005} $    &    $ -0.020^{+0.000\,+0.000}_{-0.000\,-0.000} $    &   1.25\\
\rowcolor{gray!15}
 $\sigma_{\phi G,\,\phi G}$   &    $  0.8178^{+0.2987\,+0.1799}_{-0.2032\,-0.1361} $    &    $  0.353^{+0.016\,+0.078}_{-0.014\,-0.059} $    &    $  1.5299^{+0.2443\,+0.1528}_{-0.2377\,-0.1277} $    &    $  0.553^{+0.082\,+0.055}_{-0.052\,-0.046} $    &   1.87\\
 $\sigma_{\phi G,tG}$       &    $  1.5954^{+0.5561\,+0.1352}_{-0.3835\,-0.1175} $    &    $  0.689^{+0.023\,+0.058}_{-0.020\,-0.051} $    &    $  2.4966^{+0.2465\,+0.0653}_{-0.3088\,-0.0533} $    &    $  0.902^{+0.080\,+0.024}_{-0.053\,-0.019} $    &   1.56\\
\rowcolor{gray!15}
 $\sigma_{\phi G,\,tW}$       &    $  0.2548^{+0.0807\,+0.0280}_{-0.0571\,-0.0236} $    &    $  0.110^{+0.001\,+0.012}_{-0.001\,-0.010} $    &    $  0.3980^{+0.0458\,+0.0155}_{-0.0491\,-0.0126} $    &    $  0.144^{+0.015\,+0.006}_{-0.008\,-0.005} $    &   1.56\\
 $\sigma_{tG,\,tG}$           &    $  3.4828^{+1.2967\,+0.0660}_{-0.8776\,-0.0730} $    &    $  1.503^{+0.077\,+0.028}_{-0.066\,-0.032} $    &    $  4.3464^{+0.0425\,+0.0262}_{-0.3089\,-0.0365} $    &    $  1.570^{+0.007\,+0.009}_{-0.065\,-0.013} $    &   1.25\\
\rowcolor{gray!15}
 $\sigma_{tG,\,tW}$           &    $  0.3795^{+0.1189\,+0.0079}_{-0.0843\,-0.0088} $    &    $  0.164^{+0.001\,+0.003}_{-0.001\,-0.004} $    &    $  0.4448^{+0.0074\,+0.0033}_{-0.0285\,-0.0045} $    &    $  0.161^{+0.001\,+0.001}_{-0.002\,-0.002} $    &   1.17\\
 $\sigma_{tW,\,tW}$           &    $  0.0754^{+0.0231\,+0.0023}_{-0.0165\,-0.0025} $    &    $  0.033^{+0.000\,+0.001}_{-0.000\,-0.001} $    &    $  0.0906^{+0.0019\,+0.0005}_{-0.0063\,-0.0009} $    &    $  0.033^{+0.000\,+0.000}_{-0.000\,-0.000} $    &   1.20\\
\hline
\end{tabular}
    \end{subtable}
    \caption{LO and NLO integrated cross-section predictions for $pp \to t\bar{t}H+X$ (upper table) and $pp \to e^+\nu_e\, \mu^-\bar{\nu}_{\mu}\, b\bar{b} \,H +X$ (lower table) at the LHC with $\sqrt{s}=13.6$ TeV. Results are provided for the linear, cross and  quadratic terms. They are presented for the  NNPDF3.1 PDF set and evaluated using $\mu_R = \mu_F = m_t + m_H/2$ as well as $\mu_{EFT} = m_t$. The SM results are also given for comparison purposes. The first reported uncertainty refers to the standard 7-point scale variation, while the second one refers to the $\mu_{EFT}$ scale dependence.}
    \label{tab:ttH_xsec}
\end{table}

To begin with, we can note that some quadratic and/or cross contributions are not small compared to the linear terms. This is particularly visible in the case of the $\sigma_{tG, \, tG}$ term. A comment is in order here. Even though we only have single dimension‑6 insertions at the amplitude level, the cross section automatically contains both the linear and quadratic/cross contributions of dimension‑6 operators. Having the $\sigma_{ij}$ and $\sigma_{ii}$ contribution larger than the $\sigma_i$ interference is a warning sign, but it does not  necessarily signal a breakdown of the EFT expansion. Keeping only the $\sigma_{ij}$ and $\sigma_{ii}$ terms at ${\cal O}(1/\Lambda^4)$  and dropping genuine dimension-8  contributions, stemming from the interference between the SM and dimension‑8 operators, is not a consistent truncation in strict EFT power counting. The EFT expansion is under control if the neglected terms are small compared to those included. For a dimension‑6 truncation this means that all contributions at ${\cal O}(1/\Lambda^4)$ must be smaller than the contributions included at ${\cal O}(1/\Lambda^2)$. However, there are special cases in which the behaviour of the $\sigma_{ij}$ and $\sigma_{ii}$ terms can be instructive or even helpful. In fact, there are many reasons why the $C_i^{(6)}C_{j}^{(6)}/\Lambda^4$ terms, being partial dimension-8 contributions, are of similar size or even larger than the $C_i^{(6)}/\Lambda^2$ terms, see e.g. Refs.  \cite{Contino:2016jqw,Azatov:2016sqh,BessidskaiaBylund:2016jvp}. We could have suppressed interference terms, for example, due to the regions of phase space we investigated, e.g. in the high-energy tails, where the SM amplitude is suppressed. One can imagine special phase-space cuts aimed at reducing the SM background, which can also reduce SM interference with dimension-6 operators, making the quadratic terms relatively important. There may also be accidental cancellations due to different diagrams (SM and SMEFT) which may interfere destructively, reducing the size of some linear terms while quadratic terms remain unaffected or even increase their contribution. Finally, experimental constraints are not yet that tight, so we can have a situation where some Wilson coefficient is large and we have $|C_i^{(6)}| \gg 1$. Then the quadratic contribution can dominate even if $E^2/\Lambda^2$ is small, where $E$ is the typical energy scale of the process. Therefore, we present results for all terms, but keep them separate.
\begin{figure}[t!]
\centering
\includegraphics[width=0.75\textwidth]{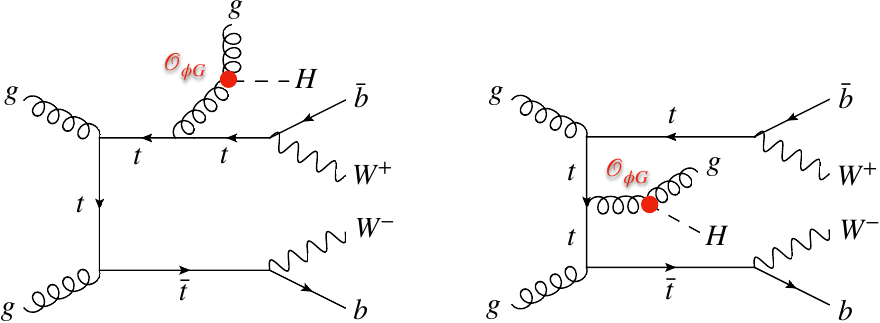} 
\caption{Examples of effective-operator insertions of ${\cal O}_{\phi G}$ for the  $pp \to t\bar{t}H+X$ process in the NWA that open up at NLO in QCD. The leptonic decays of the $W$ gauge bosons considered in our study are omitted here for simplicity.}
\label{fig:diagram_cpg_large_kfactor}
\end{figure}

We can further observe that the relative contributions due to the linear terms $\sigma_{t\phi}$ and $\sigma_{\phi G}$ are very similar for both cases $pp\to t\bar{t}H$ and $pp\to e^+ \nu_e \, \mu^- \bar{\nu}_\mu \, b\bar{b} \, H$ at LO and NLO in QCD. Thus, these two contributions do not change significantly when specific phase-space regions are explored and higher-order corrections in the decays are included.  This pattern extends further to the corresponding quadratic terms, $\sigma_{t\phi,t\phi}$ and $\sigma_{\phi G,\phi G}$.  The situation is quite different for the $\sigma_{tG}$ cross-section result, whose impact increases significantly when the full decay chain is taken into account. Moreover, in this case we also have a completely new contribution coming from the linear $\sigma_{tW}$ term, which is missing in the case of stable top quarks and whose impact is substantial. For the SMEFT contributions the ${\cal K}$-factors range from ${\cal K} = (1.25-1.99)$ depending on the operator for the stable case to ${\cal K}= (1.17-1.87)$ for the NWA case, and can be substantially different  from the corresponding SM cases. Indeed, for the SM case we receive rather moderate higher order corrections that are ${\cal K} = 1.27$ and ${\cal K}=1.19$ for $pp\to t\bar{t}H+X$ and $pp\to e^+ \nu_e \, \mu^- \bar{\nu}_\mu \, b\bar{b} \, H+X$, respectively. For all cases considered the NLO scale uncertainties are significantly reduced, highlighting the importance of adding higher-order corrections. However, also in this case, for some partial contributions, the magnitude of theoretical uncertainties resulting from scale dependence differs significantly from the SM case. This applies both to the stable case and to the NWA results. In more detail, the dependence on $\mu_R$ and $\mu_F$ scales represents the dominant source of the theoretical uncertainty for all cases both at LO and NLO. However, when the (N)LO cross-section results are normalised to the corresponding SM case, the theoretical uncertainties resulting from $\mu_{EFT}$ usually become comparable in magnitude. At the LO level, in some cases they become a dominant source of the theoretical uncertainty. It is also clear that, compared with the stable case, the relative NLO scale uncertainties for the NWA case are reduced for all the SM and SMEFT contributions. The consistent inclusion of the SMEFT operators in the production and decays, as well as the calculation of higher-order corrections in fiducial phase-space regions, brings a number of benefits. Firstly, these results can be directly compared with experimental measurements carried out by the ATLAS and CMS collaborations. Secondly, the overall level of the estimated theoretical uncertainties is substantially reduced. Finally, this may have a significant impact on the unfolding of the top quark from the fiducial to full phase space, which can be studied in a consistent manner without neglecting SMEFT effects in top-quark decays, as is currently the case.   

Comparing the different linear terms, we can observe that the contribution induced by the operator $O_{\phi G}$ has the largest ${\cal K}$-factor. Indeed, in both the stable and NWA cases, the higher-order corrections are of the order of $60\%$. This behaviour carries over to cross and quadratic terms where this operator appears. In particular, for the $\sigma_{\phi G, \, \phi G}$ case NLO QCD corrections are of the order of $90\%-100\%$. We can also note that these  contributions  are the only ones where the uncertainties in the cross-section ratios due to the $\mu_R/\mu_F$ scale variations might be larger at NLO than at LO. We can attribute this behaviour to  genuine tree-order contributions  that open up at NLO in QCD. Representative examples of Feynman diagrams belonging to this category are shown in Figure \ref{fig:diagram_cpg_large_kfactor}. These diagrams contribute to the real emission part of the NLO calculation, but no QCD splitting can connect them to any Born-level diagram. In this sense, they appear at NLO  for the first time and induce scale uncertainties comparable in size to those at LO. Such contributions are also responsible for the anomalously large ${\cal K}$-factors observed for all the terms induced by the $\mathcal{O}_{\phi G}$ operator. 
\begin{table}[t!]
\centering\renewcommand{\arraystretch}{1.5}
\begin{tabular}{|c|l|c|c|c|}
\hline
$p_T^{veto}$ [GeV] &  & $\sigma^{\rm LO}$ [fb]  & $\sigma^{\rm NLO}$ [fb]  & $\mathcal{K}$ \\
\hline
\hline
 & $\sigma_{\rm SM}$  &  $  2.3166^{+0.7068\,+0.0000}_{-0.5044\,-0.0000} $  &  $  2.7680^{+0.0560\,+0.0000}_{-0.1898\,-0.0000} $    &   1.19 \\[0.2cm]
-  & $\sigma_{\phi G}$  &  $  1.6262^{+0.5137\,+0.2089}_{-0.3638\,-0.1730} $  &  $  2.5292^{+0.2865\,+0.1154}_{-0.3098\,-0.0954} $    &   1.56 \\[0.2cm]
 & $\sigma_{\phi G, \, \phi G}$  &  $  0.8178^{+0.2987\,+0.1799}_{-0.2032\,-0.1361} $  &  $  1.5299^{+0.2443\,+0.1528}_{-0.2377\,-0.1277} $   &   1.87 \\[0.2cm]
\hline
\hline
 & $\sigma_{\rm SM}$  &  $  2.3166^{+0.7068\,+0.0000}_{-0.5044\,-0.0000} $   &   $  2.2694^{+0.0171\,+0.0000}_{-0.2326\,-0.0000} $   &   0.98\\[0.2cm]
150  & $\sigma_{\phi G}$   &  $  1.6262^{+0.5137\,+0.2089}_{-0.3638\,-0.1730} $    &   $  2.0748^{+0.0601\,+0.0583}_{-0.1642\,-0.0486} $    &   1.28\\[0.2cm]
 & $\sigma_{\phi G, \,\phi G}$  &  $  0.8178^{+0.2987\,+0.1799}_{-0.2032\,-0.1361} $    &   $  1.2207^{+0.0738\,+0.0840}_{-0.1339\,-0.0761} $  &   1.49\\[0.2cm]
\hline
\hline
 & $\sigma_{\rm SM}$  &  $  2.3166^{+0.7068\,+0.0000}_{-0.5044\,-0.0000} $  & $  1.9589^{+0.0652\,+0.0000}_{-0.3863\,-0.0000} $    &   0.85 \\[0.2cm]
100  & $\sigma_{\phi G}$   &  $  1.6262^{+0.5137\,+0.2089}_{-0.3638\,-0.1730} $   &   $  1.8239^{+0.0032\,+0.0259}_{-0.0863\,-0.0220} $    &   1.12 \\[0.2cm]
 & $\sigma_{\phi G, \, \phi G}$  & $  0.8178^{+0.2987\,+0.1799}_{-0.2032\,-0.1361} $   &  $  1.0616^{+0.0209\,+0.0483}_{-0.0815\,-0.0494} $    &   1.30 \\[0.2cm]
\hline
\end{tabular}
\caption{Integrated cross-section predictions at LO and NLO in QCD for the $pp\to e^+\nu_e\, \mu^- \bar{\nu}_\mu \, b\bar{b} \, H+X$ process in the NWA at the LHC with $\sqrt{s}=13.6$ TeV. Results are provided for the following three terms $\sigma_{\rm SM}$, $\sigma_{\phi G}$ and $\sigma_{\phi G, \, \phi G}$. They are presented for the  NNPDF3.1 PDF set and evaluated using $\mu_R = \mu_F = m_t + m_H/2$ as well as $\mu_{EFT} = m_t$. The first reported uncertainty refers to the standard 7-point scale variation, while the second one to the $\mu_{EFT}$ scale dependence. In addition to the default case, the result with two different jet-veto values, $p_T^{veto}= 150$ GeV and $p_T^{veto}= 100$ GeV, are given.}
\label{tab:jet_veto}
\end{table}

To analyse this further, we impose a jet veto on an additional resolved jet from the real emission part of the NLO computation for the following linear and quadratic terms $\sigma_{\phi G}$ and $\sigma_{\phi G, \,\phi G}$. In practice, we require that this additional jet satisfies the following condition $p_T(j) < p_T^{veto}$, so that the soft/collinear regions necessary for infrared safety remain unchanged. In Table \ref{tab:jet_veto} we present the integrated fiducial cross-section predictions at LO and NLO in QCD for the  $pp\to e^+\nu_e\, \mu^- \bar{\nu}_\mu \, b\bar{b} \, H+X$ process in NWA with and without the jet veto. In particular, we compare our previous findings with the results obtained for two different values of the  $p_T^{veto}$ cut, i.e., we use $p_T^{veto}= 150$ GeV and $p_T^{veto} = 100$ GeV. For information purposes, the SM cross-section result is also provided. In all cases, the dominant uncertainties arise from the variation of the renormalisation and factorisation scales. For $\sigma_{\phi G}$, we observe a reduction of the NLO scale uncertainty from $12\%$ to $8\%$ ($5\%$) when imposing $p_T^{veto} = 150$ ($100$) GeV. Similarly, for the  $\sigma_{\phi G, \,\phi G}$ case the uncertainties are reduced from $16\%$ to $11\%$ ($8\%$), respectively.  Also the size of NLO QCD corrections shows a reduction from $56\%$ to $28\%$ ($12\%$) for $\sigma_{\phi G}$, and from $87\%$ to $49\%$ ($30\%$) for $\sigma_{\phi G,\, \phi G}$.  It is important to remember that there is a trade-off between suppressing QCD corrections by the jet veto and perturbative stability of the results. This is obvious when looking at the SM cross-section results, whose uncertainties increase from $7\%$ to $10\%$ ($20\%$). Nevertheless, there are clear indications that the large uncertainties observed in the $\mathcal{O}_{\phi G}$ contributions are induced by contributions from the real-emissions part of the NLO calculations.
\begin{figure}[t!]
\centering
\includegraphics[width=0.65\textwidth]{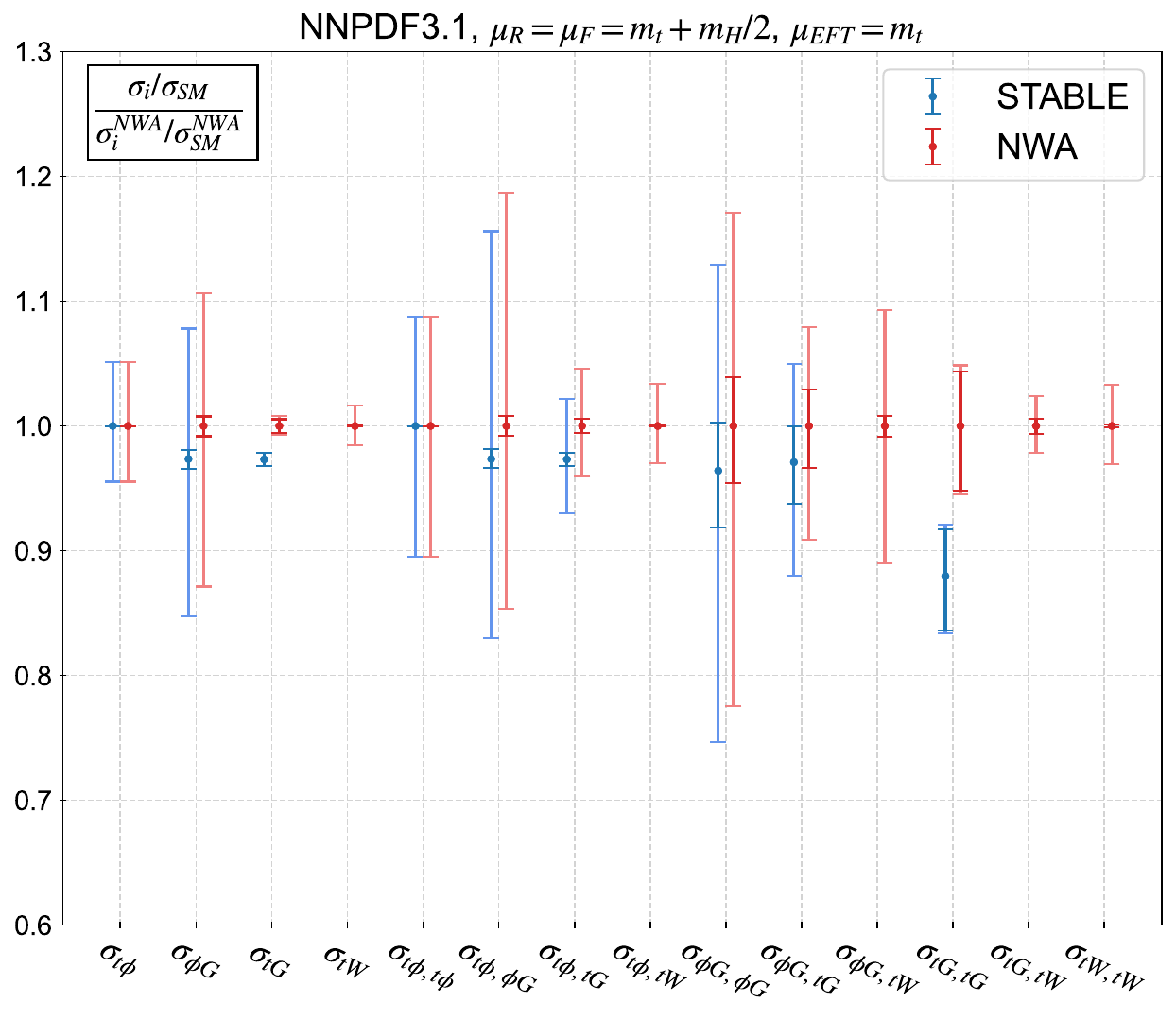}  
\includegraphics[width=0.65\textwidth]{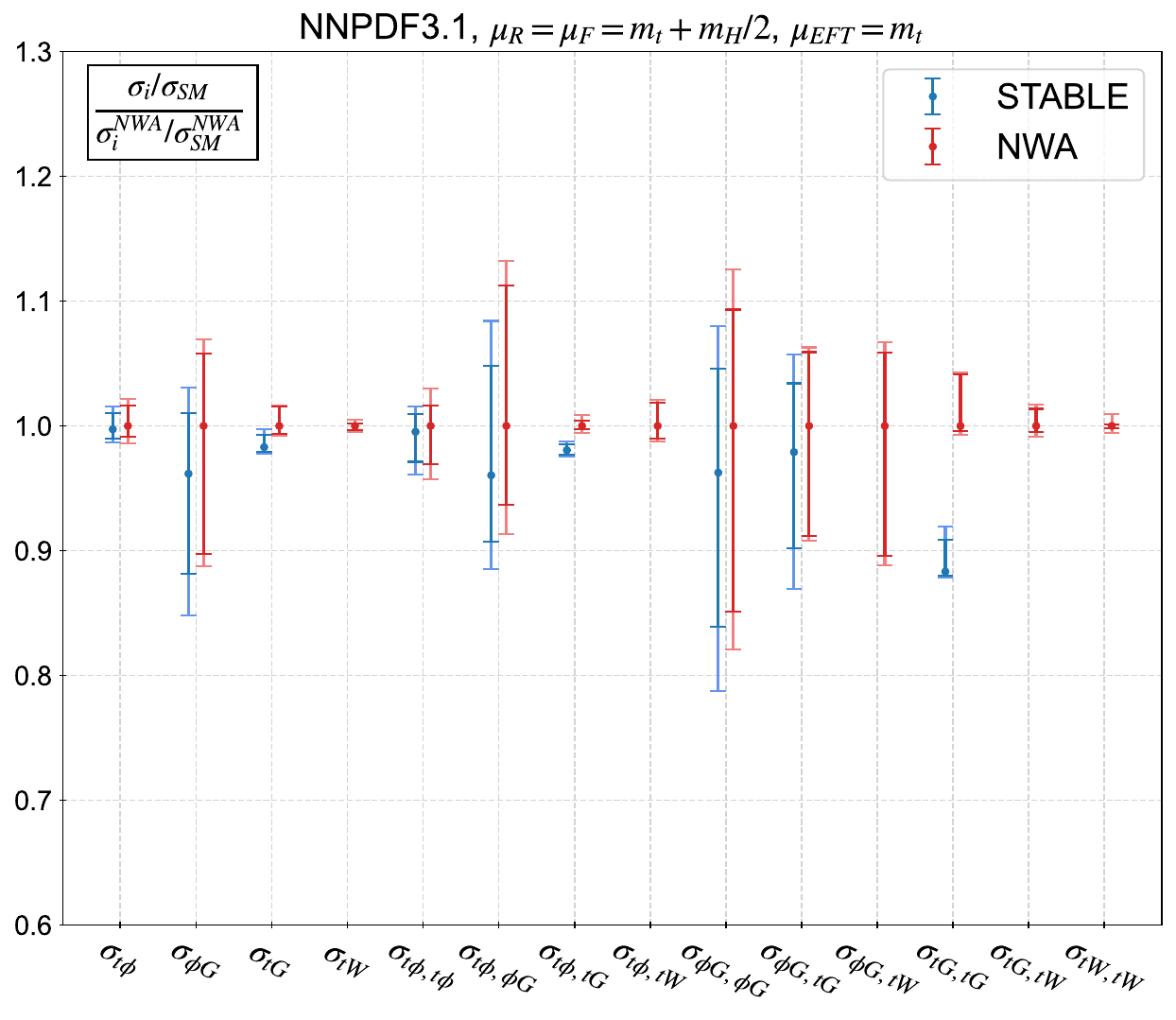} 
\caption{Integrated cross-section predictions at LO (upper plot) and NLO in QCD (lower plot) for $pp \to t\bar{t}H+X$ {\rm (STABLE)} and $pp \to e^+\nu_e\, \mu^-\bar{\nu}_{\mu}\, b\bar{b} \,H +X$ {\rm (NWA)} at the LHC with $\sqrt{s}=13.6$ TeV presented in the form of a double ratio defined according to $\mathcal{D}_i \equiv \frac{\sigma_{\rm i}/\sigma_{\rm SM}}{\sigma^{\rm NWA}_{\rm i}/\sigma^{\rm NWA}_{\rm SM}}$. Results are given for the linear, cross and quadratic terms. They are provided  for the NNPDF3.1 PDF set and evaluated using $\mu_R = \mu_F = m_t + m_H/2$ as well as $\mu_{EFT} = m_t$. Also presented are theoretical uncertainties. The inner bars represent the 7-point scale variation, while the outer bars also include the uncertainty resulting from the $\mu_{EFT}$ scale variation.}
\label{fig:double_ratio_LO_NLO}
\end{figure}
\begin{figure}[t!]
\centering
\includegraphics[width=0.65\textwidth]{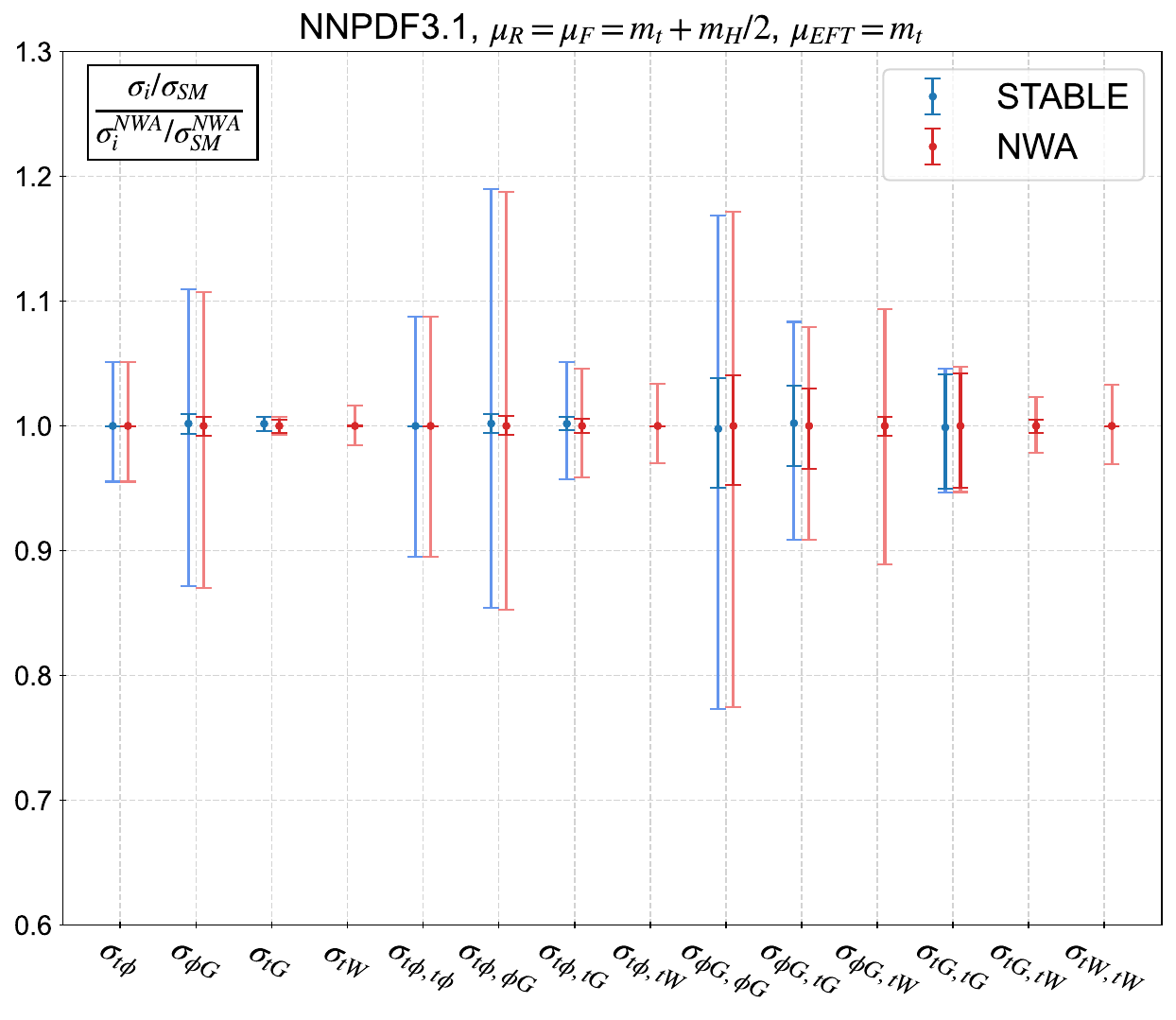} 
\caption{Integrated cross-section predictions at LO for $pp \to t\bar{t}H+X$ {\rm (STABLE)} and $pp \to e^+\nu_e\, \mu^-\bar{\nu}_{\mu}\, b\bar{b} \,H +X$ {\rm (NWA)} at the LHC with $\sqrt{s}=13.6$ TeV presented in the form of a double ratio defined according to $\mathcal{D}_i \equiv \frac{\sigma_{\rm i}/\sigma_{\rm SM}}{\sigma^{\rm NWA}_{\rm i}/\sigma^{\rm NWA}_{\rm SM}}$. In the case of NWA predictions, no kinematic cuts are applied to the top-quark decay products. Results are given for the linear, cross and quadratic terms. They are provided  for the NNPDF3.1 PDF set and evaluated using $\mu_R = \mu_F = m_t + m_H/2$ as well as $\mu_{EFT} = m_t$. Also presented are theoretical uncertainties. The inner bars represent the 7-point scale variation, while the outer bars also include the uncertainties resulting from the $\mu_{EFT}$ scale variation.}
\label{fig:double_ratio_LO_inclusive}
\end{figure}
\begin{table}[t!]
    \centering
    \renewcommand{\arraystretch}{1.5}
    \small
    \begin{subtable}[t]{\linewidth}        
\begin{tabular}{|c|rrrrc|}
 \hline
 \hline
 \multicolumn{6}{|c|}{$pp \to t\bar{t}H+X$ $\quad$ (\textsc{Stable})} \\
 \hline
 \hline
 & \multicolumn{1}{c}{$\sigma^{\textrm{LO}}$[pb]} & \multicolumn{1}{c}{$\sigma^{\textrm{LO}}/\sigma^{\textrm{LO}}_{\textrm{SM}}$} & \multicolumn{1}{c}{$\sigma^{\textrm{NLO}}$[pb]} & \multicolumn{1}{c}{$\sigma^{\textrm{NLO}}/\sigma^{\textrm{NLO}}_{\textrm{SM}}$} & $\mathcal{K}$\\
 \hline
 $\sigma_{SM}$              &    $  0.4424^{+0.1337\,+0.0000}_{-0.0956\,-0.0000} $    &    $  1.000^{+0.000\,+0.000}_{-0.000\,-0.000} $    &    $  0.5623^{+0.0327\,+0.0000}_{-0.0519\,-0.0000} $    &    $  1.000^{+0.000\,+0.000}_{-0.000\,-0.000} $    &   1.27\\
\rowcolor{gray!15}
 $\sigma_{t\phi}$           &    $ -0.0541^{+0.0117\,+0.0023}_{-0.0164\,-0.0027} $    &    $ -0.122^{+0.000\,+0.005}_{-0.000\,-0.006} $    &    $ -0.0724^{+0.0074\,+0.0008}_{-0.0056\,-0.0011} $    &    $ -0.129^{+0.001\,+0.001}_{-0.002\,-0.002} $    &   1.34\\
 $\sigma_{\phi G}$          &    $  0.3023^{+0.0945\,+0.0375}_{-0.0671\,-0.0312} $    &    $  0.683^{+0.005\,+0.085}_{-0.005\,-0.071} $    &    $  0.5095^{+0.0793\,+0.0236}_{-0.0727\,-0.0197} $    &    $  0.906^{+0.083\,+0.042}_{-0.050\,-0.035} $    &   1.69\\
\rowcolor{gray!15}
 $\sigma_{tG}$              &    $  0.4501^{+0.1393\,+0.0004}_{-0.0992\,-0.0008} $    &    $  1.017^{+0.006\,+0.001}_{-0.005\,-0.002} $    &    $  0.5638^{+0.0266\,+0.0022}_{-0.0499\,-0.0025} $    &    $  1.003^{+0.004\,+0.004}_{-0.010\,-0.004} $    &   1.25\\
 $\sigma_{t\phi,t\phi}$     &    $  0.0017^{+0.0005\,+0.0002}_{-0.0004\,-0.0001} $    &    $  0.004^{+0.000\,+0.000}_{-0.000\,-0.000} $    &    $  0.0023^{+0.0002\,+0.0001}_{-0.0003\,-0.0001} $    &    $  0.004^{+0.000\,+0.000}_{-0.000\,-0.000} $    &   1.40\\
\rowcolor{gray!15}
 $\sigma_{t\phi,\phi G}$    &    $ -0.0185^{+0.0041\,+0.0026}_{-0.0058\,-0.0033} $    &    $ -0.042^{+0.000\,+0.006}_{-0.000\,-0.007} $    &    $ -0.0324^{+0.0048\,+0.0020}_{-0.0054\,-0.0023} $    &    $ -0.058^{+0.004\,+0.004}_{-0.006\,-0.004} $    &   1.75\\
 $\sigma_{t\phi,tG}$        &    $ -0.0275^{+0.0061\,+0.0012}_{-0.0085\,-0.0013} $    &    $ -0.062^{+0.000\,+0.003}_{-0.000\,-0.003} $    &    $ -0.0363^{+0.0036\,+0.0003}_{-0.0024\,-0.0004} $    &    $ -0.065^{+0.000\,+0.000}_{-0.001\,-0.001} $    &   1.32\\
\rowcolor{gray!15}
 $\sigma_{\phi G,\phi G}$   &    $  0.1506^{+0.0548\,+0.0317}_{-0.0373\,-0.0242} $    &    $  0.340^{+0.016\,+0.072}_{-0.014\,-0.055} $    &    $  0.3122^{+0.0648\,+0.0316}_{-0.0552\,-0.0264} $    &    $  0.555^{+0.078\,+0.056}_{-0.052\,-0.047} $    &   2.07\\
 $\sigma_{\phi G,tG}$       &    $  0.2958^{+0.1027\,+0.0247}_{-0.0709\,-0.0215} $    &    $  0.669^{+0.023\,+0.056}_{-0.020\,-0.049} $    &    $  0.5070^{+0.0754\,+0.0139}_{-0.0743\,-0.0116} $    &    $  0.902^{+0.077\,+0.025}_{-0.054\,-0.021} $    &   1.71\\
\rowcolor{gray!15}
 $\sigma_{tG,tG}$           &    $  0.5852^{+0.2145\,+0.0102}_{-0.1459\,-0.0113} $    &    $  1.323^{+0.065\,+0.023}_{-0.056\,-0.025} $    &    $  0.7746^{+0.0320\,+0.0050}_{-0.0732\,-0.0066} $    &    $  1.378^{+0.007\,+0.009}_{-0.045\,-0.012} $    &   1.32\\
\hline
\end{tabular}
    \end{subtable} \\
    \vspace{0.3cm}
    \begin{subtable}[t]{\linewidth}        
\begin{tabular}{|c|rrrrc|}
 \hline
 \hline
 \multicolumn{6}{|c|}{$pp \to e^+\nu_e\,\mu^-\bar{\nu}_{\mu}\,b\bar{b} \, H + X$  $\quad$ (NWA)} \\
 \hline
 \hline
 & \multicolumn{1}{c}{$\sigma^{\textrm{LO}}$[fb]} & \multicolumn{1}{c}{$\sigma^{\textrm{LO}}/\sigma^{\textrm{LO}}_{\textrm{SM}}$} & \multicolumn{1}{c}{$\sigma^{\textrm{NLO}}$[fb]} & \multicolumn{1}{c}{$\sigma^{\textrm{NLO}}/\sigma^{\textrm{NLO}}_{\textrm{SM}}$} & $\mathcal{K}$\\
 \hline
 $\sigma_{SM}$              &    $  2.3166^{+0.7068\,+0.0000}_{-0.5044\,-0.0000} $    &    $  1.000^{+0.000\,+0.000}_{-0.000\,-0.000} $    &    $  2.7680^{+0.0560\,+0.0000}_{-0.1898\,-0.0000} $    &    $  1.000^{+0.000\,+0.000}_{-0.000\,-0.000} $    &   1.19\\
\rowcolor{gray!15}
 $\sigma_{t\phi}$           &    $ -0.2835^{+0.0617\,+0.0122}_{-0.0865\,-0.0139} $    &    $ -0.122^{+0.000\,+0.005}_{-0.000\,-0.006} $    &    $ -0.3578^{+0.0287\,+0.0033}_{-0.0117\,-0.0045} $    &    $ -0.129^{+0.002\,+0.001}_{-0.003\,-0.002} $    &   1.26\\
 $\sigma_{\phi G}$          &    $  1.6262^{+0.5137\,+0.2000}_{-0.3638\,-0.1668} $    &    $  0.702^{+0.006\,+0.086}_{-0.005\,-0.072} $    &    $  2.6114^{+0.3238\,+0.1102}_{-0.3337\,-0.0919} $    &    $  0.943^{+0.107\,+0.040}_{-0.060\,-0.033} $    &   1.61\\
\rowcolor{gray!15}
 $\sigma_{tG}$              &    $  2.4217^{+0.7568\,+0.0105}_{-0.5374\,-0.0131} $    &    $  1.045^{+0.006\,+0.005}_{-0.006\,-0.006} $    &    $  2.8196^{+0.0454\,+0.0130}_{-0.1757\,-0.0153} $    &    $  1.019^{+0.007\,+0.005}_{-0.017\,-0.006} $    &   1.16\\
 $\sigma_{tW}$              &    $  0.7238^{+0.2214\,+0.0107}_{-0.1579\,-0.0115} $    &    $  0.312^{+0.000\,+0.005}_{-0.000\,-0.005} $    &    $  0.8669^{+0.0178\,+0.0019}_{-0.0600\,-0.0033} $    &    $  0.313^{+0.000\,+0.001}_{-0.000\,-0.001} $    &   1.20\\
\rowcolor{gray!15}
 $\sigma_{t\phi,t\phi}$     &    $  0.0087^{+0.0026\,+0.0009}_{-0.0019\,-0.0007} $    &    $  0.004^{+0.000\,+0.000}_{-0.000\,-0.000} $    &    $  0.0115^{+0.0006\,+0.0003}_{-0.0010\,-0.0003} $    &    $  0.004^{+0.000\,+0.000}_{-0.000\,-0.000} $    &   1.33\\
 $\sigma_{t\phi,\phi G}$    &    $ -0.0995^{+0.0223\,+0.0140}_{-0.0314\,-0.0177} $    &    $ -0.043^{+0.000\,+0.006}_{-0.000\,-0.008} $    &    $ -0.1663^{+0.0223\,+0.0095}_{-0.0228\,-0.0110} $    &    $ -0.060^{+0.004\,+0.003}_{-0.008\,-0.004} $    &   1.67\\
\rowcolor{gray!15}
 $\sigma_{t\phi,tG}$        &    $ -0.1482^{+0.0329\,+0.0058}_{-0.0463\,-0.0064} $    &    $ -0.064^{+0.000\,+0.002}_{-0.000\,-0.003} $    &    $ -0.1822^{+0.0136\,+0.0007}_{-0.0037\,-0.0011} $    &    $ -0.066^{+0.000\,+0.000}_{-0.001\,-0.000} $    &   1.23\\
 $\sigma_{t\phi,tW}$        &    $ -0.0443^{+0.0097\,+0.0013}_{-0.0135\,-0.0014} $    &    $ -0.019^{+0.000\,+0.001}_{-0.000\,-0.001} $    &    $ -0.0559^{+0.0045\,+0.0003}_{-0.0018\,-0.0005} $    &    $ -0.020^{+0.000\,+0.000}_{-0.000\,-0.000} $    &   1.26\\
\rowcolor{gray!15}
 $\sigma_{\phi G,\phi G}$   &    $  0.8178^{+0.2987\,+0.1718}_{-0.2032\,-0.1314} $    &    $  0.353^{+0.016\,+0.074}_{-0.014\,-0.057} $    &    $  1.6003^{+0.2802\,+0.1528}_{-0.2598\,-0.1286} $    &    $  0.578^{+0.095\,+0.055}_{-0.058\,-0.046} $    &   1.96\\
 $\sigma_{\phi G,tG}$       &    $  1.5954^{+0.5561\,+0.1296}_{-0.3835\,-0.1132} $    &    $  0.689^{+0.023\,+0.056}_{-0.020\,-0.049} $    &    $  2.5519^{+0.2729\,+0.0593}_{-0.3254\,-0.0488} $    &    $  0.922^{+0.089\,+0.021}_{-0.058\,-0.018} $    &   1.60\\
\rowcolor{gray!15}
 $\sigma_{\phi G,tW}$       &    $  0.2548^{+0.0807\,+0.0268}_{-0.0571\,-0.0227} $    &    $  0.110^{+0.001\,+0.012}_{-0.001\,-0.010} $    &    $  0.4091^{+0.0508\,+0.0146}_{-0.0524\,-0.0120} $    &    $  0.148^{+0.017\,+0.005}_{-0.009\,-0.004} $    &   1.61\\
 $\sigma_{tG,tG}$           &    $  3.4828^{+1.2967\,+0.0635}_{-0.8776\,-0.0700} $    &    $  1.503^{+0.077\,+0.027}_{-0.066\,-0.030} $    &    $  4.3143^{+0.0377\,+0.0225}_{-0.2985\,-0.0320} $    &    $  1.559^{+0.009\,+0.008}_{-0.071\,-0.012} $    &   1.24\\
\rowcolor{gray!15}
 $\sigma_{tG,tW}$           &    $  0.3795^{+0.1189\,+0.0076}_{-0.0843\,-0.0084} $    &    $  0.164^{+0.001\,+0.003}_{-0.001\,-0.004} $    &    $  0.4412^{+0.0071\,+0.0029}_{-0.0274\,-0.0040} $    &    $  0.159^{+0.001\,+0.001}_{-0.003\,-0.001} $    &   1.16\\
 $\sigma_{tW,tW}$           &    $  0.0754^{+0.0231\,+0.0022}_{-0.0165\,-0.0024} $    &    $  0.033^{+0.000\,+0.001}_{-0.000\,-0.001} $    &    $  0.0895^{+0.0018\,+0.0004}_{-0.0060\,-0.0007} $    &    $  0.032^{+0.000\,+0.000}_{-0.000\,-0.000} $    &   1.19\\
\hline
\end{tabular}
    \end{subtable}
    \caption{Integrated cross-section predictions at LO and NLO in QCD for $pp \to t\bar{t}H+X$ (upper table) and $pp \to e^+\nu_e\, \mu^-\bar{\nu}_{\mu}\, b\bar{b} \,H +X$ (lower table) at the LHC with $\sqrt{s}=13.6$ TeV. Results are provided for the linear, cross and  quadratic terms. They are presented for the  NNPDF3.1 PDF set and evaluated using $\mu_R = \mu_F = \mu_{EFT} = m_t + m_H/2$. The SM results are also given for comparison purposes. The first reported uncertainty refers to the standard 7-point scale variation, while the second one refers to the $\mu_{EFT}$ scale dependence.}
    \label{tab:ttH_xsec_muEFT_eq_mu0}
\end{table}

Subsequently, we turn our attention to a graphical representation of the LO and NLO integrated cross-section predictions for $pp \to t\bar{t}H+X$ and $pp \to e^+\nu_e\, \mu^-\bar{\nu}_{\mu}\, b\bar{b} \,H +X$ to investigate their similarities and differences. In Figure \ref{fig:double_ratio_LO_NLO} these theoretical predictions are given in the form of a double ratio  $\mathcal{D}_{i}$ that facilitates the comparison and also has the advantage of offering a more synoptic view. The  double ratio  is defined according to 
\begin{equation}
\mathcal{D}_i \equiv \frac{\sigma_{\rm i}/\sigma_{\rm SM}}{\sigma^{\rm NWA}_{\rm i}/\sigma^{\rm NWA}_{\rm SM}} \,,
\label{eq:double_ratio}
\end{equation}
where $i$ spans the various SMEFT contributions (linear, cross and quadratic terms). The upper and lower plots show LO and NLO results, respectively. For each $\mathcal{D}_{i}$, two error bars are displayed along with the central value. The inner bar denotes the 7-point scale variation, while the outer bar includes the uncertainties coming from the $\mu_{EFT}$ scale variation as well. They are added in quadrature. Note here that the uncertainties are calculated solely based on the numerator in $D_i$. The denominator $\sigma_i^{\rm NWA}/\sigma_{\rm SM}^{\rm NWA}$ simply serves as the normalisation factor. We further emphasize that the contributions induced by $\mathcal{O}_{tW}$ only enter the top-quark decay matrix elements and therefore do not appear in the stable top-quark case.  

We can observe that in several cases, for example for $\sigma_{t\phi}$ and $\sigma_{\phi G}$ as well as for some cross terms and all quadratic terms in which they occur, the fully inclusive and fiducial cross-section results agree surprisingly well, despite the phase-space cut employed in the latter case. This applies to both LO and NLO theoretical predictions.  In other cases, especially for $\sigma_{tG}$ and $\sigma_{tG,\, tG}$, substantial differences between the two approaches can be found. Different aspects contribute the observed differences at LO and NLO. At the LO level, the picture is quite simple. Indeed, the  $\mathcal{O}_{t\phi}$, $\mathcal{O}_{\phi G}$ and $\mathcal{O}_{tG}$ operators enter only in the production part of the calculation, while $\mathcal{O}_{tW}$ appears only in the top-quark decays. Thus, the differences we observed are of a purely kinematic nature and originate from the fiducial phase-space cuts.  We emphasize, however, that the cuts we use are very inclusive, so the differences may be even larger in some specific phase-space regions, e.g. for boosted top quarks. To illustrate this even better, we present in Figure \ref{fig:double_ratio_LO_inclusive} the LO results again, but this time we compare the stable case with the NWA case, where no cuts on the top-quark decay products are applied.  As expected, both results agree well with each other. At the NLO QCD level, various effects are interrelated and influence the final result.  In this case, the following factors should be taken into account simultaneously: (i) kinematic cuts, (ii) the SMEFT contributions in top-quark decays, (iii) NLO QCD corrections to top-quark decays. Distinguishing between all these aspects naturally requires more detailed comparisons of NLO results based on different approaches to modelling top-quark decays and/or different kinematic constraints. Nevertheless, it is already clear that an analysis of the fiducial phase-space regions for the $pp \to t\bar{t}H+X$ process leads to noticeable differences in the interpretation of the cross-section results obtained within the  SMEFT framework. 

We also present in Table \ref{tab:ttH_xsec_muEFT_eq_mu0} the LO and NLO cross-section results  for $pp \to t\bar{t}H+X$ and $pp \to e^+\nu_e\, \mu^-\bar{\nu}_{\mu}\, b\bar{b} \,H +X$  evaluated  using $\mu_R = \mu_F = \mu_{EFT} = m_t + m_H/2$. A comparison of these results with our default predictions obtained with $\mu_R = \mu_F =  m_t + m_H/2$ and $\mu_{EFT} = m_t$ allows us to estimate the effect of logarithms of the form $\log(\mu_{EFT}^2/\mu_R^2)$.  We would like to remind the reader that, in accordance with the parametrisation given in Eq. \eqref{eq:parametrization_cross-section}, the Wilson coefficients have been factored out from the results reported in the Tables. Consequently no difference can be observed in the central value of the LO cross sections. Because the one-loop matrix elements are directly sensitive to $\mu_{EFT}$ via UV vertices, we can see differences in the NLO results up to a few percent. These differences are within the estimated theoretical uncertainties. The largest deviations, amounting to approximately $5\%$, have been observed in the case of  $\sigma_{\phi G}$ and $\sigma_{\phi G, \, \phi G}$. This is hardly surprising, as the difference between $\mu_{EFT} = m_t$ and $\mu_{EFT} = m_t+m_H/2$ is not very large.  The extent of this effect may vary significantly if various dynamic scale settings are used instead.


\section{Differential cross-section predictions}
\label{sec:differential_cross_sections_fiducial}

%
\begin{figure}[t!]
    \centering
    \begin{tabular}{cc}
         \includegraphics[width=0.49\linewidth]{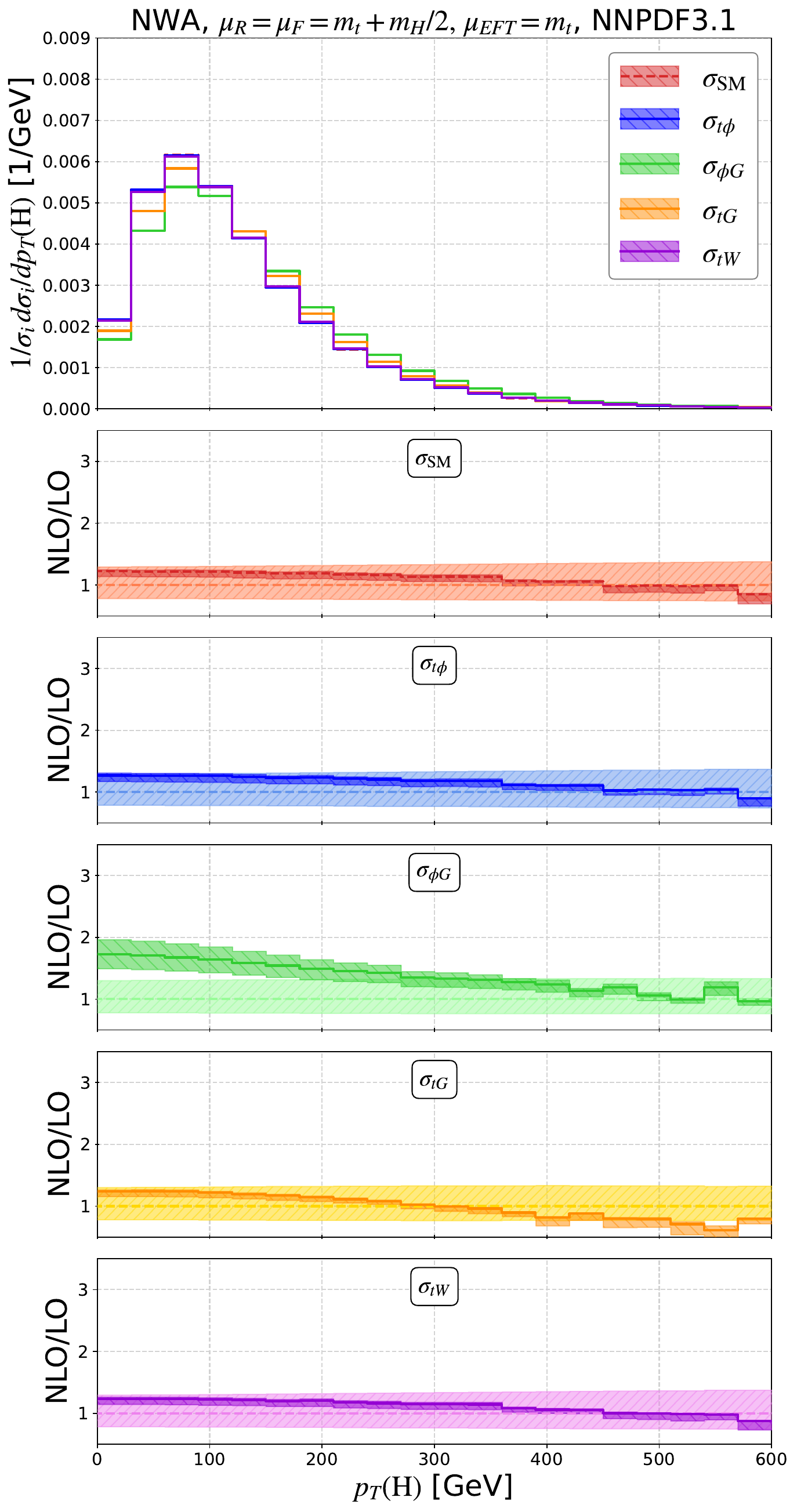} &
         \includegraphics[width=0.49\linewidth]{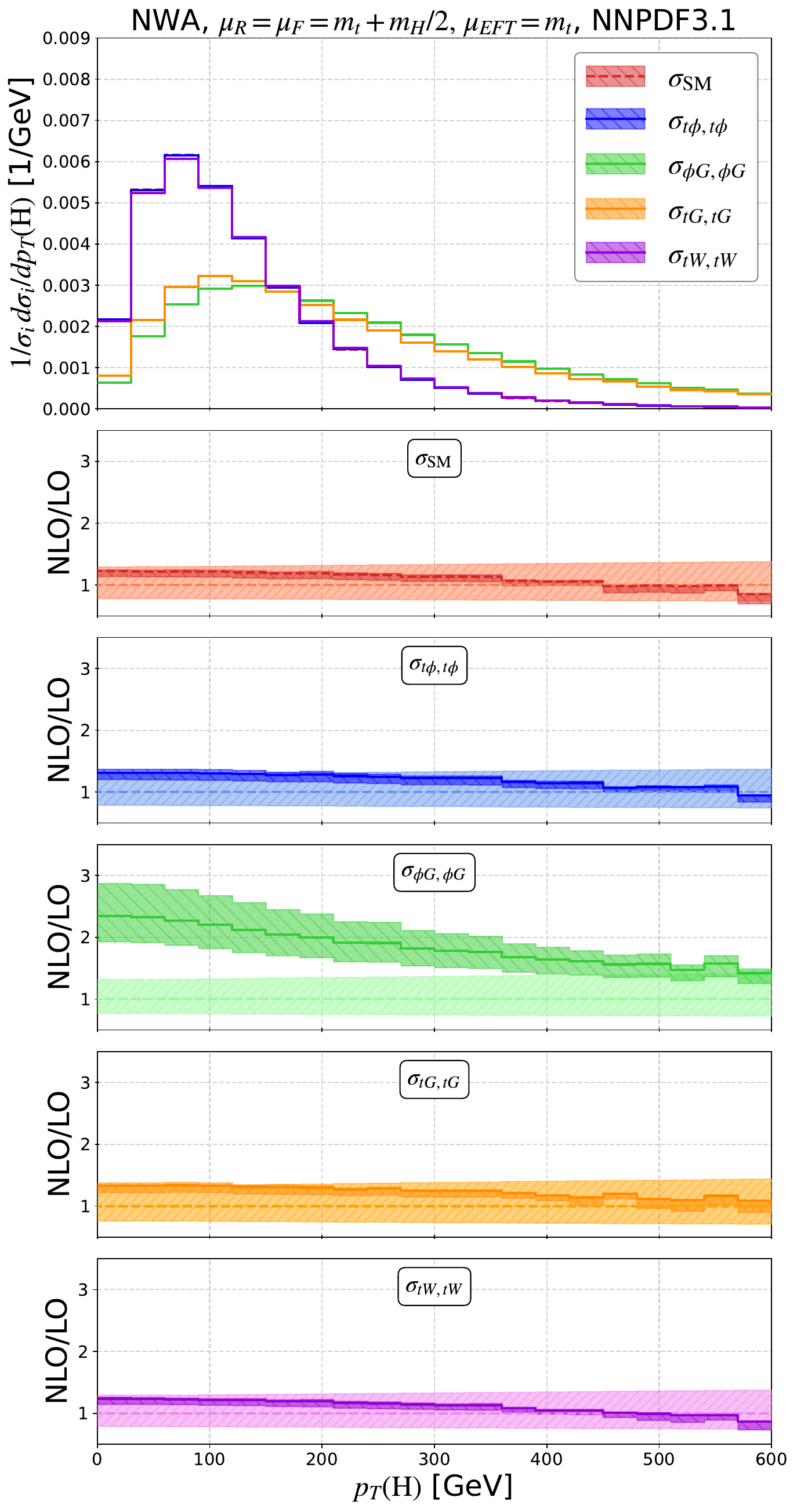}
    \end{tabular}
    \caption{Normalised differential cross-section distributions at NLO in QCD  as a function of $p_T(H)$ for the $pp \to e^+\nu_e\, \mu^-\bar{\nu}_{\mu}\, b\bar{b} \,H +X$ process at the LHC with $\sqrt{s}=13.6$ TeV. The upper panels show the linear $\sigma_i$ (left) and quadratic $\sigma_{ii}$ (right) terms. The lower panels display the differential ${\cal K}$-factors together with their uncertainty bands estimated with the help of the  7-point scale variation and the relative scale uncertainties of the LO cross sections, separately for each operator. The SM result is also plotted for comparison purposes. The $\mu_{EFT}$  scale uncertainties are not shown. Results are presented for the NNPDF3.1 PDF set and evaluated using $\mu_R=\mu_F=m_t+m_H/2$ as well as  $\mu_{EFT}=m_t$.}
    \label{fig:normalized_differential_pT(H)}
\end{figure}
\begin{figure}[t!]
    \centering
    \begin{tabular}{cc}
         \includegraphics[width=0.49\linewidth]{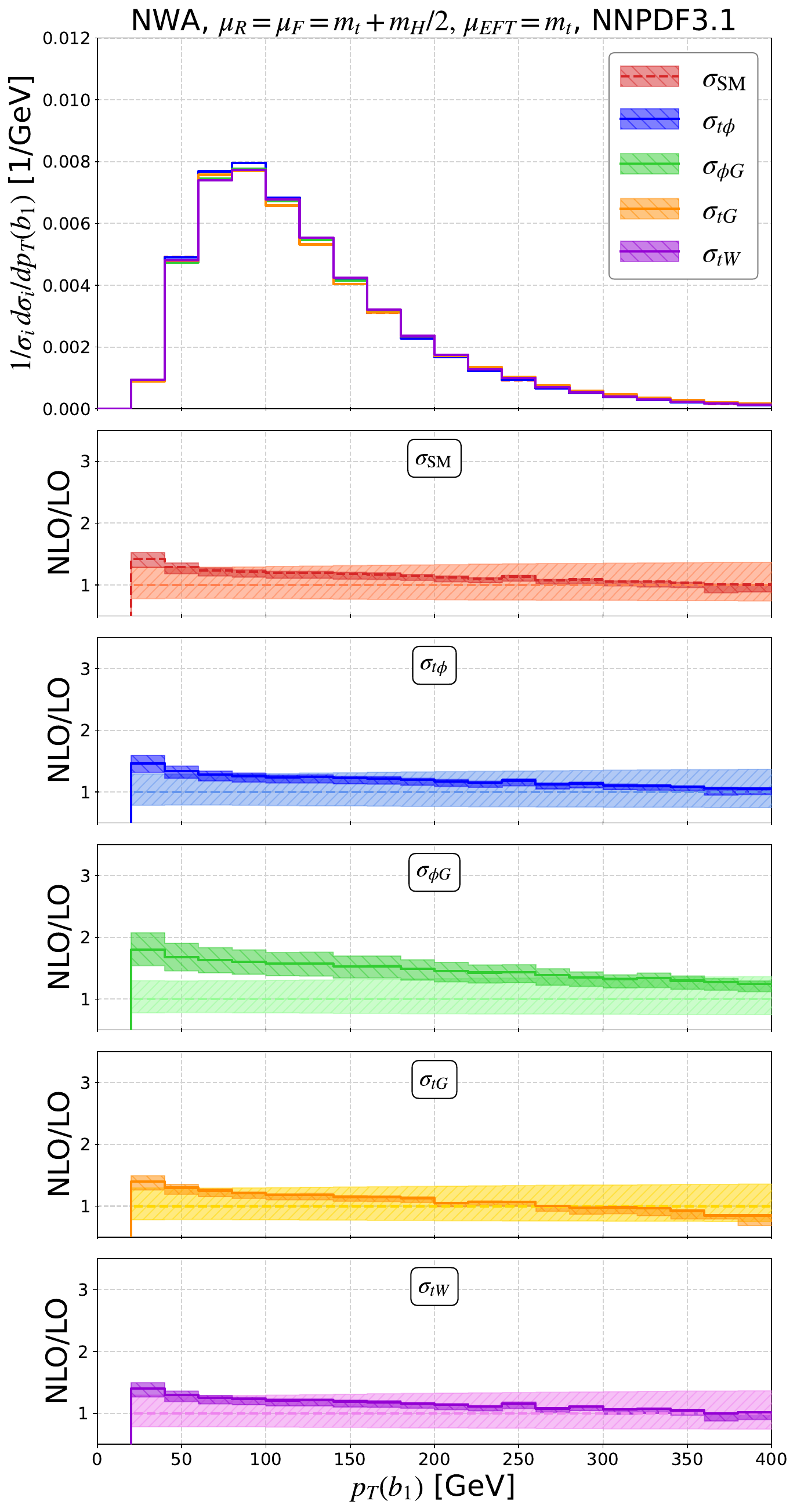} &
         \includegraphics[width=0.49\linewidth]{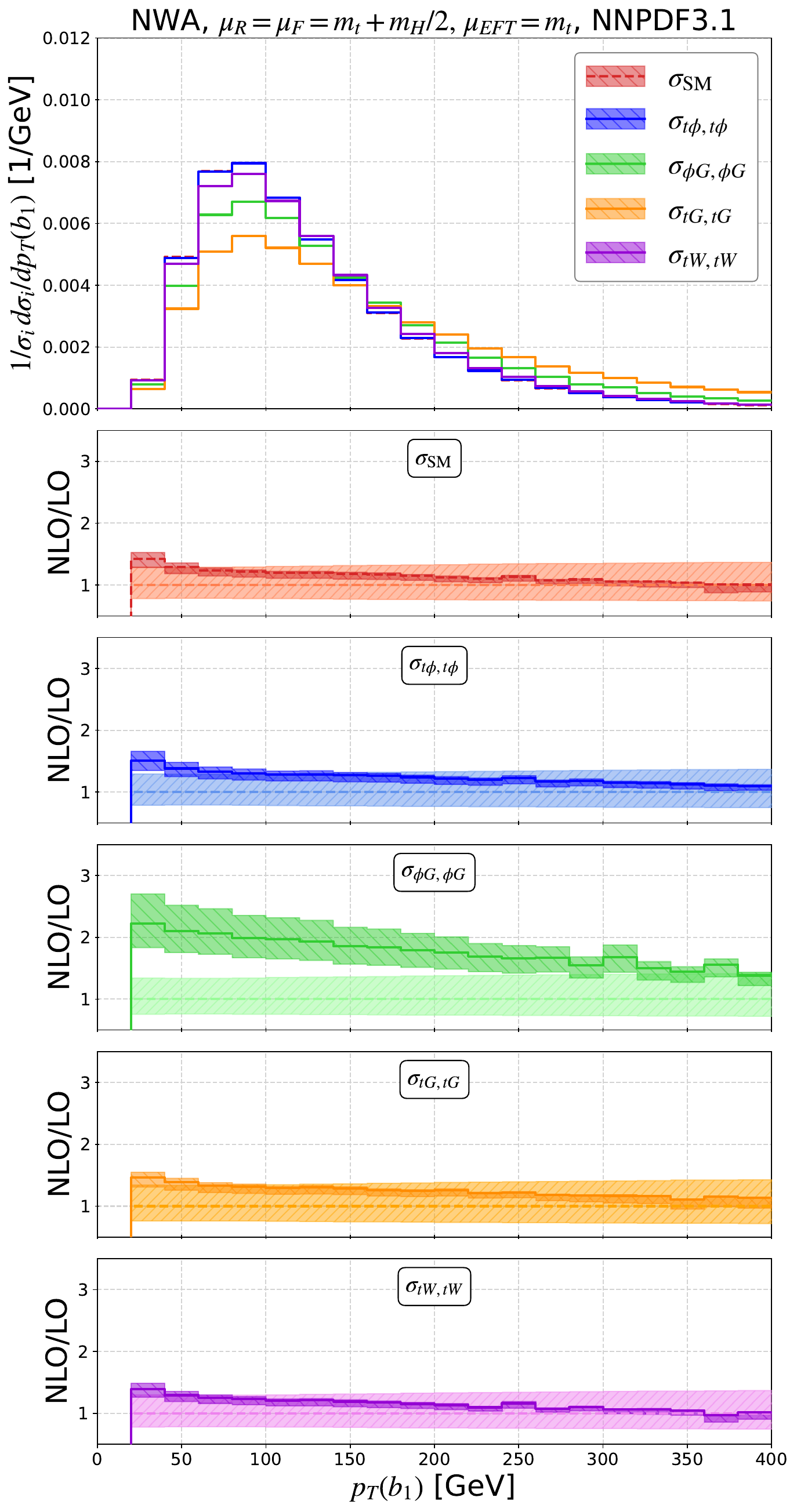}
    \end{tabular}
    \caption{Normalised differential cross-section distributions at NLO in QCD  as a function of $p_T(b_1)$ for the $pp \to e^+\nu_e\, \mu^-\bar{\nu}_{\mu}\, b\bar{b} \,H +X$ process at the LHC with $\sqrt{s}=13.6$ TeV. The upper panels show the linear $\sigma_i$ (left) and quadratic $\sigma_{ii}$ (right) terms. The lower panels display the differential ${\cal K}$-factors together with their uncertainty bands estimated with the help of the  7-point scale variation  and the relative scale uncertainties of the LO cross sections, separately for each operator. The SM result is also plotted for comparison purposes. The $\mu_{EFT}$  scale uncertainties are not shown. Results are presented for the NNPDF3.1 PDF set and evaluated using $\mu_R=\mu_F=m_t+m_H/2$ as well as  $\mu_{EFT}=m_t$.}
    \label{fig:normalized_differential_pT(b1)}
\end{figure}
\begin{figure}[!ht]
    \centering
    \begin{tabular}{cc}
         \includegraphics[width=0.49\linewidth]{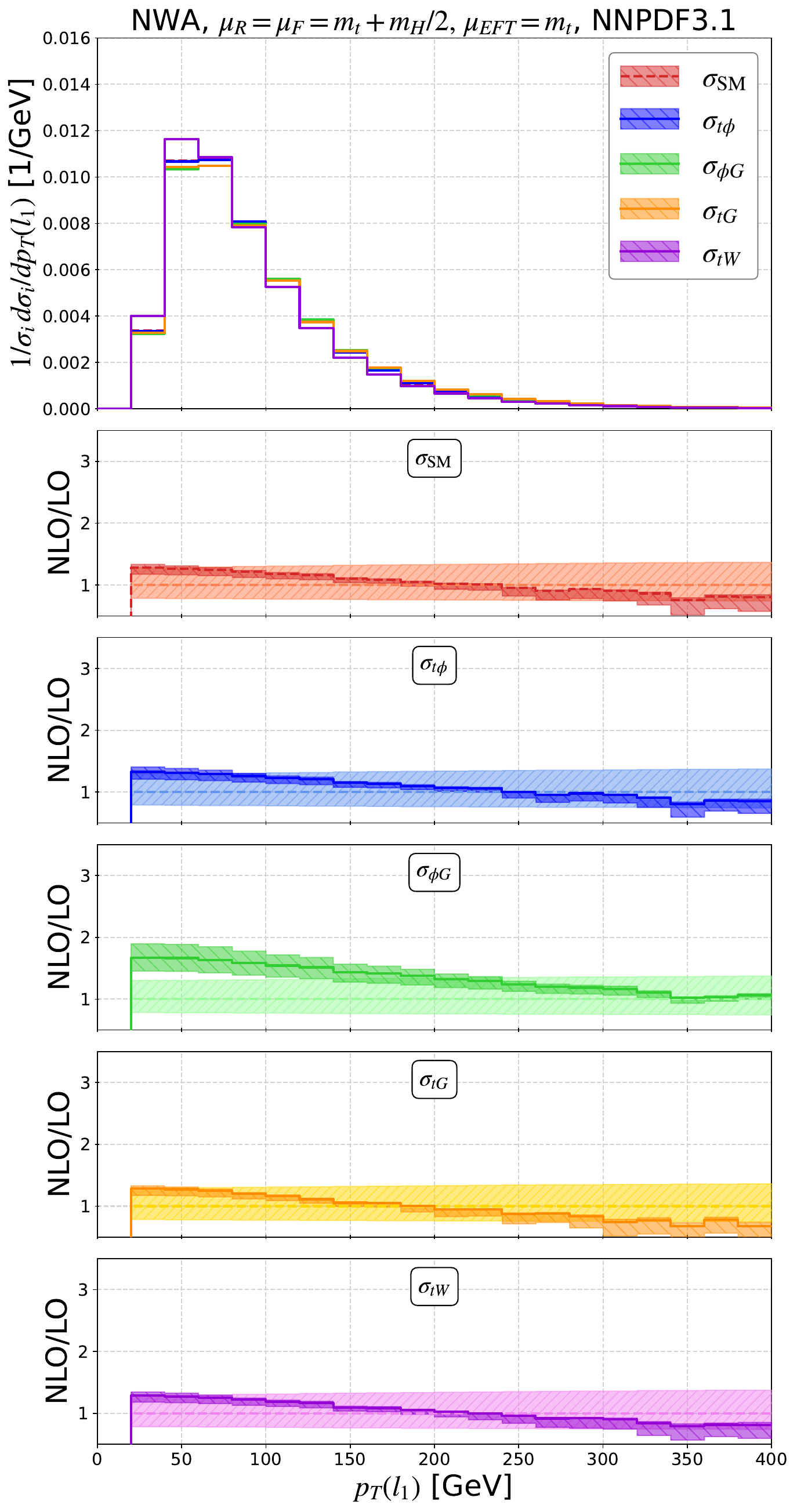} &
         \includegraphics[width=0.49\linewidth]{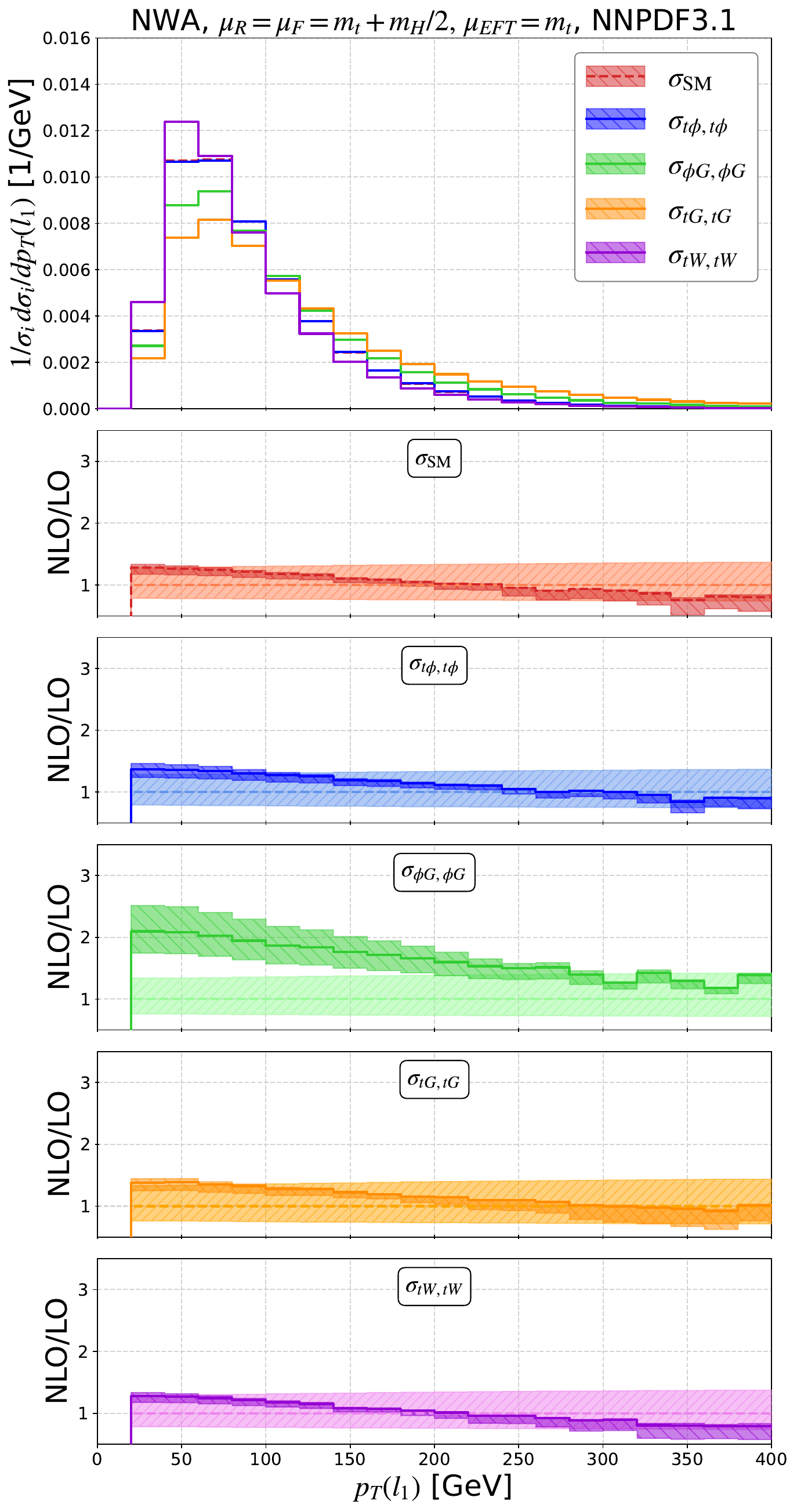}
    \end{tabular}
    \caption{Normalised differential cross-section distributions at NLO in QCD  as a function of $p_T(\ell_1)$ for the $pp \to e^+\nu_e\, \mu^-\bar{\nu}_{\mu}\, b\bar{b} \,H +X$ process at the LHC with $\sqrt{s}=13.6$ TeV. The upper panels show the linear $\sigma_i$ (left) and quadratic $\sigma_{ii}$ (right) terms. The lower panels display the differential ${\cal K}$-factors together with their uncertainty bands estimated with the help of the  7-point scale variation  and the relative scale uncertainties of the LO cross sections, separately for each operator. The SM result is also plotted for comparison purposes. The $\mu_{EFT}$  scale uncertainties are not shown. Results are presented for the NNPDF3.1 PDF set and evaluated using $\mu_R=\mu_F=m_t+m_H/2$ as well as  $\mu_{EFT}=m_t$.}
    \label{fig:normalized_differential_pT(l1)}
\end{figure}
\begin{figure}[t!]
    \centering
    \begin{tabular}{cc}
         \includegraphics[width=0.49\linewidth]{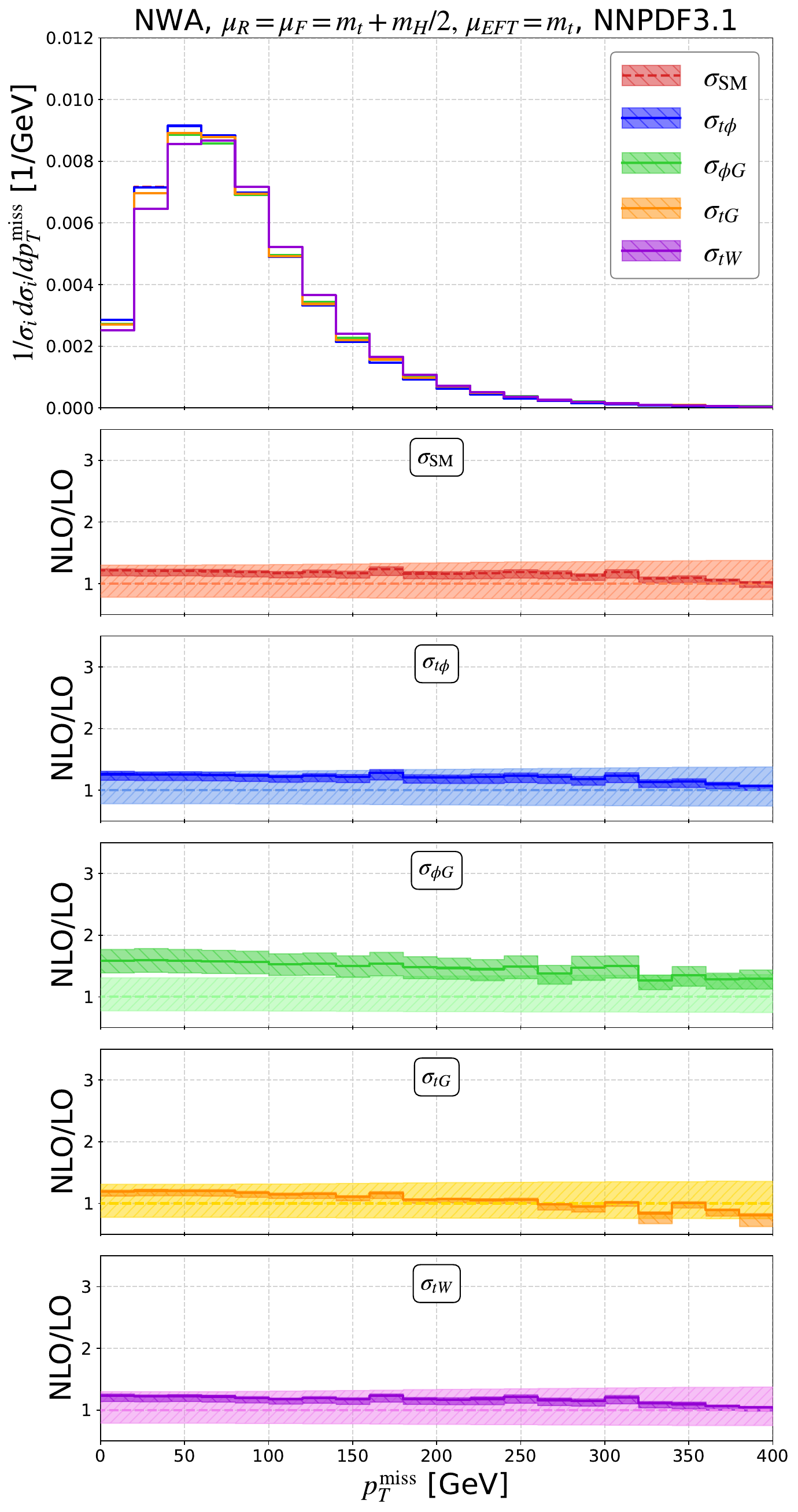} &
         \includegraphics[width=0.49\linewidth]{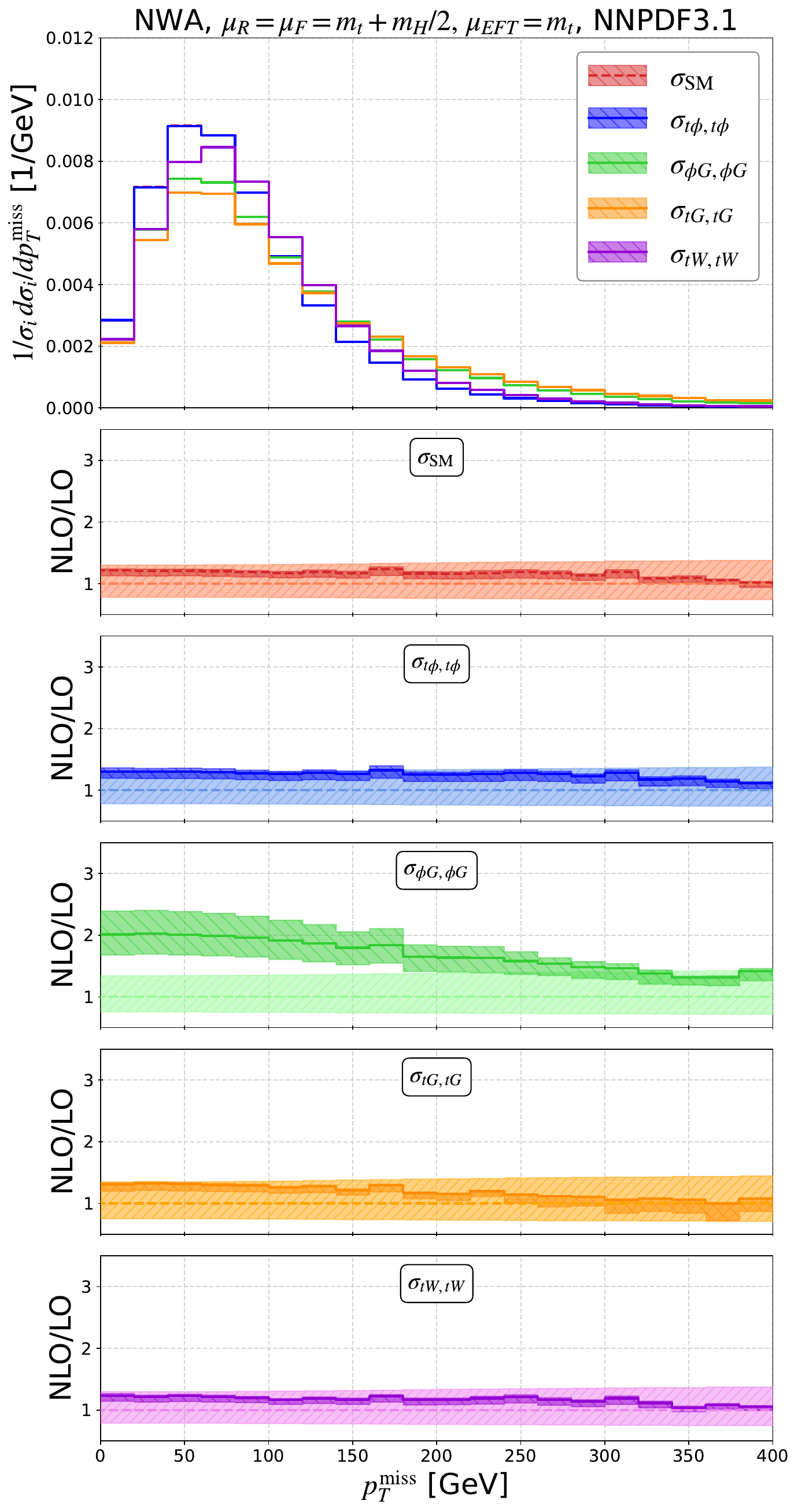}
    \end{tabular}
    \caption{Normalised differential cross-section distributions at NLO in QCD  as a function of $p_T^{miss}$ for the $pp \to e^+\nu_e\, \mu^-\bar{\nu}_{\mu}\, b\bar{b} \,H +X$ process at the LHC with $\sqrt{s}=13.6$ TeV. The upper panels show the linear $\sigma_i$ (left) and quadratic $\sigma_{ii}$ (right) terms. The lower panels display the differential ${\cal K}$-factors together with their uncertainty bands estimated with the help of the  7-point scale variation  and the relative scale uncertainties of the LO cross sections, separately for each operator. The SM result is also plotted for comparison purposes. The $\mu_{EFT}$  scale uncertainties are not shown. Results are presented for the NNPDF3.1 PDF set and evaluated using $\mu_R=\mu_F=m_t+m_H/2$ as well as  $\mu_{EFT}=m_t$.}
    \label{fig:normalized_differential_pT(miss)}
\end{figure}
\begin{figure}[t!]
    \centering
    \begin{tabular}{cc}
         \includegraphics[width=0.49\linewidth]{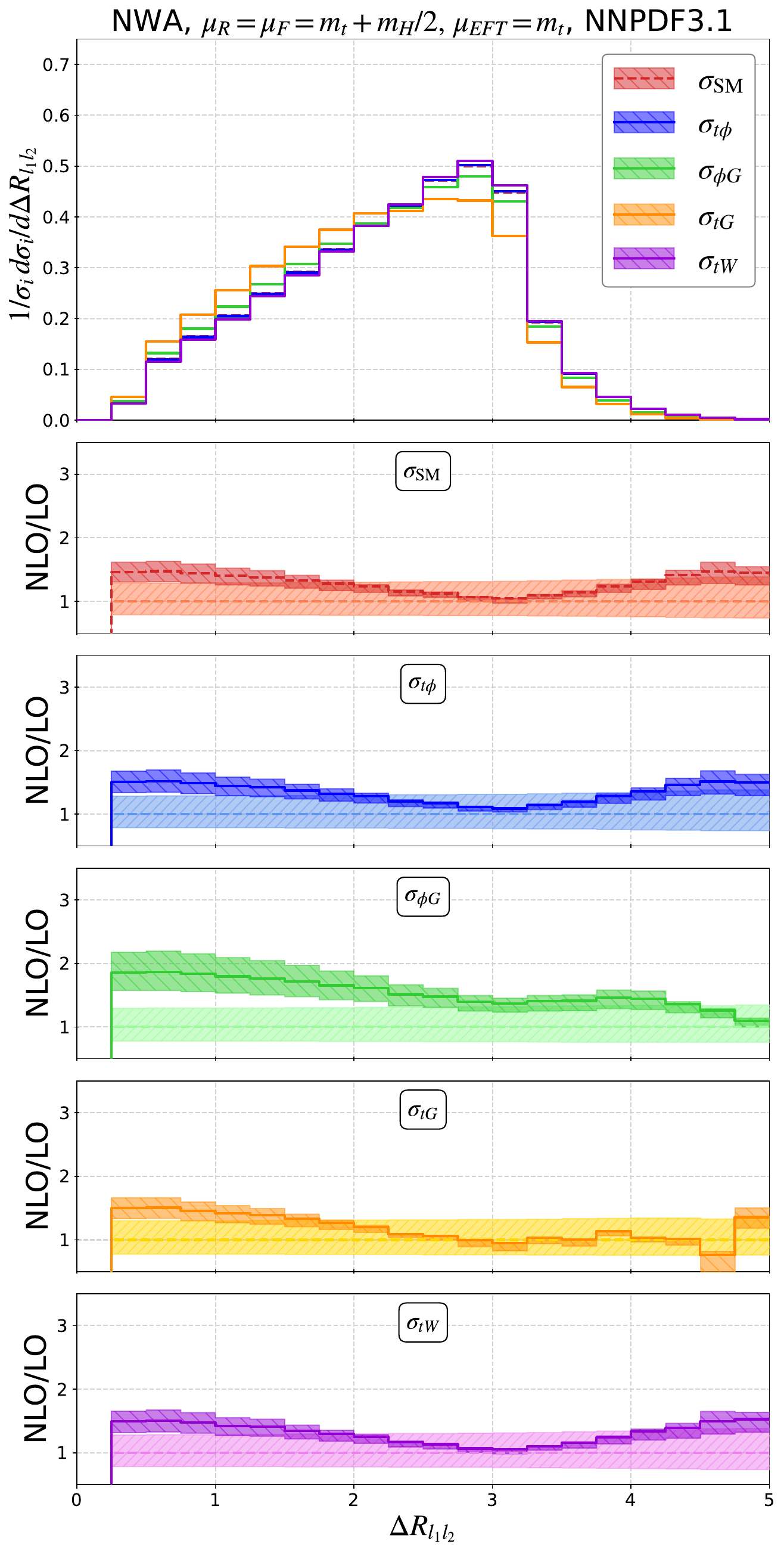} &
         \includegraphics[width=0.49\linewidth]{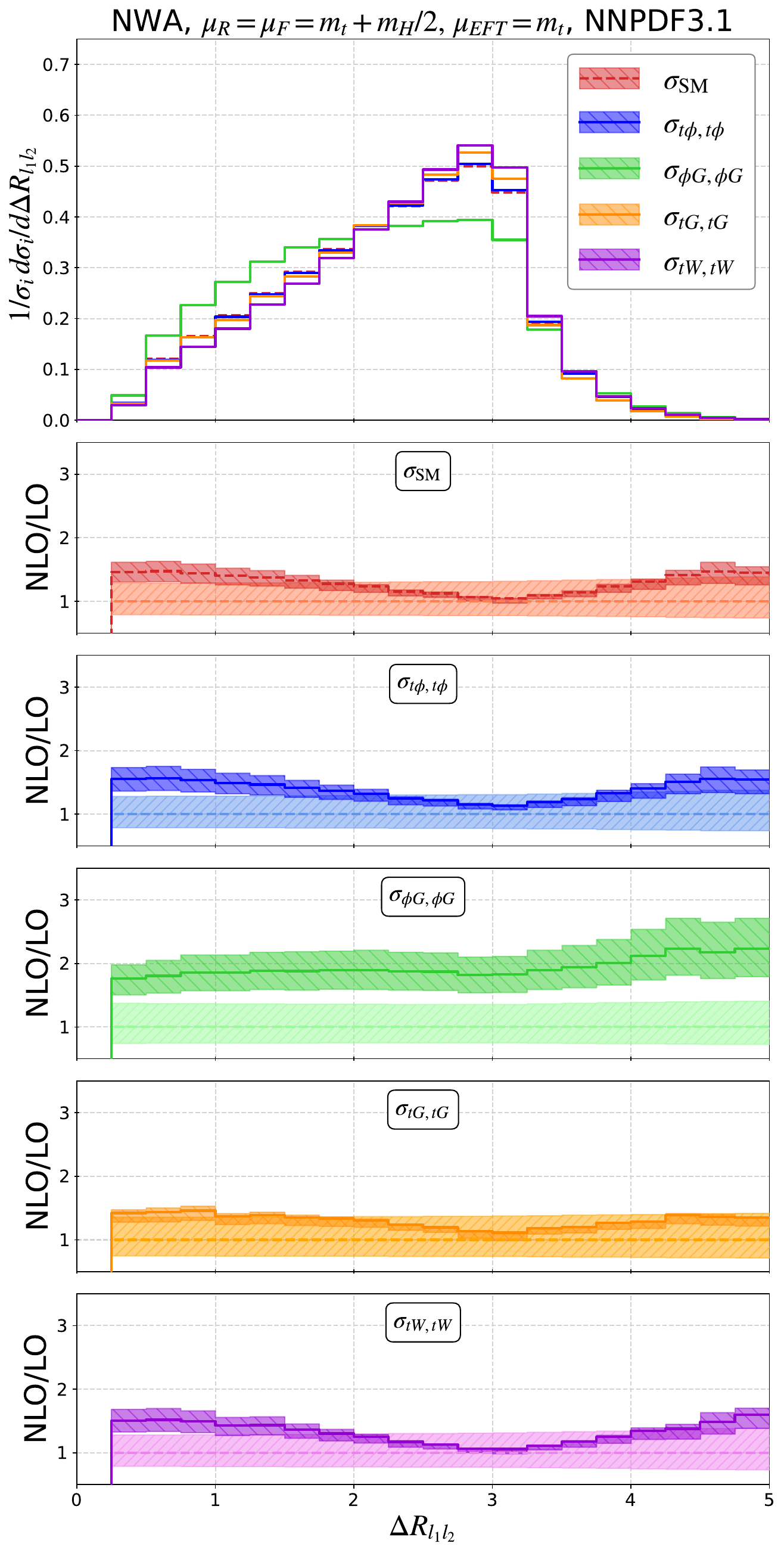}
    \end{tabular}
    \caption{Normalised differential cross-section distributions at NLO in QCD  as a function of $\Delta R_{\ell_1 \ell_2}$ for the $pp \to e^+\nu_e\, \mu^-\bar{\nu}_{\mu}\, b\bar{b} \,H +X$ process at the LHC with $\sqrt{s}=13.6$ TeV. The upper panels show the linear $\sigma_i$ (left) and quadratic $\sigma_{ii}$ (right) terms. The lower panels display the differential ${\cal K}$-factors together with their uncertainty bands estimated with the help of the  7-point scale variation and the relative scale uncertainties of the LO cross sections, separately for each operator. The SM result is also plotted for comparison purposes. The $\mu_{EFT}$  scale uncertainties are not shown. Results are presented for the NNPDF3.1 PDF set and evaluated using $\mu_R=\mu_F=m_t+m_H/2$ as well as  $\mu_{EFT}=m_t$.}
    \label{fig:normalized_differential_dR(l1l2)}
\end{figure}
\begin{figure}[t!]
    \centering
    \begin{tabular}{cc}
         \includegraphics[width=0.49\linewidth]{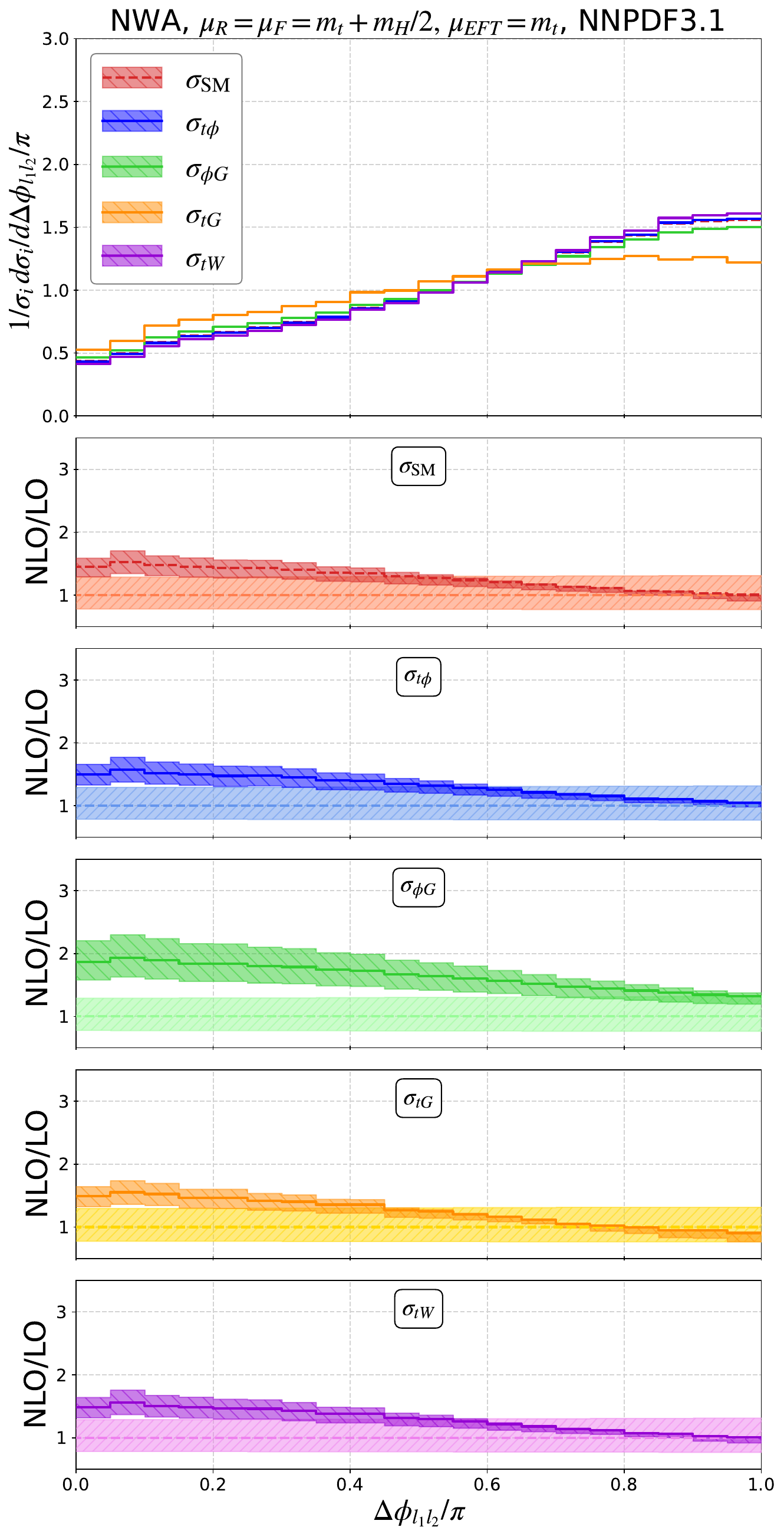} &
         \includegraphics[width=0.49\linewidth]{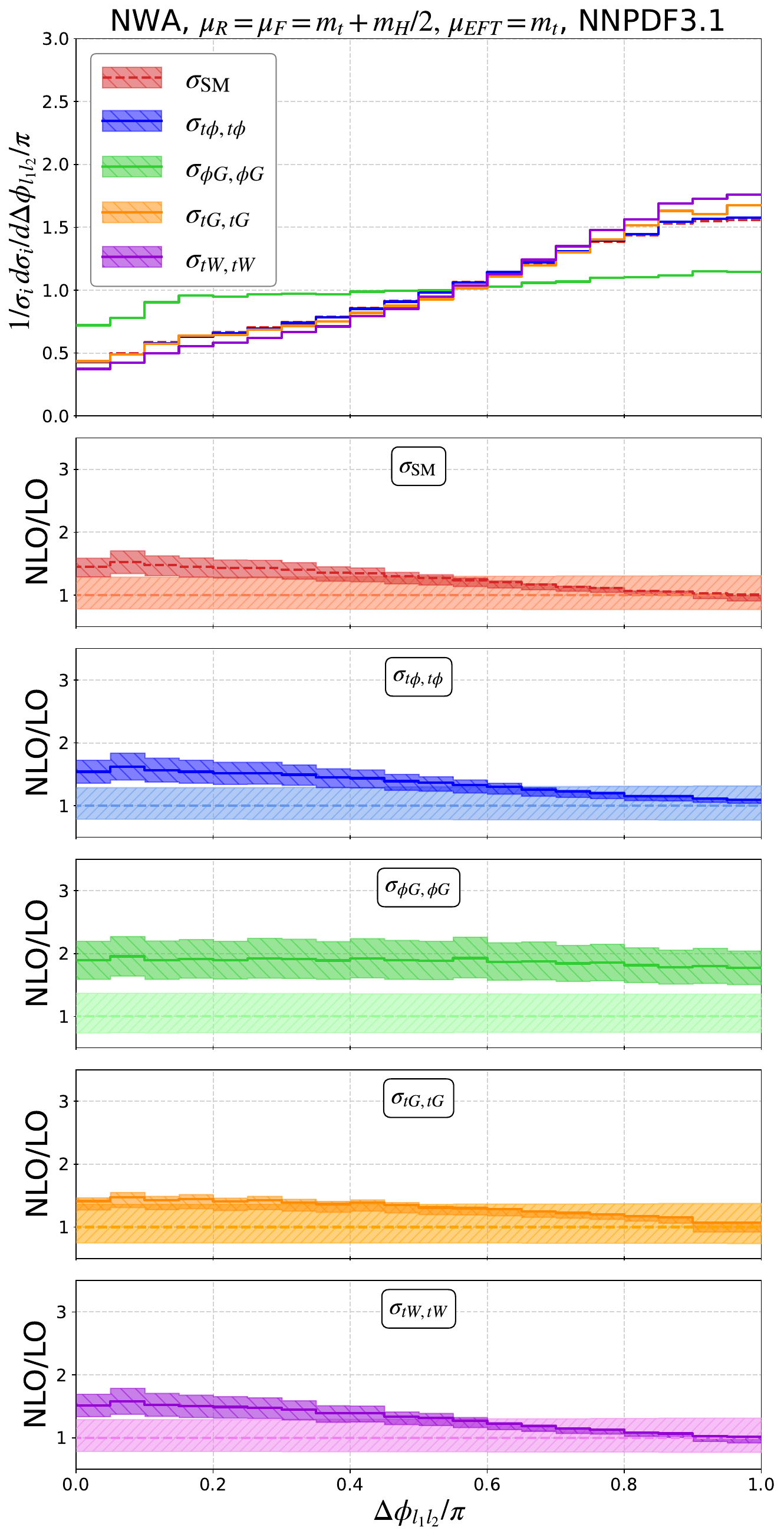}
    \end{tabular}
    \caption{Normalised differential cross-section distributions at NLO in QCD  as a function of $\Delta \phi_{\ell_1 \ell_2}/\pi$ for the $pp \to e^+\nu_e\, \mu^-\bar{\nu}_{\mu}\, b\bar{b} \,H +X$ process at the LHC with $\sqrt{s}=13.6$ TeV. The upper panels show the linear $\sigma_i$ (left) and quadratic $\sigma_{ii}$ (right) terms. The lower panels display the differential ${\cal K}$-factors together with their uncertainty bands estimated with the help of the  7-point scale variation  and the relative scale uncertainties of the LO cross sections, separately for each operator. The SM result is also plotted for comparison purposes. The $\mu_{EFT}$  scale uncertainties are not shown. Results are presented for the NNPDF3.1 PDF set and evaluated using $\mu_R=\mu_F=m_t+m_H/2$ as well as  $\mu_{EFT}=m_t$.}
    \label{fig:normalized_differential_dPhi(l1l2)}
\end{figure}
\begin{figure}[!ht]
    \centering
    \begin{tabular}{cc}
         \includegraphics[width=0.49\linewidth]{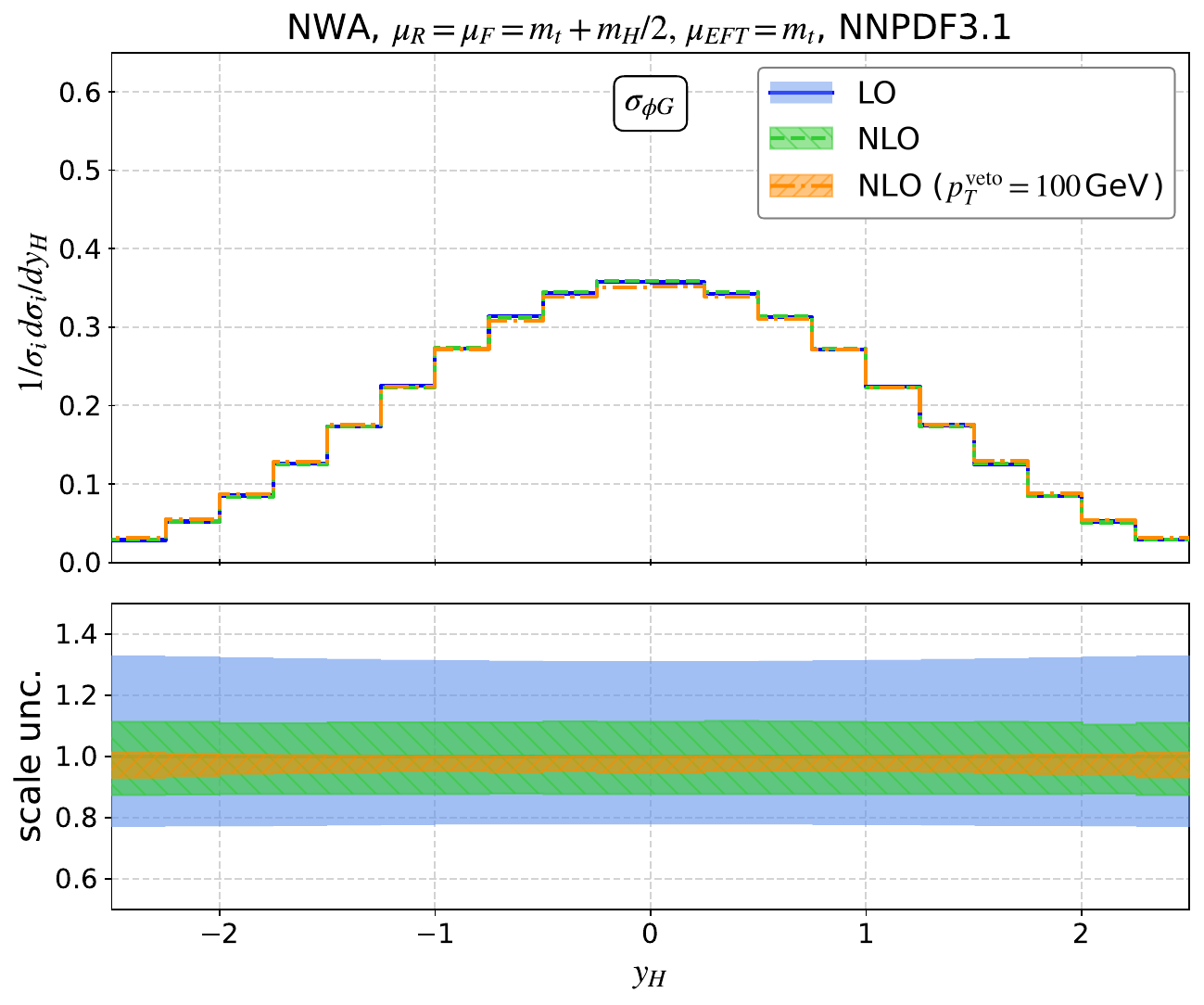} &
         \includegraphics[width=0.49\linewidth]{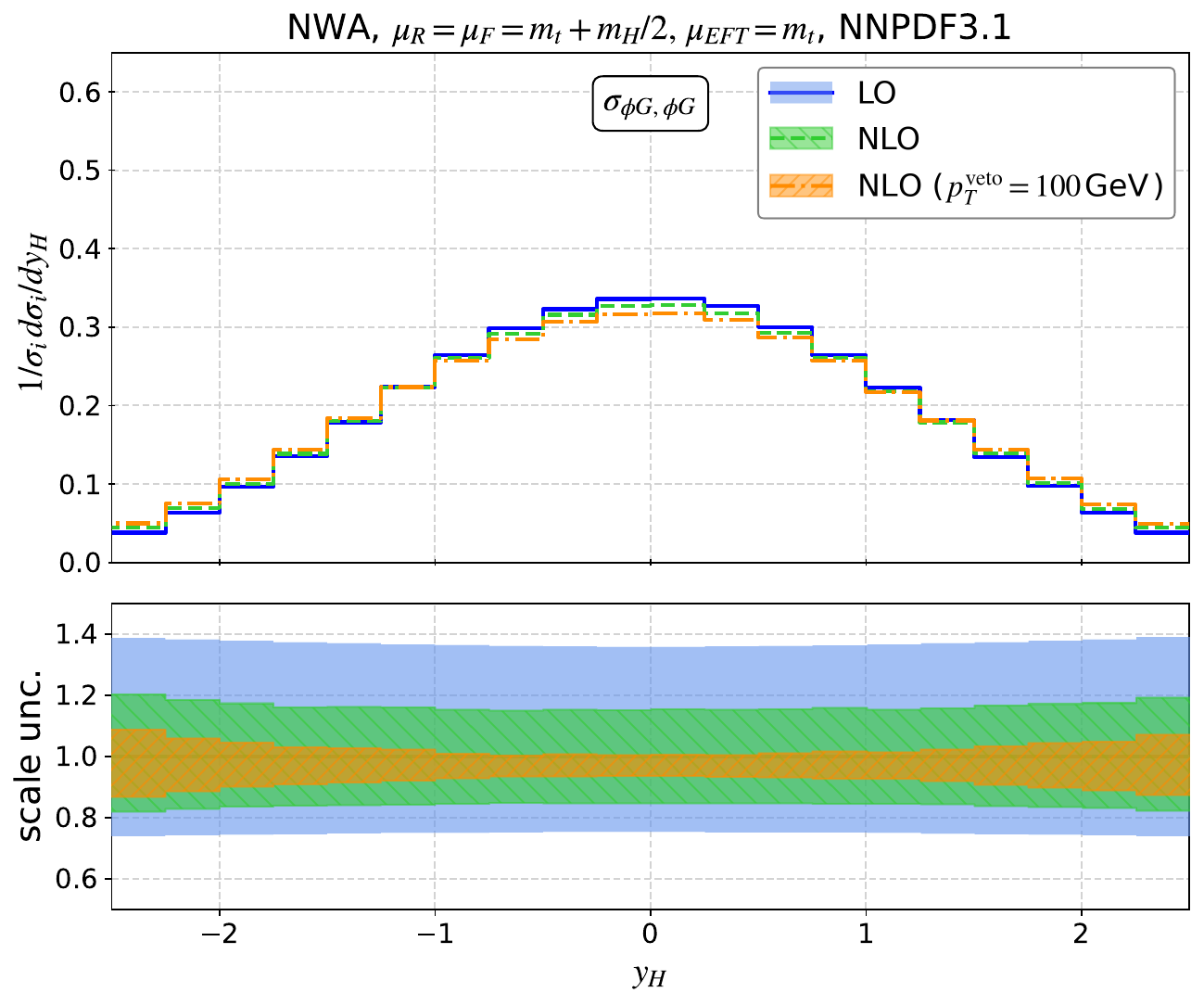}
    \end{tabular}
    \caption{Normalised differential cross-section distributions at LO and NLO in QCD as a function of $y_H$ for the $pp \to e^+\nu_e\, \mu^-\bar{\nu}_{\mu}\, b\bar{b} \,H +X$ process at the LHC with $\sqrt{s}=13.6$ TeV. The impact of the jet veto with $p_{T}^{veto}=100$ GeV on the contributions  induced by the $\mathcal{O}_{\phi G}$ operator is presented.  The  results for $\sigma_{\phi G}$ (left) and $\sigma_{\phi G, \,\phi G}$ (right) are displayed. The lower panels show the corresponding scale uncertainties for all cases.  Results are presented for the NNPDF3.1 PDF set and evaluated using $\mu_R=\mu_F=m_t+m_H/2$ as well as  $\mu_{EFT}=m_t$.}
    \label{fig:normalized_differential_jetveto_y(H)}
\end{figure}
\begin{figure}[!ht]
    \centering
    \begin{tabular}{cc}
         \includegraphics[width=0.49\linewidth]{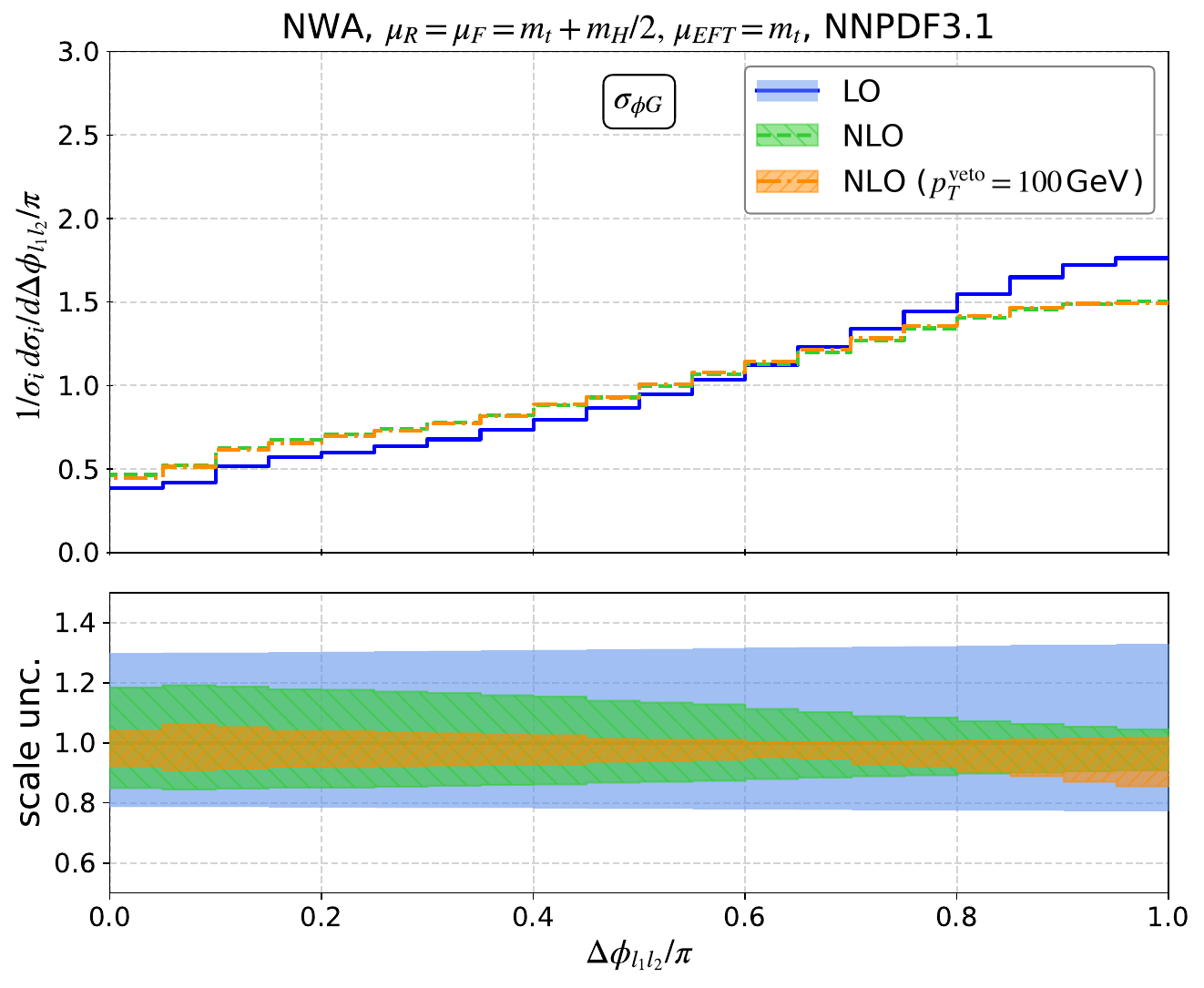} &
         \includegraphics[width=0.49\linewidth]{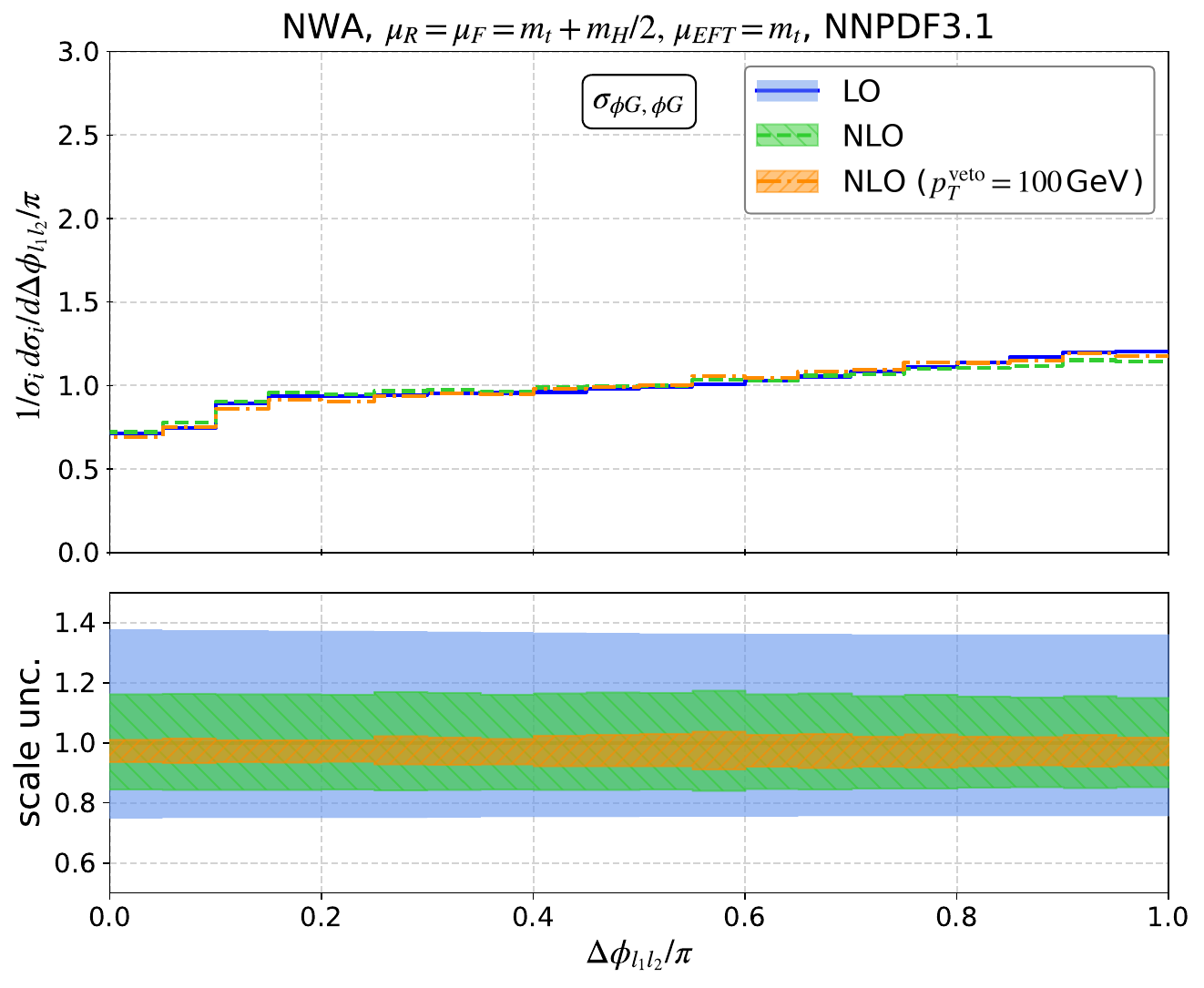}
    \end{tabular}
    \caption{Normalised differential cross-section distributions at LO and NLO in QCD as a function of $\Delta \phi_{\ell_1 \ell_2}/\pi$ for the $pp \to e^+\nu_e\, \mu^-\bar{\nu}_{\mu}\, b\bar{b} \,H +X$ process at the LHC with $\sqrt{s}=13.6$ TeV. The impact of the jet veto with $p_{T}^{veto}=100$ GeV on the contributions  induced by the $\mathcal{O}_{\phi G}$ operator is presented.  The results for $\sigma_{\phi G}$ (left) and $\sigma_{\phi G, \,\phi G}$ (right) are displayed. The lower panels show the corresponding scale uncertainties for all cases. Results are presented for the NNPDF3.1 PDF set and evaluated using $\mu_R=\mu_F=m_t+m_H/2$ as well as  $\mu_{EFT}=m_t$.}
\label{fig:normalized_differential_jetveto_Dphi(l1l2)}
\end{figure}

The impact of NLO QCD corrections can be significantly increased in specific regions of phase space. To investigate this, we present various differential cross-section distributions describing the kinematics of the Higgs boson, leptons, and $b$-jet. Furthermore, we want to understand whether the shape of these observables changes in a significant way when SMEFT effects are included. To this end, we present a few examples for both dimensionful and dimensionless observables for the $pp \to e^+\nu_e\, \mu^-\bar{\nu}_{\mu}\, b\bar{b} \,H +X$ process at the LHC with $\sqrt{s}=13.6$ TeV. In each case we display normalised differential cross-section distributions at NLO in QCD for the linear $\sigma_i$ (left) and quadratic $\sigma_{ii}$ (right) terms. Furthermore, in the lower panels we provide the differential ${\cal K}$-factors together with their uncertainty bands estimated with the help of the  7-point scale variation  and the relative scale uncertainties of the LO cross sections, separately for each operator. The SM result is also plotted for comparison purposes. Results are reported for the NNPDF3.1 PDF set and calculated using $\mu_R=\mu_F=m_t+m_H/2$ as well as $\mu_{EFT}=m_t$.

In Figure \ref{fig:normalized_differential_pT(H)} we present the transverse momentum of the Higgs boson, denoted $p_{T}(H)$. We can note that this observable has a discriminating power for the operators ${\cal O}_{\phi G}$ and ${\cal O}_{tG}$ already for the interference contributions. These differences are further enhanced in the quadratic terms, where much harder $p_T$ spectra can be observed. The latter effect can be explained by the fact that in both cases we are dealing with effective interactions involving the Higgs boson and gluon, the strength of which depends on the momentum of this gluon. They can be used to distinguish the contributions from ${\cal O}_{\phi G}$ and ${\cal O}_{tG}$ from the SM process and the remaining operators. The size of the NLO QCD corrections, as shown by the differential ${\cal K}$-factors, varies depending on the contribution analysed. However, for the operator ${\cal O}_{\phi G}$, these effects are particularly large at the beginning of the spectrum, where they can reach $70\%$ and $130\%$ for the linear and quadratic terms, respectively. In addition, in these phase-space regions, the NLO uncertainties increase significantly compared to the other operators. For ${\cal O}_{\phi G}$ and ${\cal O}_{tG}$ the NLO uncertainties are not within the LO ones. A judicious choice of dynamic scale can help to flatten the differential ${\cal K}$-factors  and mitigate this behaviour at least in the case of $\sigma_{tG}$.

In Figure \ref{fig:normalized_differential_pT(b1)}  we show the transverse momentum  of the hardest $b$-jet, $p_T(b_1)$. In the case of the $\sigma_i$ terms, very small differences in shape can be observed, suggesting that the SM part plays a dominant role in almost all interference terms we analysed for this observable. In the case of $\sigma_{tG, \,tG}$ and $\sigma_{\phi G, \, \phi G }$ the harder $p_T$ spectra are again preferred. Similarly to the $p_T(H)$ case, particularly large ${\cal K}$-factors are obtained for the terms $\sigma_{\phi G}$ $(80\%)$ and $\sigma_{\phi G, \, \phi G}$ $(120\%)$.

In Figure \ref{fig:normalized_differential_pT(l1)} and Figure \ref{fig:normalized_differential_pT(miss)} we plot the transverse momentum of the hardest charged lepton, $p_T(\ell_1)$, and the missing transverse momentum, $p_{T}^{miss}$. In both cases, we observe shape differences for $\sigma_{tW}$ up to $10\%-20\%$ in a large part of the plotted ranges. Also in these cases, the importance of higher-order corrections is clearly visible.

Of particular interest are various angular distributions of the charged leptons, which are very well measured by both the ATLAS and CMS collaborations.  We show two examples. Figure  \ref{fig:normalized_differential_dR(l1l2)} presents the $\Delta R_{\ell_1 \ell_2}$ separation defined in the rapidity $(\Delta y_{\ell_1 \ell_2})$ and azimuthal angle ($\Delta \phi_{\ell_1 \ell_2}$) plane, whereas Figure \ref{fig:normalized_differential_dPhi(l1l2)} displays the $\Delta \phi_{\ell_1 \ell_2}$ observable itself.  The latter cross-section distribution is widely studied in various analyses due to its  inherent sensitivity to spin correlations and possible new physical effects that may arise. For both observables, we see large effects for the terms $\sigma_{t G}$ and  $\sigma_{\phi G, \, \phi G}$. In the case of $\sigma_{t G}$, these effects are of the order of $20\%-30\%$ in the phase-space regions where $\Delta \phi_{\ell_1 \ell_2} \approx 0$ or $\Delta \phi_{\ell_1 \ell_2} \approx \pi$. However, they are no longer visible for the quadratic term $\sigma_{t G, \, t G}$. This can be explained as follows. For the linear term, we have chirality-flipping contributions $\sim(\bar{Q} \,\sigma^{\mu\nu} \, T_a \, t)\,G^a_{\mu\nu}$ which interfere with the SM. This is not the case for the quadratic term, where the SM contributions are absent. For the $\sigma_{\phi G}$ linear term only minor shape deviations are noted. However, they are amplified for the $\sigma_{\phi G, \, \phi G}$ term, where the differences from the SM background result are of the order of $20\%-40\%$ and $30\%- 65\%$ for $\Delta R_{\ell_1 \ell_2}$ and $\Delta\phi_{\ell_1 \ell_2}$, respectively. 

It is also interesting to check the impact of the jet veto on the $\sigma_{\phi G}$ and  $\sigma_{\phi G,\, \phi G}$ contributions  at the differential cross-section level. In Figure \ref{fig:normalized_differential_jetveto_y(H)} and Figure \ref{fig:normalized_differential_jetveto_Dphi(l1l2)} we show two representative examples, that is, the rapidity of the Higgs boson, $y_H$,  and  $\Delta \phi_{\ell_1 \ell_2}$.  We compare the normalised cross-section distributions at  LO and NLO in QCD together with their corresponding scale uncertainties. For the NLO case, two results are presented, i.e. without and with $p_T^{veto} = 100$ GeV. For these two dimensionless observables, we see no substantial change in their shape when the jet veto is employed. However, significant modifications can be observed in the size of the theoretical error estimated from the scale variation. In particular, the LO uncertainties for the linear (quadratic) term are of the order of $30\%$ ($40\%$). They are reduced to $10\%-15\%$ ($20\%$) at NLO in QCD and further to $5\%$ ($10\%$) when the jet veto is used. Similarly, for $\Delta \phi_{\ell_1 \ell_2}$ we obtain a reduction of the theoretical error from $40\%-30\%$ to $10\%-15\%$ for $\sigma_{\phi G}$ and $\sigma_{\phi G, \, \phi G}$. 
%

\section{Reconstruction of top-quark kinematics}
\label{sec:top_kin}

To better understand the impact of fiducial cuts on the top quark kinematics in the presence of SMEFT effects, we present a comparison of results for $pp\to t\bar{t}H+X$ and $pp \to e^+\nu_e\, \mu^-\bar{\nu}_{\mu}\, b\bar{b} \,H +X$. To be able to compare these two processes at the differential cross-section level, in the NWA case the momenta of the top and anti-top quarks are reconstructed from the available decay products. Because we work in the NWA, this reconstruction is rather straightforward. Among the various observables that can be produced from the top quark momenta, we focus on the top quark transverse momentum, $p_T(t)$, and the scalar sum of the transverse momenta of the top quark, the anti-top quark and the Higgs boson, $H_T^{\rm reco}$. The latter observable is defined as follows 
\begin{equation}
H_T^{\rm reco}=p_T(t)+p_T(\,\bar{t}\,) +p_T(H)\,.
\end{equation}
We further investigate how changes in the top-quark kinematics affect the differential cross-section results for the Higgs boson. Therefore, we also present the transverse momentum and rapidity of the Higgs boson, $p_T(H)$ and $y(H)$, respectively. In each case, we provide the normalised cross-section predictions for the NWA case and with the stable top quarks for the  linear and quadratic terms, along with the corresponding scale uncertainties. In the upper panels, we always present the ratios of these two predictions. The flatter the ratio distributions are, the more similar the kinematics of the reconstructed and stable top quarks are for a given observable. In addition, for comparison, we provide the SM predictions.
\begin{figure}[t!]
    \centering
    \begin{tabular}{cc}
\includegraphics[width=0.49\linewidth]{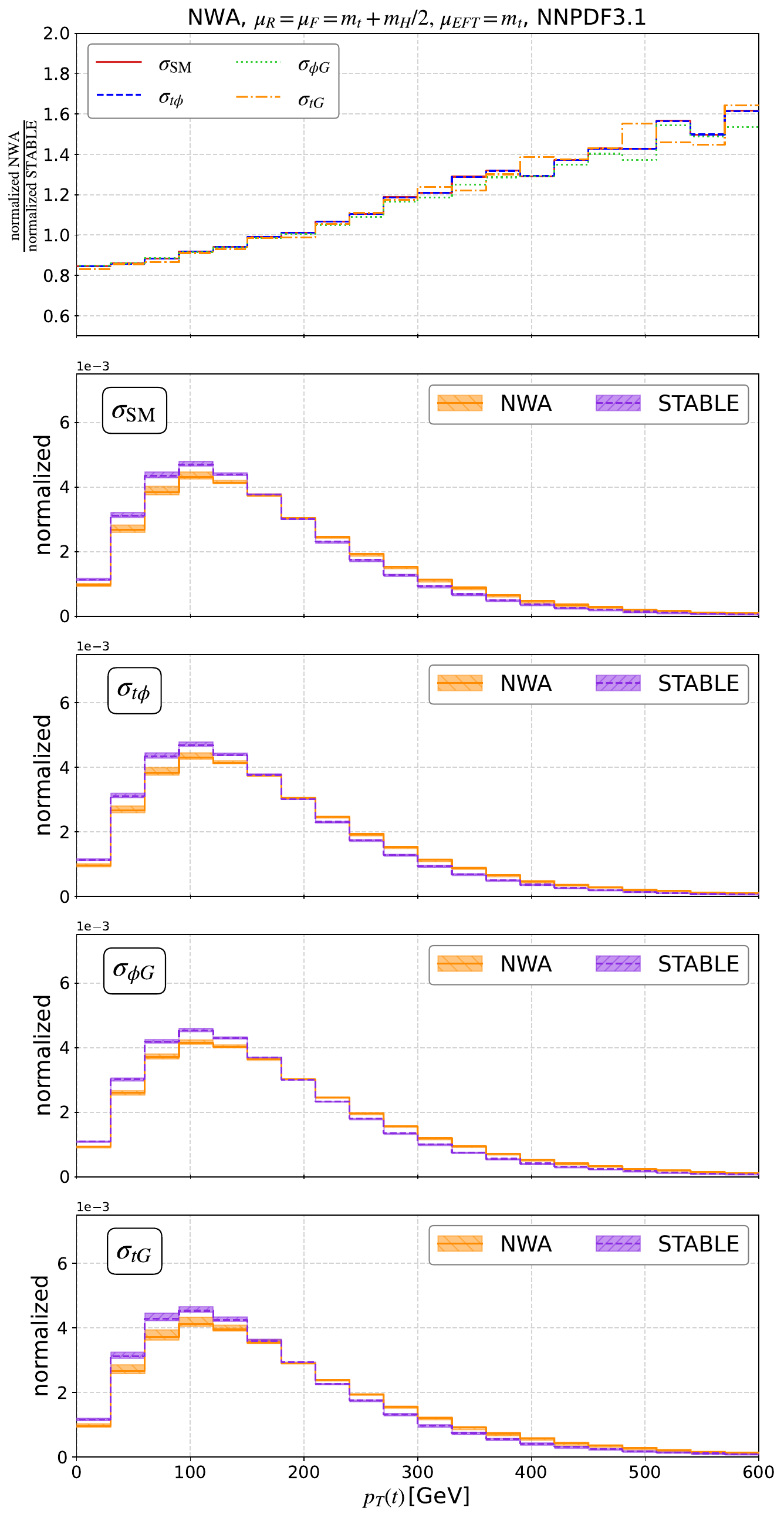} 
\includegraphics[width=0.49\linewidth]{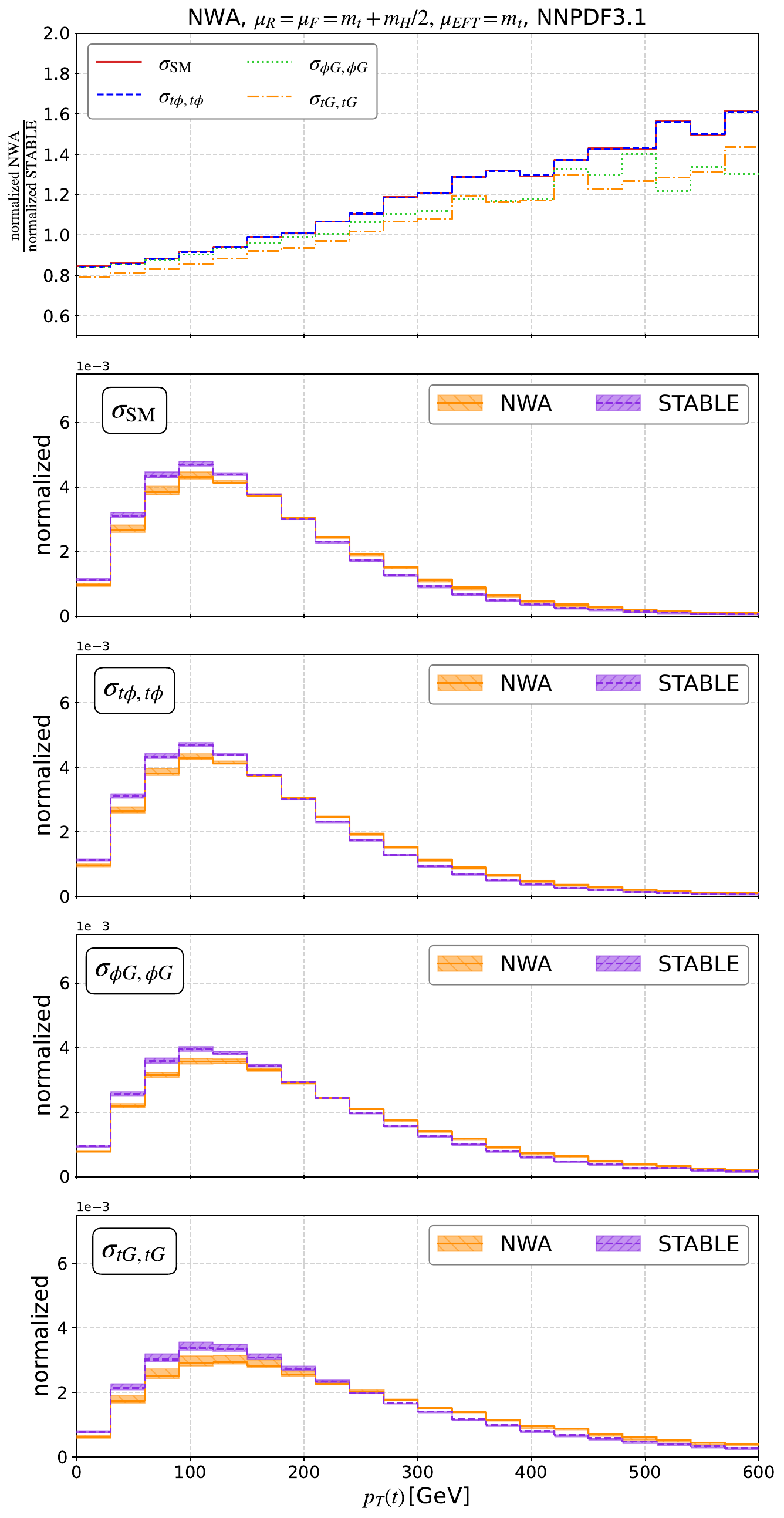}
    \end{tabular}
\caption{Normalised differential cross-section distributions at NLO in QCD as a function of $p_T(t)$ for $pp\to t\bar{t}H+X$ (Stable) and $pp \to e^+\nu_e\, \mu^-\bar{\nu}_{\mu}\, b\bar{b} \,H +X$ (NWA) at the LHC with $\sqrt{s}=13.6$ TeV. Results are provided for the linear and quadratic terms together with the corresponding scale uncertainties. The SM result is also plotted for comparison purposes. The upper panels display the ratio of the two predictions. Results are presented for the NNPDF3.1 PDF set and evaluated using $\mu_R=\mu_F=m_t+m_H/2$ as well as  $\mu_{EFT}=m_t$.}
\label{fig:differential_nwa_vs_stable_pT(t)}
\end{figure}
\begin{figure}[t!]
    \centering
    \begin{tabular}{cc}
\includegraphics[width=0.49\linewidth]{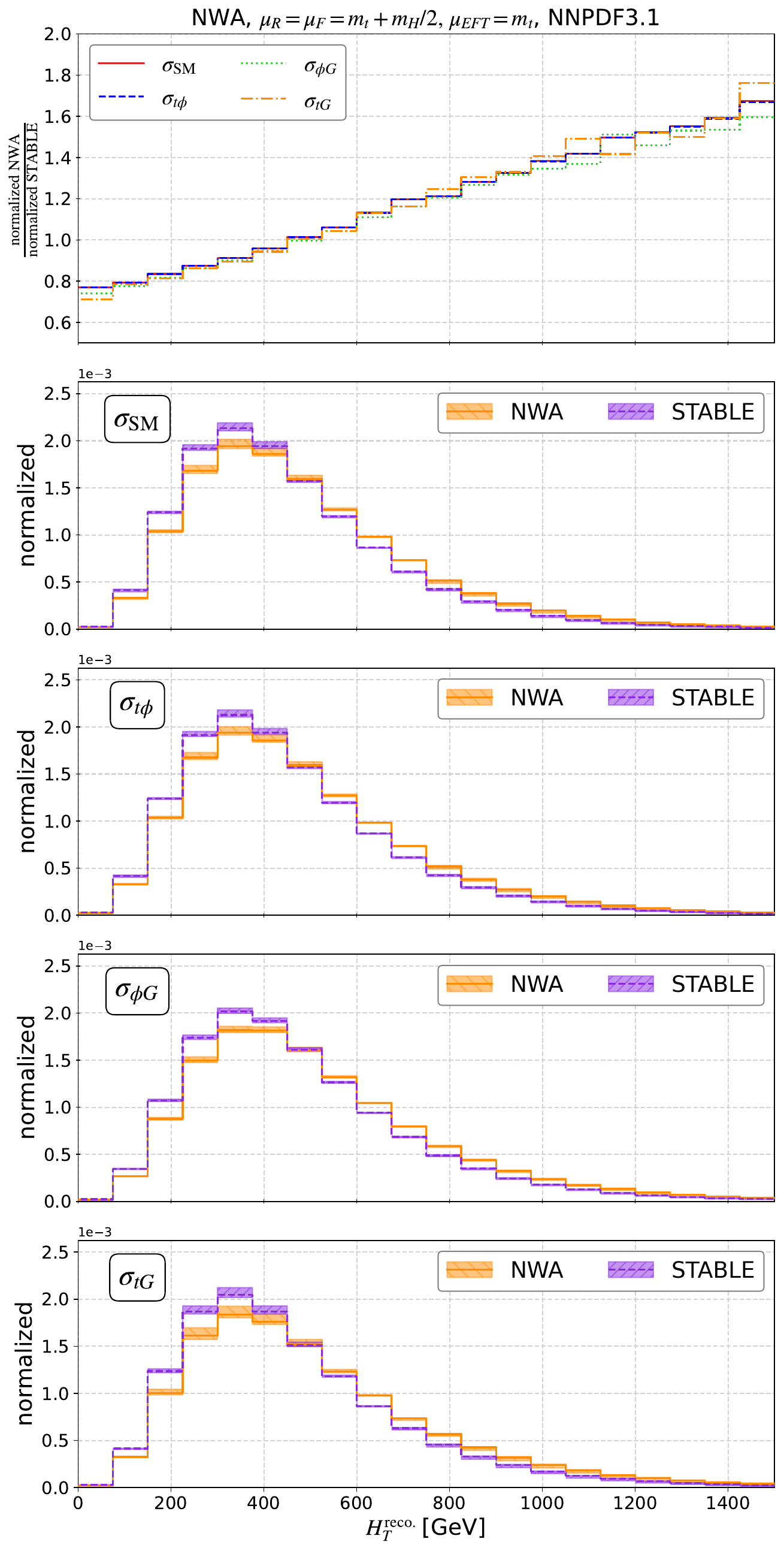} 
\includegraphics[width=0.49\linewidth]{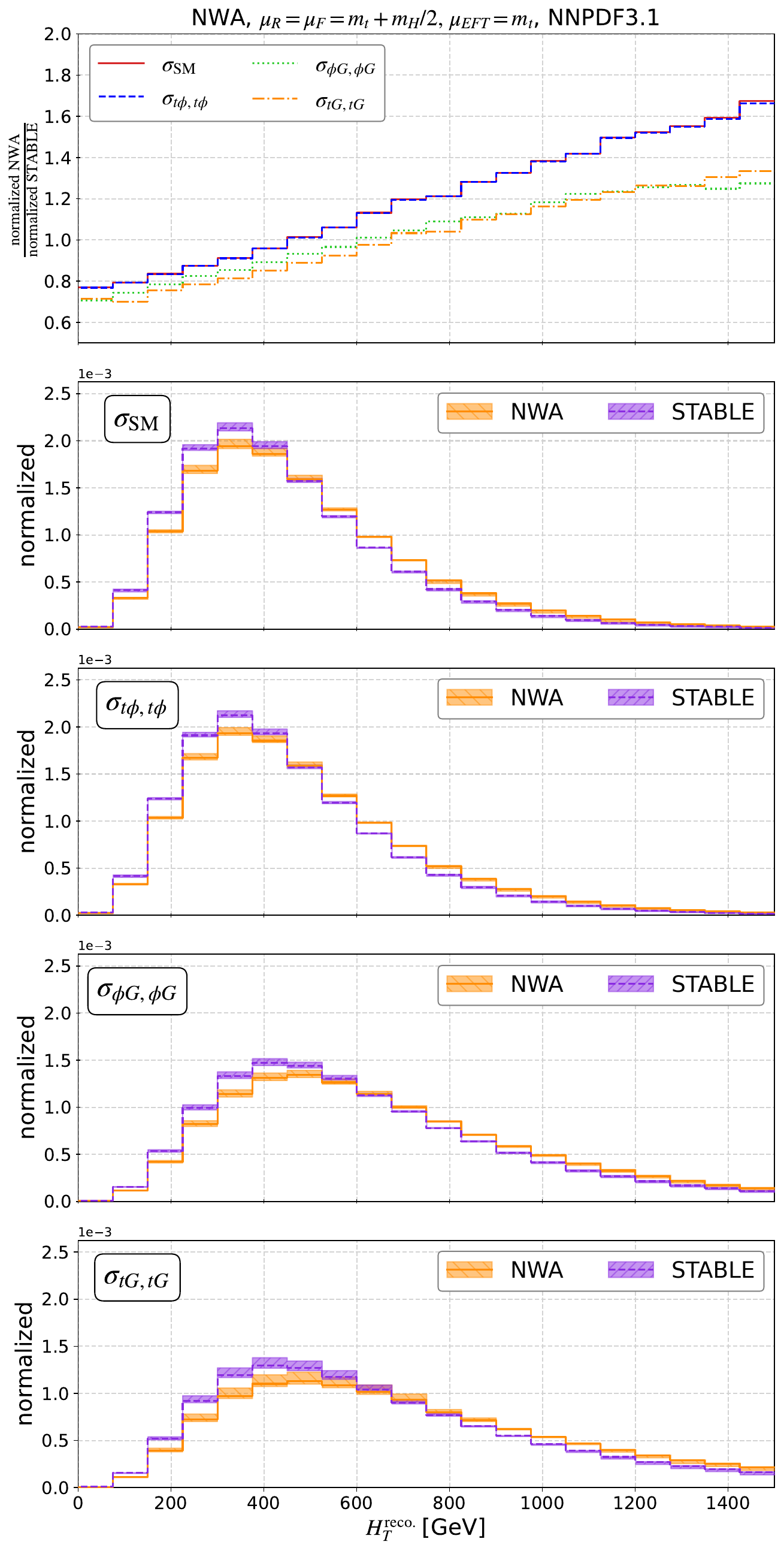}
    \end{tabular}
\caption{Normalised differential cross-section distributions at NLO in QCD as a function of $H_T^{\rm reco}$ for $pp\to t\bar{t}H+X$ (Stable) and $pp \to e^+\nu_e\, \mu^-\bar{\nu}_{\mu}\, b\bar{b} \,H +X$ (NWA) at the LHC with $\sqrt{s}=13.6$ TeV. Results are provided for the linear and quadratic terms together with the corresponding scale uncertainties. The SM result is also plotted for comparison purposes. The upper panels display the ratio of the two predictions. Results are presented for the NNPDF3.1 PDF set and evaluated using $\mu_R=\mu_F=m_t+m_H/2$ as well as  $\mu_{EFT}=m_t$.}
\label{fig:differential_nwa_vs_stable_Ht}
\end{figure}
\begin{figure}[t!]
    \centering
    \begin{tabular}{cc}
\includegraphics[width=0.49\linewidth]{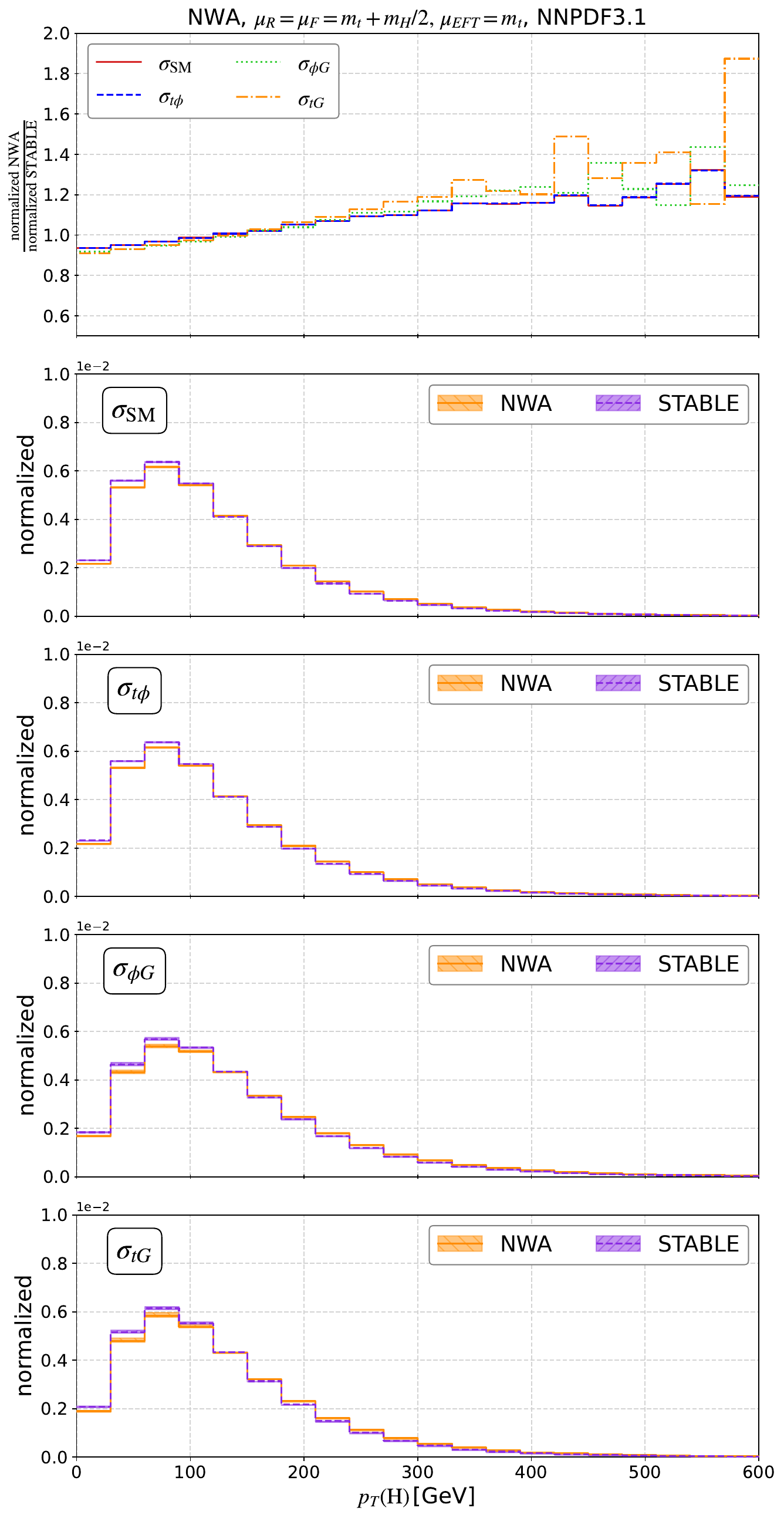} 
\includegraphics[width=0.49\linewidth]{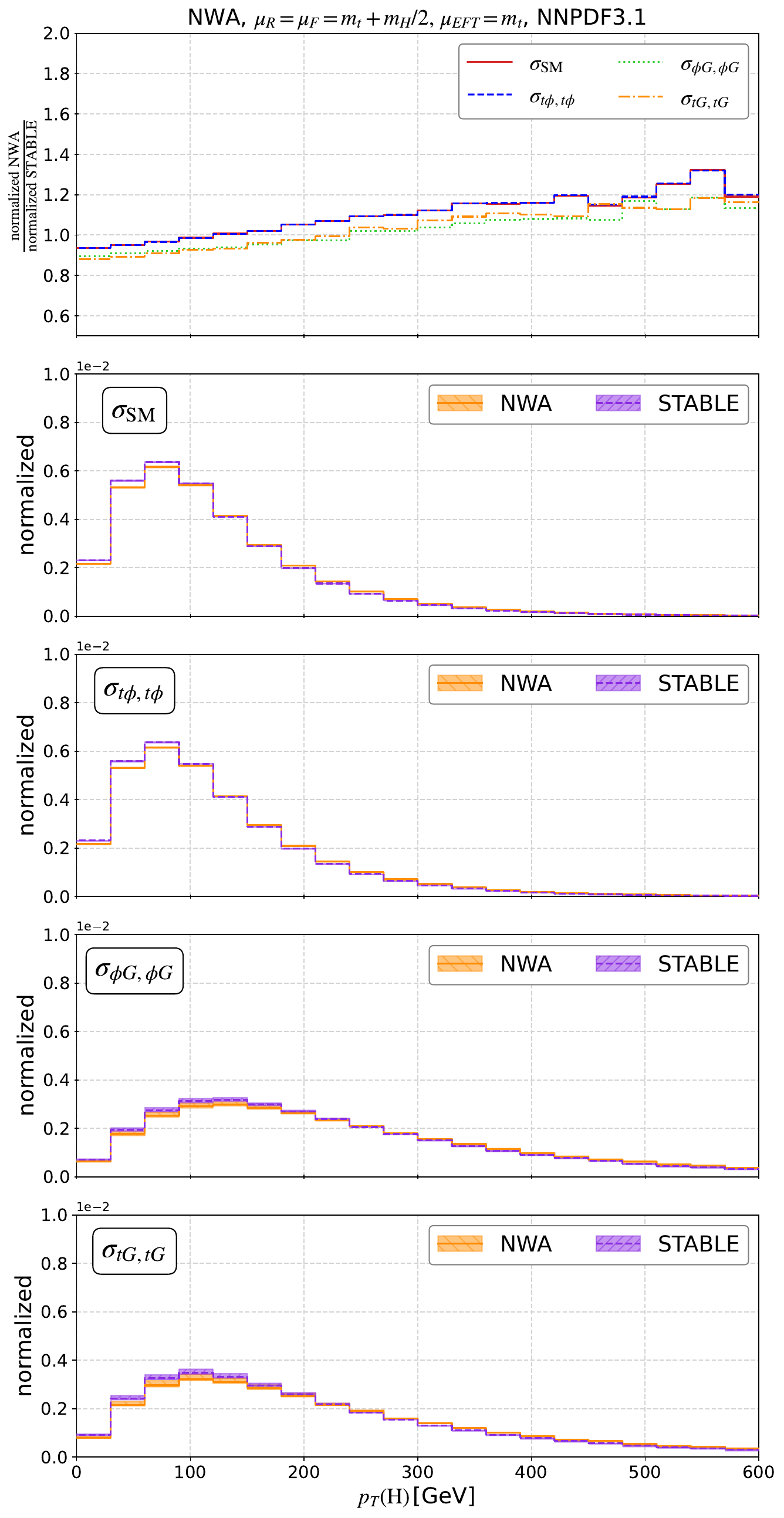}
    \end{tabular}
\caption{Normalised differential cross-section distributions at NLO in QCD as a function of $p_T(H)$ for $pp\to t\bar{t}H+X$ (Stable) and $pp \to e^+\nu_e\, \mu^-\bar{\nu}_{\mu}\, b\bar{b} \,H +X$ (NWA) at the LHC with $\sqrt{s}=13.6$ TeV. Results are provided for the linear and quadratic terms together with the corresponding scale uncertainties. The SM result is also plotted for comparison purposes. The upper panels display the ratio of the two predictions. Results are presented for the NNPDF3.1 PDF set and evaluated using $\mu_R=\mu_F=m_t+m_H/2$ as well as  $\mu_{EFT}=m_t$.}
\label{fig:differential_nwa_vs_stable_pT(H)}
\end{figure}
\begin{figure}[t!]
    \centering
    \begin{tabular}{cc}
\includegraphics[width=0.49\linewidth]{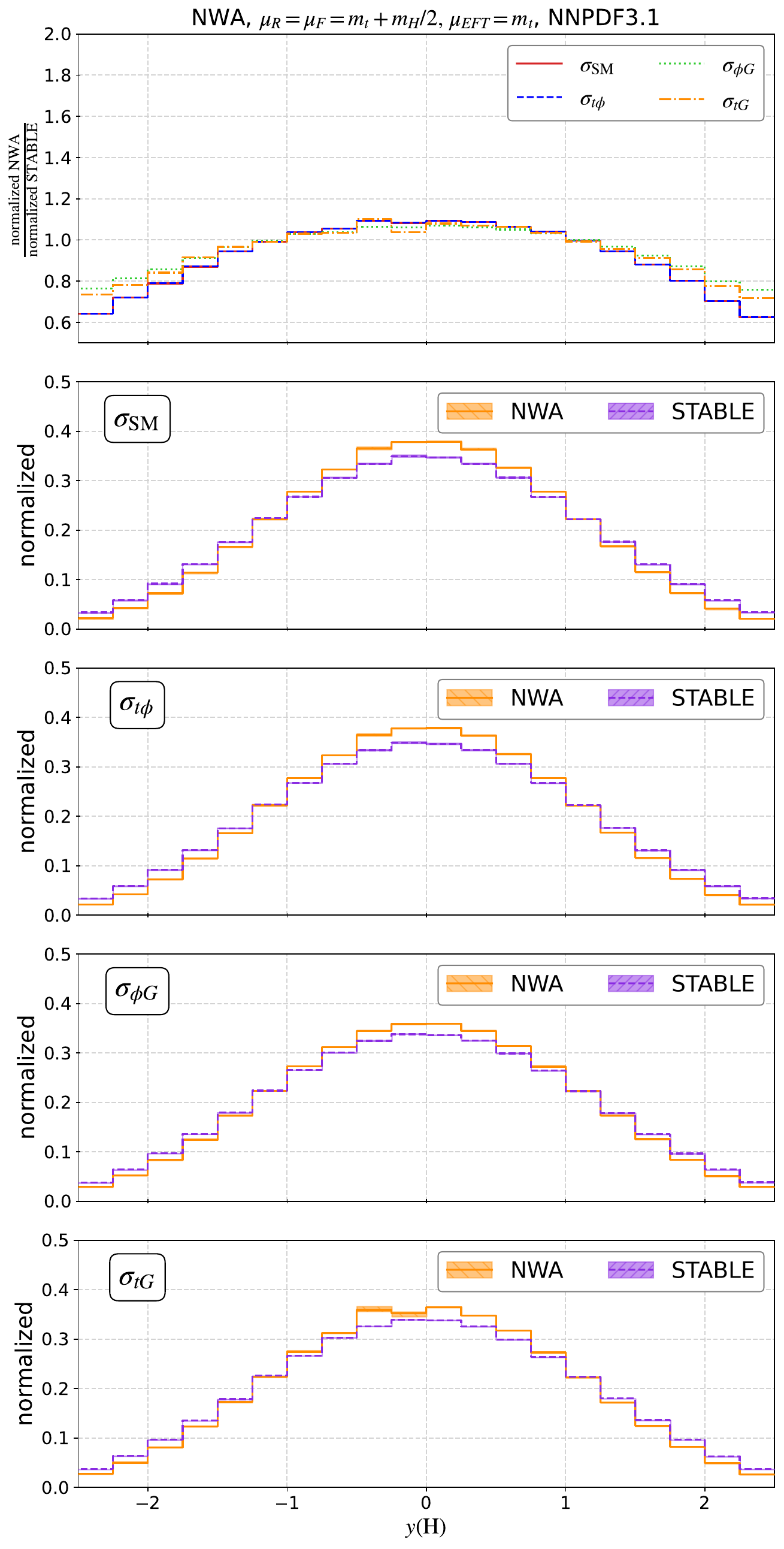} 
\includegraphics[width=0.49\linewidth]{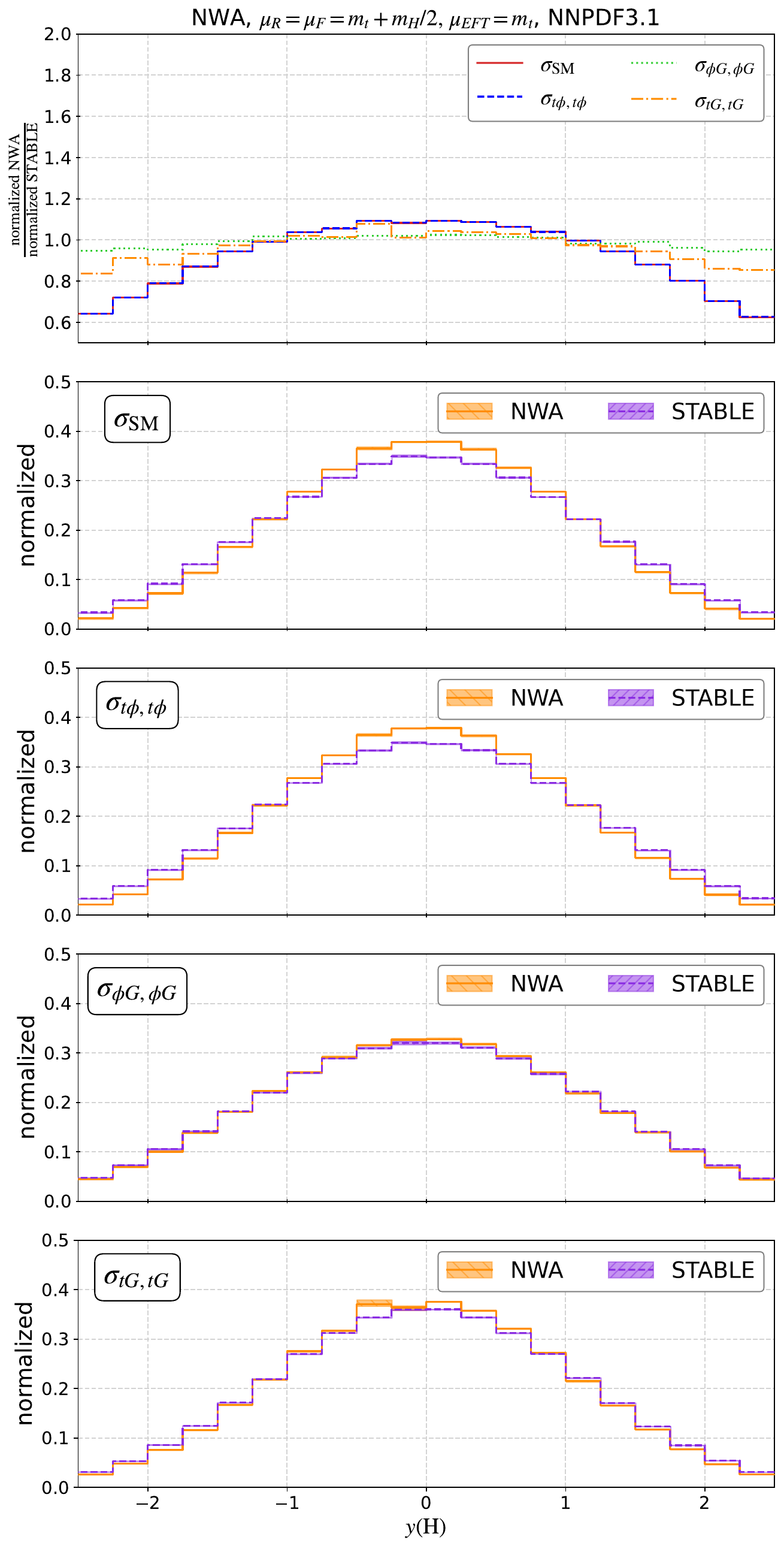}
    \end{tabular}
\caption{Normalised differential cross-section distributions at NLO in QCD as a function of $y(H)$ for $pp\to t\bar{t}H+X$ (Stable) and $pp \to e^+\nu_e\, \mu^-\bar{\nu}_{\mu}\, b\bar{b} \,H +X$ (NWA) at the LHC with $\sqrt{s}=13.6$ TeV. Results are provided for the linear and quadratic terms together with the corresponding scale uncertainties. The SM result is also plotted for comparison purposes. The upper panels display the ratio of the two predictions. Results are presented for the NNPDF3.1 PDF set and evaluated using $\mu_R=\mu_F=m_t+m_H/2$ as well as  $\mu_{EFT}=m_t$.}
\label{fig:differential_nwa_vs_stable_y(H)}
\end{figure}

We start with the $p_T(t)$ observable that is displayed in Figure \ref{fig:differential_nwa_vs_stable_pT(t)}. In this case for the $\sigma_i$ terms we observe very large shape distortions  of the order of $75\%$. Indeed, the differences between the two spectra range from $-15\%$ at the beginning of the $p_T(t)$ distribution to $+60\%$ in the tails, significantly exceeding the theoretical uncertainties estimated in these phase-space regions. This can be easily seen by examining the other panels in Figure \ref{fig:differential_nwa_vs_stable_pT(t)}. Indeed, in the bulk of the distribution theoretical uncertainties stemming from the scale dependence are of the order of $10\%$. They are similar in size for the SM case as well as for the SMEFT contributions. Such large shape differences, which are already present for this set of very inclusive cuts that we have imposed on the final states, tell us that there is a large difference in what we call the stable and reconstructed top quark. This behaviour is common to all the linear terms we have analysed. For quadratic terms the situation is more complex, since $\sigma_{\phi G, \, \phi G}$ and $\sigma_{tG, \, tG}$ deviate more significantly from the SM  case as well as from  other partial contributions.  The differences between the reconstructed and stable top quark due to SMEFT decrease in the tail of the distribution for these two contributions, but are also more pronounced for small values of $p_T(t)$ for the $\sigma_{tG, \, tG}$ contribution.

For the normalised cross-section distribution $H_T^{\rm reco}$, given in Figure \ref{fig:differential_nwa_vs_stable_Ht}, theoretical uncertainties are of the order of $5\%-15\%$ depending on the phase-space regions   for all shown predictions. In this case even larger differences between the reconstructed case and the full phase space can be observed. The shape distortions induced by the two approaches are of the order of $90\%-100\%$ for $\sigma_{t \phi}$, $\sigma_{\phi G}$ and $\sigma_{t G}$. For quadratic contributions, on the other hand, a similar situation occurs as in the case of the transverse momentum distribution of the top quark. For both $\sigma_{\phi G, \, \phi G}$ and $\sigma_{tG, \, tG}$ the differences in the tail of the distribution compared to the SM case  amount to  $25\%-30\%$.

It is interesting to examine the momentum and rapidity of the Higgs boson because in this case the quantity being compared is exactly the same for the both approaches analysed.  For the transverse momentum of the Higgs bosons, shown in  Figure \ref{fig:differential_nwa_vs_stable_pT(H)}, the distortions in the shape of the distribution are much smaller (up to $25\%$), but they are still clearly visible.  These differences are again larger than the estimated theoretical uncertainties. The latter uncertainties are less than $5\%$ at the beginning of the $p_T(H)$ spectrum for the SM and linear contributions and less than $10\%$ for the quadratic terms. 

Finally, we discuss the rapidity distribution of the Higgs boson, presented in Figure \ref{fig:differential_nwa_vs_stable_y(H)}.  In this case, the theoretical uncertainties are almost negligible and amount to about $1\%$ for all contributions. On the other hand, the differences between the NWA and stable top quark cases are much larger, especially for $|y(H)| \approx 2.5$. In these phase-space regions they differ from the SM case already for the linear terms $\sigma_{\phi G}$ and $\sigma_{t G}$. This effect is amplified for their quadratic contributions. In this case, the observed differences are of the order of $20\%$ and $30\%$ for $\sigma_{t G, \, tG}$ and $\sigma_{\phi G, \, \phi G}$, respectively.

In summary, we can conclude this section by stating that kinematic cuts, along with higher-order effects and the SMEFT operators in top-quark decays that we incorporated into our study, are important and should be consistently considered together. As we have shown in several cases, they have a significant impact on various observables, such as those that can be constructed from the top quark momenta, or simply affect the most important differential  cross-section distributions of the Higgs boson.  Both dimensionful and dimensionless observables are subject to non-trivial changes by these effects, as are different phase-space regions. These changes in the shape of the distributions can increase or decrease depending on which cuts are set on the top-quark decay products. This could have substantial impact on the extrapolation of the fiducial measurement to the full phase space in various SMEFT studies. This conclusion should not be surprising, since extrapolation to full phase space is typically performed using various Monte Carlo programs that simply do not consistently take all these effects into account.

\section{Summary and conclusions}
\label{sec:summary}

In this paper, we have presented a NLO study of the $pp \to t\bar{t}H$ process in the di-lepton decay channel at the LHC.  Relevant dimension-6 operators  $({\cal O}_{t\phi}, \, {\cal O}_{\phi G},\, {\cal O}_{tG}, \, {\cal O}_{tW})$ from SMEFT have been incorporated in the process with the help of the NWA. Furthermore, we have calculated NLO corrections to the $pp\to t\bar{t}H+X$ process with stable top quarks, aiming to use these results for comparisons with the predictions obtained in the NWA.  All results have been obtained with the help of the package \textsc{Helac-Smeft} for the LHC Run III energy of $\sqrt{s}=13.6$ TeV. We have shown predictions at the integrated fiducial cross-section level for the linear, cross and quadratic terms together with their uncertainties, including renormalisation group effects. In our study, we have simultaneously considered kinematic cuts, SMEFT contributions, and NLO QCD corrections in top-quark decays. The analysis of the fiducial phase-space regions for the $pp \to t\bar{t}H+X$ process has shown noticeable differences in the interpretation of some cross-section results obtained in the SMEFT framework compared to the $pp \to t\bar{t}H$ process with stable top quarks. Even for such a small set of effective operators, substantial  differences in the size of the SMEFT  contributions, ${\cal K}$-factors and the magnitude of theoretical uncertainties have been observed. To distinguish all these aspects, it is necessary to perform more detailed comparisons of NLO results based on different approaches to modelling top-quark decays and to study various more exclusive fiducial cuts. 

Furthermore, we have examined a variety of differential cross-section distributions of phenomenological interest. In this case we have examined  the linear and quadratic terms. We have observed that the size of the NLO QCD corrections, as shown by the differential ${\cal K}$-factors, has varied depending on the contribution analysed. However, for the linear  $\sigma_{\phi G}$ and quadratic $\sigma_{\phi G, \, \phi G}$ terms, these effects have been particularly large. We have noticed huge effects for dimensionful observables  at the beginning of the spectrum, where the NLO QCD effects could easily reach $100\%-150\%$ for both the linear and quadratic terms. However, the angular distributions have  also been affected by the large higher-order QCD corrections. These gigantic ${\cal K}$-factors could be significantly reduced using a jet veto.

To better understand the impact of fiducial cuts on the top-quark kinematics in the presence of SMEFT effects, we have presented  a comparison of differential cross-section results for $pp\to t\bar{t}H+X$ and $pp \to e^+\nu_e\, \mu^-\bar{\nu}_{\mu}\, b\bar{b} \,H +X$. In the latter case the top-quark momenta have been reconstructed from their decay products. Additionally, we have investigated how changes in the kinematics of the top quark affect the differential cross-section results for the Higgs boson. To this end, we have also examined the transverse momentum and rapidity of the Higgs boson. Our results have shown that kinematic cuts, as well as higher-order effects and SMEFT operators in top-quark decays, should be considered together because they have a significant impact on the shape of standard observables that are measured at the LHC for the $pp\to t\bar{t}H+X$ process.

In the next step, we plan to extend our study to include various dynamical scale settings for the renormalisation, factorisation and effective scales, SMEFT effects in the total top quark width, and more exclusive fiducial cuts to understand how robust our predictions are with respect to these effects. We would also like to use our theoretical findings to investigate how the combination of these effects would impact the constraints  on the operator coefficients in current and future LHC runs. Finally, we would like to include full off-shell effects, where finite top-width effects are consistently included along with  the single- and non-resonant top-quark  contributions and all interference effects.

\acknowledgments{
We would like to thank Jonathan Hermann for his contribution during the initial phase of this study, and Micha\l{} Czakon for very fruitful discussions during the final phase of the project.  

The work of M.R. and M.W. was supported by the German Research Foundation  (Deutsche Forschungsgemeinschaft - DFG) under grant 396021762 - TRR 257: \textit{Particle Physics Phenomenology after the Higgs Discovery}. 

The research of G.B. was supported by the Hellenic Foundation for Research and Innovation (H.F.R.I.) under the \textit{2nd Call for H.F.R.I. Research Projects to support Faculty Members $\&$ Researchers}. Project Number: 02674 HOCTools-II.

The authors gratefully acknowledge the computing time provided to them at the NHR Center NHR4CES at RWTH Aachen University (project number \texttt{p0020216}). This is funded by the Federal Ministry of Research, Technology and Space, and the state governments participating on the basis of the resolutions of the GWK for national high performance computing at universities.}

\appendix

%
\section{UV counterterms from the anomalous 
dimension matrix}
\label{appendix:a}
%

\begin{figure}[t!]
    \centering
    \includegraphics[width=0.9\linewidth]{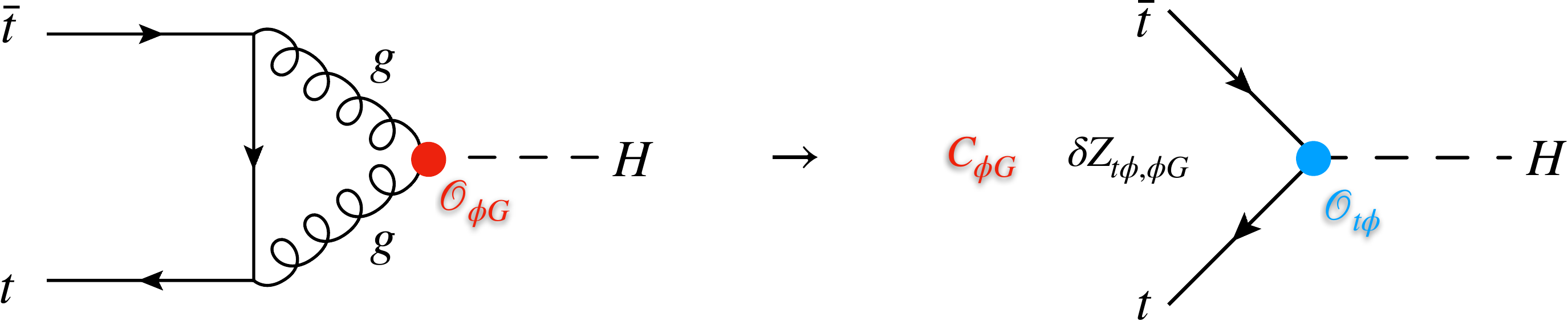}
    \caption{An example of operator mixing in SMEFT. The $ggH$ vertex is generated by the ${\cal O}_{\phi G}$ operator, but the overall diagram represents a one-loop vertex correction to the $t\bar{t}H$ vertex with the structure generated by the ${\cal O}_{t\phi}$ operator.}
    \label{fig:app_a:operator_mixing}
\end{figure}

In this Appendix we summarize the basic steps for deriving the Feynman rules for UV counterterms arising from the $\mathcal{O}_{\phi G}$ operator. We restrict our attention to the dimension-6 part of the SMEFT Lagrangian, $\mathcal{L}^{(6)}$, which is expressed in terms of a basis of operators $\mathcal{O}^{(6)}_i$ and corresponding Wilson coefficients $C_i^{(6), \,B}$:
\begin{equation}
\mathcal{L}^{(6)}=\sum_{i=1}^N \frac{C_i^{(6), \, B}}{\Lambda^2} \mathcal{O}^{(6)}_i(\{\phi_k^B\},\{\lambda_k^B\}) \,.
\label{eq:L_dim_six}
\end{equation} 
Each operator is built from a set of SM fields ($\{\phi_k^B\}$) and SM parameters ($\{\lambda_k^B\}$). We explicitly use the superscript $B$ to indicate \textit{bare} quantities. 
Consequently, we have two contributions to the renormalisation. One originates from the renormalisation constants of the SM fields and parameters, and the other is generated by the Wilson coefficients, which are expressed as
\begin{equation}
C_i^{(6),\, B} = Z_{ij} \, C_j^{(6)}  = \left( \delta_{ij}+\delta Z_{ij} \right) \, C_j^{(6)}  \,.
\label{eq:renorm_C}
\end{equation} 
The mixing of SMEFT operators requires the usage of a non-diagonal $\delta Z_{ij}$ to absorb all appearing UV singularities in the renormalisation procedure. To better illustrate this mechanism we show a simple example in Figure \ref{fig:app_a:operator_mixing}. The one-loop diagram on the left is proportional to $C_{\phi G}$. The UV divergence of the latter is cancelled by the counterterm on the right, which has the form of a UV $t\bar{t}H$ vertex. The UV vertex counterterm, generated by the operator $\mathcal{O}_{t\phi}$, carries a $C_{\phi G}$ coefficient.  This can only be accommodated by Eq. \eqref{eq:renorm_C} provided that $\delta Z_{ij}$ is non-diagonal. In general, the completeness of the operator basis ensures that for each UV counterterm required to make a given one-loop amplitude UV-finite one can always find a $\mathcal{O}^{(6)}_i$ in Eq. \eqref{eq:L_dim_six}, which generates the corresponding UV vertex.
In $\overline{\rm MS}$ scheme, the renormalisation matrix takes the form 
\begin{equation}
\delta Z_{ij} =  \frac{1}{\epsilon} \, \Delta(\mu_R,\mu_{EFT})\,\frac{1}{2}\, \gamma_{ij}  = \frac{\alpha_s}{4\pi}\frac{1}{\epsilon}\Delta(\mu_R,\mu_{EFT})\,\frac{1}{2}\, \gamma^{\textrm{QCD},1}_{ij} \,,
\label{eq:definition_deltaZij}
\end{equation} 
where 
\begin{equation}
\Delta(\mu_R,\mu_{EFT}) = \Gamma(1+\epsilon)\left(\frac{4\pi\mu_R^2}{\mu_{EFT}^2}\right)^{\epsilon}  \,,
\label{eq:definition_delta}
\end{equation}
and $\gamma_{ij}$ is the anomalous dimension matrix entering Eq. \eqref{eq:RGE_evolution_1}. In practice,  $\gamma^{\textrm{QCD},1}_{ij}$ represents the $\mathcal{O}(\alpha_s)$ term in the QCD expansion. Given the SMEFT operators considered in our NLO QCD study, we can restrict our attention to the set $\{\mathcal{O}_{t\phi}, \mathcal{O}_{\phi G}, \mathcal{O}_{tG}, \mathcal{O}_{tW}, \mathcal{O}_{tB}\}$ which is closed under renormalisation.

Some additional caution has to be exercised due to dimension-4 pieces in the SMEFT operator definitions. For example, the following operator $\mathcal{O}_{\phi G} = (\phi^\dagger\phi-\frac{v^2}{2})\,G^{a}_{\mu\nu}G^{a \, \mu\nu}$ is used rather than $\tilde{\mathcal{O}}_{\phi G} = \phi^\dagger\phi\,G^{a}_{\mu\nu}G^{a \, \mu\nu}$. The additional piece in the $\tilde{{\cal O}}_{\phi G}$ operator simply rescales the SM operator $G^a_{\mu\nu}G^{a \, \mu\nu}$ by a dimensionless number $v^2/\Lambda^2$. However, such a term is still needed to renormalise ${\cal O}_{tG}$ due to the ${\cal O}_{tG} \to {\cal O}_{\phi G}$ mixing. One can reintroduce this contribution by adding additional terms proportional to $1/\Lambda^2$ to the SM renormalisation. In Ref. \cite{Maltoni:2016yxb} the needed modification for the specific subset of $\mathcal{O}_{t\phi}$, $\mathcal{O}_{\phi G}$ and $\mathcal{O}_{tG}$ can be found. In this case all additional terms are proportional to $C_{tG}$. We are specifically interested in counterterms for ${\cal O}_{\phi G}$. Therefore, we do not have to compute such contributions and can instead rely on the counter-terms provided in the \textsc{Smeft@NLO} library for all terms proportional to $C_{tG}$. Otherwise one would have to implement the SM Lagrangian and compute the terms induced by the SMEFT contribution to the SM renormalisation as well.

\bibliographystyle{JHEP}

\providecommand{\href}[2]{#2}\begingroup\raggedright\endgroup

\end{document}